\documentclass[lettersize,journal]{IEEEtran}
\usepackage{amsmath,amsfonts}
\usepackage{algorithmic}
\usepackage{algorithm}
\usepackage{array}
\usepackage[caption=false,font=normalsize,labelfont=sf,textfont=sf]{subfig}
\usepackage{textcomp}
\usepackage{stfloats}
\usepackage{url}
\usepackage{verbatim}
\usepackage{graphicx}
\usepackage{cite}
\usepackage{xcolor} 
\usepackage{multirow}
\usepackage{booktabs}

\newtheorem{remark}{Remark}%

\begin{document}

\title{Nearest Neighbor CCP-Based Molecular Sequence Analysis}

\author{Sarwan Ali$^{1,*}$, Prakash Chourasia${^2}$, Bipin Koirala$^3$, Murray Patterson$^2$
\\
$^1$Columbia University, Irving Medical Center, New York, NY, USA
\\
$^2$Georgia State University, Atlanta, GA, USA
\\
$^3$Georgia Institute of Technology, Atlanta, GA, USA 
\\
sa4559@cumc.columbia.edu, pchourasia1@student.gsu.edu, bkoirala3@gatech.edu, mpatterson30@gsu.edu
\\
$^{*}$ Corresponding author
}



\markboth{Journal of \LaTeX\ Class Files,~Vol.~14, No.~8, August~2021}%
{Shell \MakeLowercase{\textit{et al.}}: A Sample Article Using IEEEtran.cls for IEEE Journals}


\maketitle

\begin{abstract}
Molecular sequence analysis is crucial for understanding several biological processes, including protein-protein interactions, functional annotation, and disease classification. The large number of sequences and the inherently complicated nature of protein structures make it challenging to analyze such data. Finding patterns and enhancing subsequent research requires the use of dimensionality reduction and feature selection approaches. 
Recently, a method called Correlated Clustering and Projection (CCP) has been proposed as an effective method for biological sequencing data.  
The CCP technique remains computationally expensive, despite its effectiveness for sequence visualization. Furthermore, its utility for classifying molecular sequences is still uncertain. To solve these two problems, we present a Nearest-Neighbor Correlated Clustering and Projection (CCP-NN)-based technique for efficiently preprocessing molecular sequence data. 
To group related molecular sequences and produce representative supersequences, CCP makes use of sequence-to-sequence correlations. As opposed to conventional methods, CCP does not rely on matrix diagonalization, therefore, it can be applied to a range of machine-learning problems. 
We estimate the density map and compute the correlation using a nearest-neighbor search technique. 
We perform a molecular sequence classification using CCP and CCP-NN representations to assess the efficacy of our proposed approach. Our findings show that CCP-NN considerably improves the accuracy of the classification task and significantly outperforms CCP in terms of computational runtime. 

\end{abstract}



\section{Introduction}
Molecular sequences are a crucial part of the dynamic changes in sequence composition that control biological processes. Researchers have analyzed these molecular sequences and have made progress toward a better understanding of physiology, biological development, and disease~\cite{zheng2017massively}. 
To understand various biological mechanisms and disorders, it is essential to understand how proteins interact with each other and carry out certain functions~\cite{wei2013microchip}. Unfortunately, because of their vastness, complexity, and lack of distinct patterns, it is still difficult to analyze thousands of molecular sequences at once. These obstacles hamper the development of medicines and treatments for human diseases. Deciphering the functions that proteins serve in various physiological and pathological circumstances is the goal of the field of proteomics~\cite{pandey2000proteomics}. Large sets of molecular sequences can now be made feasible by recent technical advancements to be analyzed to find patterns that could ultimately create brand new medications and vaccines~\cite{buchfink2015fast}. However, processing such massive amounts of data requires the use of cutting-edge computing tools and statistical techniques to extract pertinent information.

Furthermore, it is well known that data with a highly dimensional feature space will become sparse, making it difficult for statistical analysis to identify statistical significance and key factors~\cite{ali2025hist2vec}. To facilitate prediction, analysis, and visualization, it is preferable to minimize the dimensionality of high-dimensional data. Due to these challenges, a variety of dimensionality reduction (DR) techniques have been developed that can accurately reflect the inherent correlations in the original data in a low-dimensional space. Several linear and non-linear DR methods have been proposed, such as principal component analysis (PCA), linear discriminant analysis (LDA), Multidimensional Scaling (MDS)~\cite{mead1992review}, 
LargeVis~\cite{tang2016visualizing} is a few linear dimensionality reduction methods. Whereas kernel PCA~\cite{scholkopf1998nonlinear,ali2024elliptic,ali2023pcd2vec}, Sammon mapping~\cite{sammon1969nonlinear} and spectral embedding~\cite{van2009dimensionality} are non-linear dimensionality reduction techniques.

Correlated Clustering and Projection (CCP), a non-linear dimensionality reduction approach, computes the pairwise correlation matrix of samples and imposes a cutoff distance to prevent the global summation during the projection to increase computational efficiency~\cite{hozumi2022ccp,hozumi2023analyzing}. CCP has several benefits to offer, such as handling the dimensionality reduction of high sample sizes (because it avoids matrix diagonalization and instead solves a matrix to lower the dimensionality), employing statistical metrics such as covariances to quantify the high-level dependence between random feature vectors~\cite{kriegel2009clustering}, and it can be used in conjunction with a frequency domain method for secondary dimensionality reduction to improve the preservation of data's global structures and increase accuracy~\cite{achtert2006deriving}.

This research is focused on creating a pipeline for analyzing molecular sequences using a method based on Nearest Neighbor CCP-NN to preprocess the data and produce condensed representations that accurately reflect the original sequences. Here, instead of computing the distance between data points, we use Nearest Neighbor (NN) to compute the nearest neighbor distance. We propose a method, which we name CCP-NN, that is based on CCP to give an extra edge over the original CCP. 
The main contributions of our research are as follows:
\begin{enumerate}
    \item We propose a novel approach based on CCP to preprocess molecular sequence data, which leverages sequence-sequence correlations to generate representative super-sequences.
    \item We demonstrate the effectiveness of CCP and CCP-NN by evaluating their performance in molecular sequence classification tasks. Our results indicate that CCP-NN significantly improves the accuracy of classification compared to other methods.
    \item We offer a thorough framework for examining molecular sequence data using the capabilities of CCP. Our method facilitates effective feature selection, dimensionality reduction, and visualization.
\end{enumerate}


\section{Related Work}
\label{sec_RW}
Sequence classification is a well-researched issue in bioinformatics~\cite{Krishnan2021PredictingVaccineHesitancy,chourasia2023empowering,murad2023spike2cgr}. A phylogenetic approach is frequently used in more conventional techniques of analyzing sequencing data~\cite{minh_2020_iqtree2}, however, they are not scalable due to higher computational cost. 
To counter the issue, some machine learning (ML) approaches, including alignment-based~\cite{kuzmin2020machine,ali2022pwm2vec,ali2023solvent,tayebi2023t} and alignment-free~\cite{ali2021spike2vec} embedding approaches, have become popular for ML tasks such as classification and clustering.  Due to the extremely high dimensionality of the feature vector, these techniques do, however, also have scalability issues. 
The classification of biological sequences also makes use of the kernel matrix technique~\cite{ali2022efficient}. The Wasserstein distance (WD) is used in~\cite{shen2018wasserstein} to extract the features.   
Some efforts have been made to improve computational performance, such as Locality Sensitive Hashing (LSH)~\cite{shi2019vector}, which can train models faster and more accurately. The hash function in~\cite{argerich2016hash2vec} is used to generate an approximate word embedding for language processing. However, collisions might occur in the resulting vectors, which reduces the embedding's effectiveness. The use of bloom filters for the correction of errors in raw read data to aid a de novo assembly was demonstrated by the authors in~\cite{ellis2019dibella}. However, these methods are prone to information loss. 
Some of the most popular methods are principal component analysis (PCA)~\cite{kambhatla1997dimension}, Multidimensional Scaling~\cite{mead1992review}, kernel selection methods~\cite{chourasia2023enhancing}, and LargeVis~\cite{tang2016visualizing}. 
The curse of dimensionality and the difficulties with the analysis of outliers are another issue~\cite{indyk1998approximate}. When it comes to data noise, missing data, and poor-quality data, it can be quite unstable~\cite{indyk1998approximate}. 
Sequence analysis is still difficult to perform despite extensive effort due to the high dimensionality and quantity of data~\cite{lahnemann2020eleven}. Therefore, methods that can manage the dimensionality reduction of high sample numbers and do not involve matrix diagonalization are required.

\section{Proposed Approach}\label{sec_PA}
We divide this section into two parts, where we first discuss the original Correlated Clustering and Projection (CCP) method proposed by~\cite{hozumi2022ccp,hozumi2023analyzing}. After that, we discuss the proposed method, called Nearest Neighbor CCP-NN, in detail.

\subsection{Correlated Clustering and Projection (CCP) Algorithm}

The Correlated Clustering and Projection (CCP) algorithm~\cite{hozumi2022ccp,hozumi2023analyzing} is a data clustering and dimensionality reduction technique that identifies and groups correlated features within a high-dimensional dataset. The algorithm operates by partitioning the features into clusters based on their correlation patterns and then projecting the data onto the subspace spanned by the identified clusters. The purpose of this algorithm is to capture the underlying structure of the data by focusing on feature subsets that exhibit strong correlations, thus facilitating meaningful analysis and visualization.
Given a dataset with $N$ samples and $M$ features represented by the matrix $\mathbf{X}$, the CCP proceeds as follows: 

\paragraph{Step 1 (\textbf{Data Preprocessing})} The algorithm begins by calculating the variance of each feature to identify non-zero variance features, which are essential for meaningful clustering. 

\paragraph{Step 2 (\textbf{Selecting Features for Clustering})} The next step is to select a subset of features for clustering based on variance. The algorithm chooses the top $numCutoff$ features, which is a user-defined parameter representing the percentage of non-zero variance features to retain.

\paragraph{Step 3 (\textbf{K-Means Clustering})} The selected features are clustered using the K-Means algorithm with $n\_components - 1$ clusters, where $n\_components$ (a hyperparameter) is the desired number of clusters.

\paragraph{Step 4 (\textbf{Partitioning Non-Clustered Features})} The features that were not assigned to any cluster due to low variance are grouped into a separate cluster. This creates a new cluster that contains the remaining features.

\paragraph{Step 5 (\textbf{Computing Density Map})} For each cluster, the algorithm computes a density map to capture the correlation between features within the cluster (see Algorithm~\ref{algo_ccp_org}). The density map is estimated using either an exponential kernel or a Lorentz kernel, which are defined as follows:

    \begin{equation}
        \text{Exponential Kernel} => K(x) = e^{-\left(\frac{x}{\text{scale}}\right)^{\text{power}}}
    \end{equation}
    
    \begin{equation}
        \text{Lorentz Kernel} => K(x) = \frac{1}{1 + \left(\frac{x}{\text{scale}}\right)^{\text{power}}}
    \end{equation}

    where $x$ represents the pairwise distance between two features, \text{scale} is a scaling factor, and \text{power} is a user-defined parameter.

    In Algorithm~\ref{algo_ccp_org}, the CCP performs several steps to compute the correlation and estimate the density map based on the inputs and parameters given. It begins by calculating the pairwise distances between the selected features. If the transformation flag is set to true, it calculates the distances between the features in the input data and the reference data. Otherwise, it calculates the distances between the features in the reference data. 
    Next, if the scaling factor is not already calculated for the specified component, it proceeds to compute the average minimum distance, which is important for scaling the density estimation.
    Similarly, if the cutoff value is not already set for the specified component, it computes the average and standard deviation (SD) of the pairwise distances. The cutoff is then defined as the average plus three times the SD. It helps determine the threshold beyond which correlation values are considered negligible.
    The algorithm calculates the scaling factor by multiplying the user-defined scaling parameter with the previously computed average minimum distance. The scaling factor is used to adjust the scale of the density estimation.
    Finally, the algorithm estimates the density map, also known as the correlation, based on the calculated scaling factor and cutoff value. This estimation is done by computing the density of the pairwise distances using a density estimation function. The resulting density map represents the correlation between the selected features.

\begin{algorithm}[h!]
\centering
    \begin{algorithmic}[1]
    \scriptsize
\STATE {computeCorrelations}$(index\_component, index\_Feat, X, transform)$
    \IF{$transform$}
        \STATE $corr \gets 
         \textsc{pairwise\_distances}(X[:, index\_Feat], \text{self.X}[:, index\_Feat], \text{self.metric})$
    \ELSE
        \STATE $corr \gets 
         \textsc{pairwise\_distances}(\text{self.X}[:, index\_Feat], \text{self.metric})$
    \ENDIF
    
    \IF{$\text{self.avgmindist}[index\_component] == 0$}
        \STATE $\text{self.avgmindist}[index\_component] \gets  \textsc{computeAvgMinDistance}(corr)$
    \ENDIF
    
    \IF{$\text{self.cutoff}[index\_component] == 0$}
        \STATE $avg \gets \textsc{mean}(corr)$
        \STATE $std \gets \textsc{std}(corr)$
        \STATE $\text{self.cutoff}[index\_component] \gets avg + 3 \times std$
    \ENDIF
    
    \STATE $Scale \gets \text{self.scale} \times \text{self.avgmindist}[idx\_comp]$
    \STATE cutOffVal $\gets $ self.cutoff[ind\_comp]
    \STATE $corr \gets \textsc{computeDensity}(corr, Scale, cuttOffVal)$
    
    \STATE \textbf{return} $corr$
\end{algorithmic}
\caption{Pseudocode for computing Correlation for CCP.}
\label{algo_ccp_org}
\end{algorithm}

    
    
    
    

\paragraph{Step 6 (\textbf{Density-based Clustering})} The density map obtained for each cluster is used for a final density-based clustering. Features are assigned to clusters based on their density values, where higher density indicates a stronger correlation.

\paragraph{Step 7 (\textbf{Projection})} Finally, the data is projected onto the subspace spanned by the identified clusters. Each sample is represented as a vector of density values corresponding to each cluster. This new projected representation into the subspace, called $\phi_{CCP}$ is used as the low dimensional embedding representation for the given data point.

\begin{remark}
\small
    For a detailed description of the original CCP algorithm, readers are referred to~\cite{hozumi2022ccp,hozumi2023analyzing}.
\end{remark}

\subsection{Nearest Neighbors Based CCP}
In the nearest neighbor (NN) version of CCP (our proposed method), all steps from 1 to 7 are followed from the original CCP as described in the above subsection. The main modification is made in Step 5, where we compute the density map using the NN algorithm for efficient and fast computation of the density map. The pseudocode for computing the density map is given in Algorithm~\ref{algo_ccp_approx}, where the NearestNeighborComputeCorr function incorporates the use of an NN search technique, specifically the AnnoyIndex data structure~\cite{AnnoyIndex_website_url}, to calculate the correlation and estimate the density map. The steps involved in this process are as follows:
An AnnoyIndex is created with the specified metric, and the features from the reference data are added to the index. 
The function checks if the transformation flag is set to true, then we find a correlation by vector; else if it is set to false, we find a correlation by item.


\begin{algorithm}[h!]
\centering
    \begin{algorithmic}[1]
    \scriptsize
\STATE \textsc{NearestNeighborComputeCorr}$(idx\_component, idx\_Feat, X, transform)$
\STATE $idx \gets \textsc{Annoyidx}(\text{len}(idx\_Feat),\text{self.metric})$
\FOR{$i \gets 0$ to $\text{len}(\text{self.X}[:, idx\_Feat])$}
    \STATE $idx.add\_item(i, \text{self.X}[:, idx\_Feat][I])$
\ENDFOR
\STATE $idx.build(-1)$
        
    \IF{$transform$}
        \STATE $corr \gets [\textsc{idx.get\_nns\_by\_vector}(\text{Feat}, 1)
            \text{for Feat in } X[:, idx\_Feat]]$
    \ELSE
        \STATE $corr \gets [\textsc{idx.get\_nns\_by\_item}(i, 1) \text{ for } i \text{ in range}(\text{len}(\text{self.X}[:, idx\_Feat]))]$
    \ENDIF
    
    \STATE $corr \gets \textsc{reshape}(corr, (-1, 1))$
    
    \IF{$\text{self.avgmindist}[idx\_component] == 0$}
        \STATE $\text{self.avgmindist}[idx\_component] \gets \textsc{computeAvgMinDistance}(corr)$
    \ENDIF
    
    \IF{$\text{self.cutoff}[idx\_component] == 0$}
        \STATE $avg \gets \textsc{mean}(corr)$
        \STATE $std \gets \textsc{std}(corr)$
        \STATE $\text{self.cutoff}[idx\_component] \gets avg + 3 \times std$
    \ENDIF

    \STATE $Scale \gets \text{self.scale} \times \text{self.avgmindist}[idx\_component]$
    \STATE cuttOffVal $\gets $ self.cutoff[idx\_component]
    \STATE $corr \gets \textsc{computeDensity}(corr, Scale, cuttOffVal)$
    
    \STATE \textbf{return} $corr$

\end{algorithmic}
\caption{Pseudocode for computing correlation for Nearest Neighbor CCP.}
\label{algo_ccp_approx}
\end{algorithm}

\begin{remark}
\small
    In NN-CCP, we utilize the AnnoyIndex data structure, which is an efficient implementation of NN. Instead of computing the exact distances between data points, AnnoyIndex builds an index that allows for fast retrieval of NN. This significantly speeds up the computation of pairwise distances, making it useful for large datasets and high-dimensional spaces.
\end{remark}

Once the AnnoyIndex is constructed, the function retrieves the NN for each feature in the input data. If the transformation flag is true, it retrieves the NN based on the features in the input data. If the flag is false, it retrieves the NN based on the features in the reference data.
The retrieved nearest neighbors (NN) are reshaped into a proper format for further processing.
Similar to Algorithm~\ref{algo_ccp_org}, the function checks if the scaling factor needs to be computed for the specified component. If the scaling factor is not already calculated, it proceeds to compute the average minimum distance, which is crucial for scaling the density estimation.

If the cutoff value is not set, it computes the average and SD of the correlation values obtained from the NN. The cutoff is then defined as the average plus three times the SD. This cutoff value helps determine the threshold beyond which correlation values are considered negligible.
The scaling factor is calculated by multiplying the user-defined scaling parameter with the previously computed average minimum distance. This scaling factor is used to adjust the scale of the density estimation.

Finally, the function estimates the density map, also known as the correlation, by applying the density estimation function to the correlation values obtained from the NN. This density map represents the correlation between the selected features, taking into account the scaling factor and the cutoff value.
In summary, Algorithm~\ref{algo_ccp_approx} incorporates the use of NN to calculate the correlation and estimate the density map. It involves constructing an AnnoyIndex, retrieving the NN, reshaping the obtained correlations, computing the scaling factor and cutoff value, and estimating the density map based on these values. This approach allows for efficient computation of correlations and density estimation, particularly for high-dimensional data.

After computing the density map, Step 6 and Step 7 are followed similarly to the original CCP as described above (i.e., used to compute $\phi_{CCP}$). After Step 7, we get the new projected representation into the subspace, called $\phi_{CCP\_NN}$ (where NN stands for Nearest Neighbor), which is used as the low dimensional embedding representation for the given data point.


\subsection{Algorithm Complexity}
In this section, we discuss the time and space complexity of CCP and CCP-NN.
\subsubsection{CCP}
To begin, the computation of variance along each feature axis takes \(\mathcal{O}(N \cdot M)\), where \(N\) is the number of samples and \(M\) is the number of features. Sorting the variance values to select the top \(f \ (\text{where} \ f \leq M)\) features takes \(\mathcal{O}(f \log f)\). Following this, the K-Means clustering step generally depends on the number of iterations \(n_{\text{iter}}\), the number of features \(f\), the number of clusters \(n_c\), and the number of samples \(N\). This step has a time complexity of \(\mathcal{O}(n_c \cdot f \cdot n_{\text{iter}} \cdot N)\).

To this end, the overall time complexity of the algorithm can be expressed as:
\begin{equation}
    \mathcal{O}(N \cdot M + f \log f + n_c \cdot f \cdot n_{\text{iter}} \cdot N)
\end{equation}

Since the K-Means setup dominates the variance computation and sorting steps, this simplifies to:
\begin{equation}
    \mathcal{O}(n_c \cdot n_{\text{iter}} \cdot f \cdot N)
\end{equation}

The total space complexity is dominated by the size of the input matrix and the memory used for K-Means clustering. Therefore, the overall space complexity is:
\begin{equation}
    \mathcal{O}(N \cdot M)
\end{equation}

Furthermore, in order to compute the density map to capture the correlation between features within a cluster, a pairwise distance calculation is required, which is the most computationally expensive operation. If \(f_i\) is the number of features selected for the \(i\)-th cluster, the pairwise distance calculation takes \(\mathcal{O}(N^2 \cdot f_i)\). This is because the pairwise distance calculation compares each of the \(N\) samples with every other sample across the \(f_i\) features. Summing over all \(n_c\) components, the total time complexity becomes:

\begin{equation}
    \mathcal{O}\left(\sum_{i=1}^{n_c} N^2 \cdot f_i \right) = \mathcal{O}(N^2 \cdot f)
\end{equation}
where \(f = \sum_{i=1}^{n_c} f_i\) is the total number of features in all components. The space complexity associated with this procedure is \(\mathcal{O}(N^2)\). Therefore, the overall complexity is; 
\begin{equation}
    \mathcal{O}\big(N\cdot f (n_c \cdot n_{iter} + N)\big)
\end{equation}
and the space complexity is; 
\begin{equation}
    \mathcal{O}\big(N(M+N)\big)
\end{equation}

\subsubsection{CCP-NN}
On the other hand, if we were to use Approximate Nearest Neighbor (ANN) via the Annoy Index as a proxy for pairwise distance computation, we would benefit significantly in terms of computation time. 
Building an Annoy Index takes \(\mathcal{O}(N \log N \cdot f)\), because the algorithm builds a forest of random projection trees. Querying the Annoy Index for nearest neighbors is approximately \(\mathcal{O}(\log N)\) per query, and since this is repeated for all \(N\) samples, the complexity of querying is \(\mathcal{O}(N \log N)\).


With $n_c$ total clusters, the total time complexity becomes 
\begin{equation}
    \mathcal{O}(n_c \cdot n_{\text{iter}} \cdot f \cdot N) + \mathcal{O}\big(\sum_{i=1}^{n_c} N \log N \cdot f_i \big) = \mathcal{O}\big(N\cdot f (n_c \cdot n_{iter} + \log N)\big)
\end{equation}
and the space complexity is; 
\begin{equation}
    \mathcal{O} \big(N (M + \log N \cdot f)\big)
\end{equation}
Clearly, CCP-NN has an advantage over CCP in terms of speed and memory requirements. 

\subsection{Convergence Analysis}
The key step in CCP-NN is the estimation of the density map, which is used to capture the correlation between features within a cluster. This is done using the nearest neighbor search, especially using the Annoy Index data structure~\cite{AnnoyIndex_website_url}.

Let $X \in \mathbb{R}^{N \times M}$ be a dataset with $N$ samples and $M$ features (dimension). Given a set of features $\{x_i\}_{i=1}^M$ within a cluster, the density estimate at each point $x_i$ is determined by the proximity of its nearest neighbors. The Annoy Index facilitates the retrieval of the nearest neighbors, denoted by $\mathcal{N}_k(x_i)$, where $k$ is the number of neighbors considered.

Assume the following;
\begin{itemize}
    \item $X \sim \mathbb{P}(X)$ with a well defined density function $p(x)$
    \item Nearest neighbor search in CCP-NN provides a close approximation to the true nearest neighbors, with an error margin $\epsilon$
\end{itemize}

Now, let $\tilde{\mathcal{N}}_k(x_i)$ represent the nearest neighbors returned by the Annoy Index, and let $\mathcal{N}_k$ represent the true nearest neighbors. The accuracy of the Annoy Index guarantees that: 
\begin{equation}
    \mathbb{E}[\vert\vert \mathcal{N}_k(x_i) - \tilde{\mathcal{N}}_k(x_i)\vert\vert] \leq \epsilon
\end{equation}
where $\epsilon > 0$ depends on the dimensionality of the data and the parameters of the Annoy Index.

Given the convergence of the nearest neighbor search, we can now analyze the consistency of the density estimation process. Let $\hat{p}(x_i)$ be the density estimate at $x_i$ obtained by CCP-NN, and let $p(x_i)$ be the true density. The density estimate is given by; 

\begin{equation}
    \hat{p}(x_i) = \frac{1}{hk} \sum_{x_j \in \hat{\mathcal{N}}_k(x_i)} K\bigg(\frac{x_i - x_j}{h}\bigg)
\end{equation}

where $K$ is a kernel function and $h$ is a bandwidth parameter.
Using Triangle-Inequality:

\begin{equation}
    \vert \hat{p}(x_i) - p(x_i) \vert \leq \vert \hat{p}(x_i) - p(x_i, \hat{N}_k(x_i)) \vert + \vert p(x_i, \hat{N}_k(x_i)) - p(x_i) \vert
\end{equation}
where $p(x_i, \hat{N}_k(x_i))$ denotes the density estimate using the true nearest-neighbor. We take the expectation on both sides, and the first term of the R.H.S in the inequality above can be bounded by $\mathcal{O}(\epsilon)$ due to the accuracy of the Annoy Index. The second term can be bounded using the convergence results of the standard kernel density estimation~\cite{yenchic_lec7_url}.

Therefore, the upper bound on the error estimate is:
\begin{equation}
    \mathbb{E}\big[\vert \hat{p}(x_i) - p(x_i) \vert \big] \leq \mathcal{O}(\epsilon + h^4 + 1/kh)
\end{equation}

Our method produces a number of features equal to 20, which are selected using standard cross-validation. Each CCP component corresponds to one output feature, where each feature represents the aggregated correlation/density within a specific subset of the original feature space. The original high-dimensional input is therefore reduced to 20 CCP-derived features that capture the local density structure of different feature groupings.

\section{Experimental Setup}
\label{sec_ES}
In this section, we describe the different datasets used for the experiments. We also go through the baseline methods and evaluation metrics we use for the classification. 
A Windows 10 $64$-bit machine with an Intel(R) Core i5 processor operating at $2.10$ GHz and $32$ GB of memory is used for all experiments. 
We used standard cross-validation to tune the hyperparameters.

\begin{figure}[!t]
\centering
\subfloat[\raggedright \scriptsize OHE (CCP)]{\includegraphics[scale=0.065]{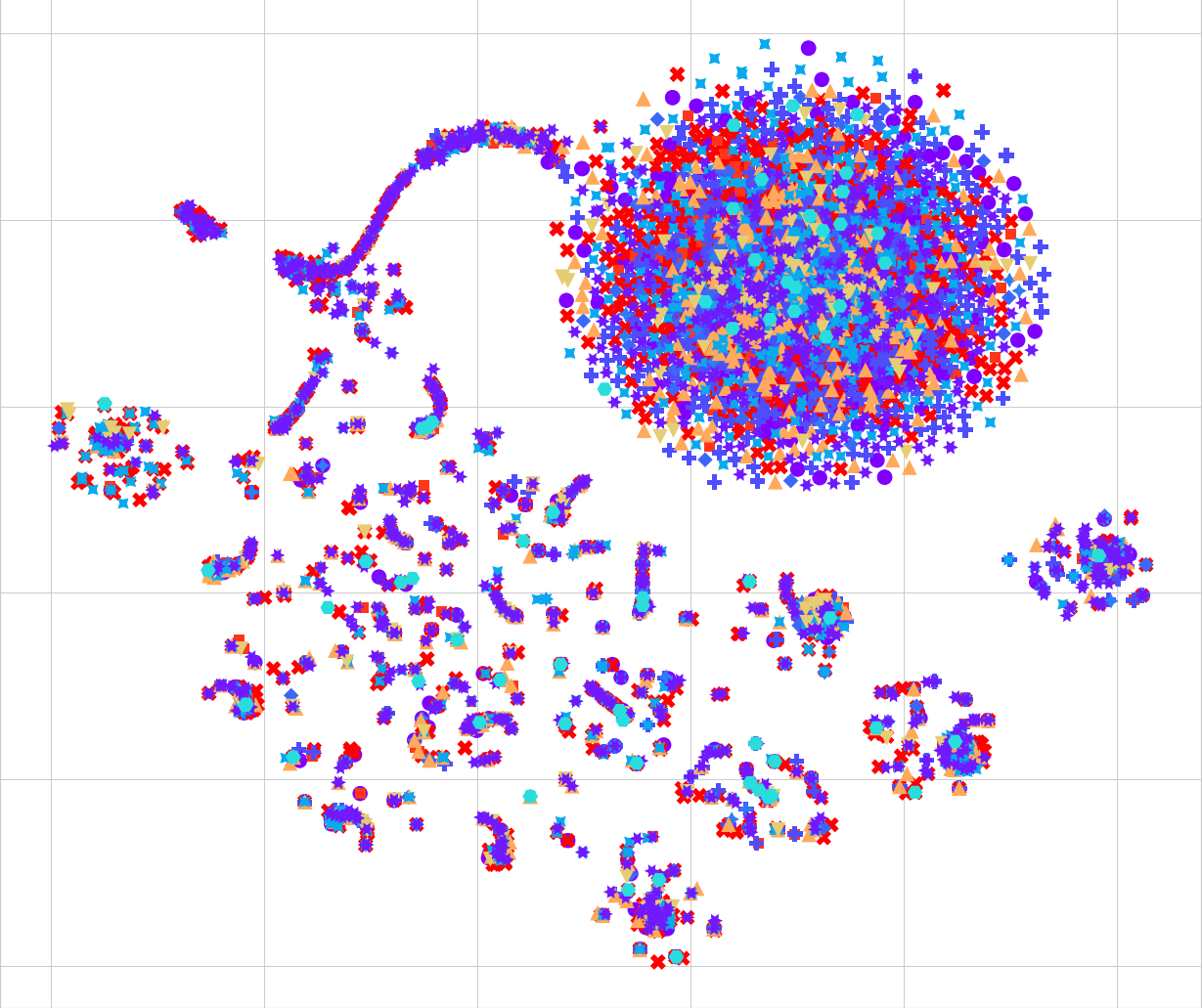}%
\label{OHE_CCP_protein_subcellular}}
\hfil
\subfloat[\raggedright \scriptsize Spike2Vec (CCP)]{\includegraphics[scale=0.065]{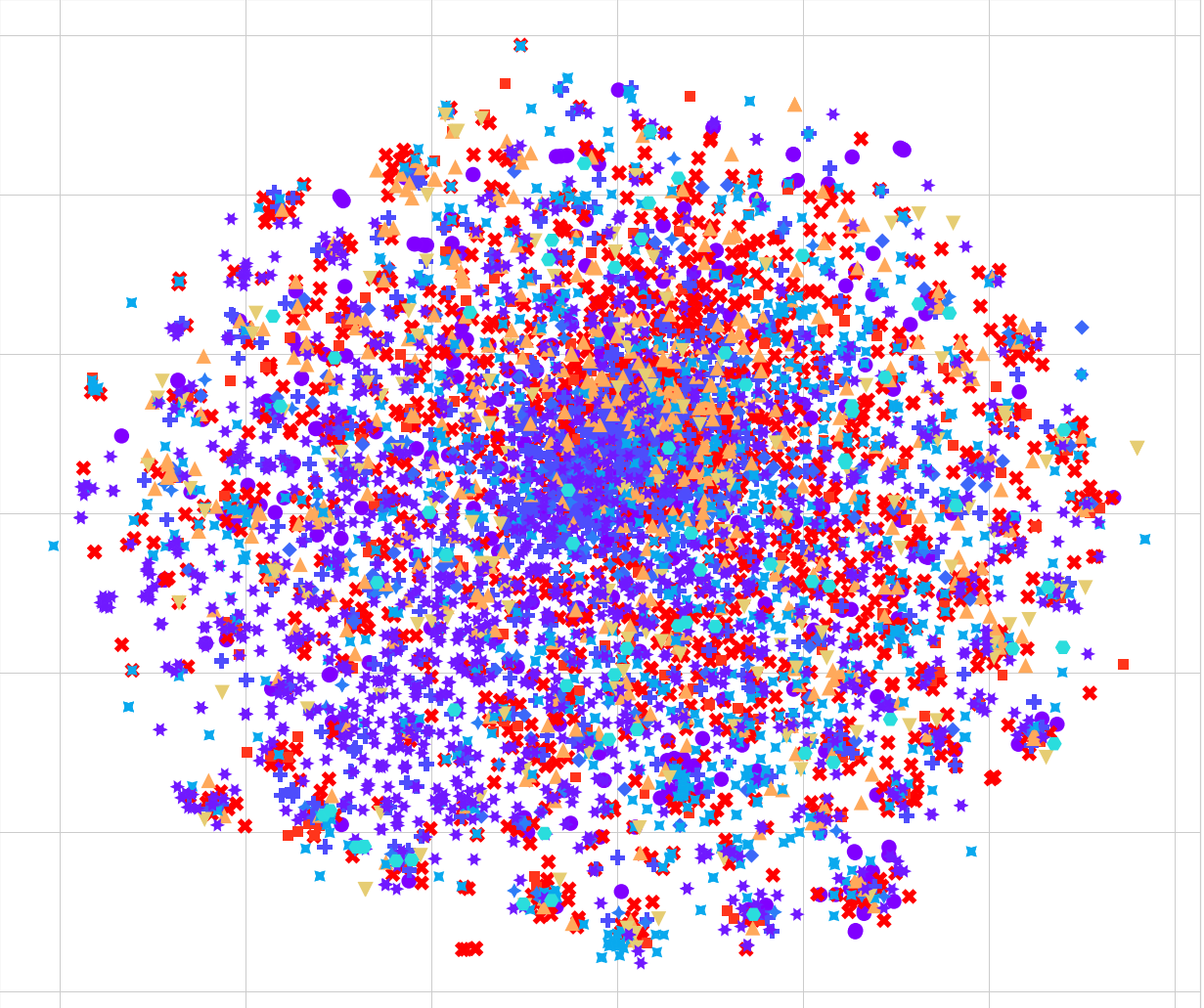}%
\label{Spike2Vec_CCP_protein_subcellular}}
\hfil
\subfloat[\raggedright \scriptsize PWM2Vec (CCP)]{\includegraphics[scale=0.065]{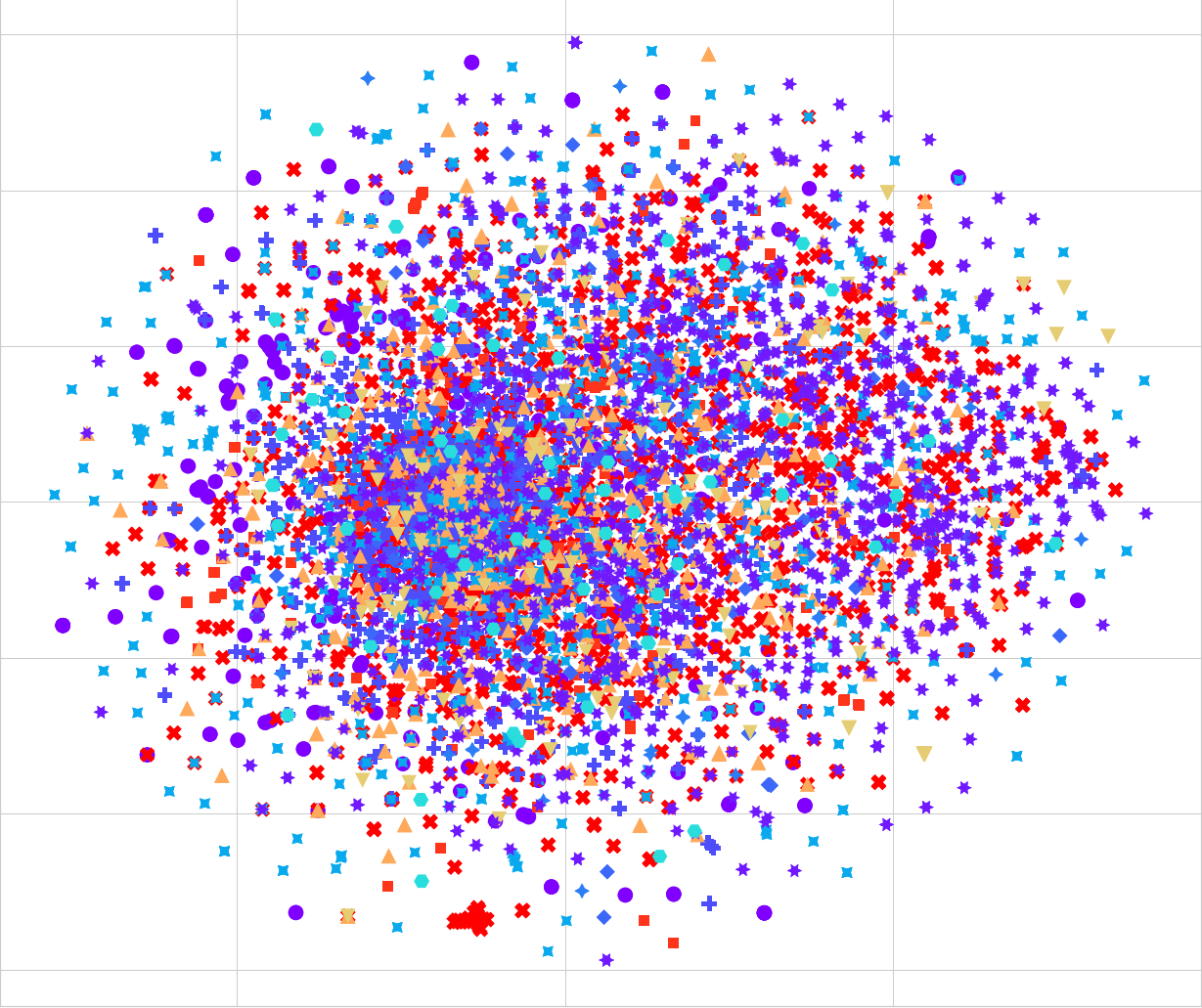}%
\label{PWM2Vec_CCP_protein_subcellular}}
\\
\subfloat[\raggedright \scriptsize Autoencoder (CCP)]{\includegraphics[scale=0.065]{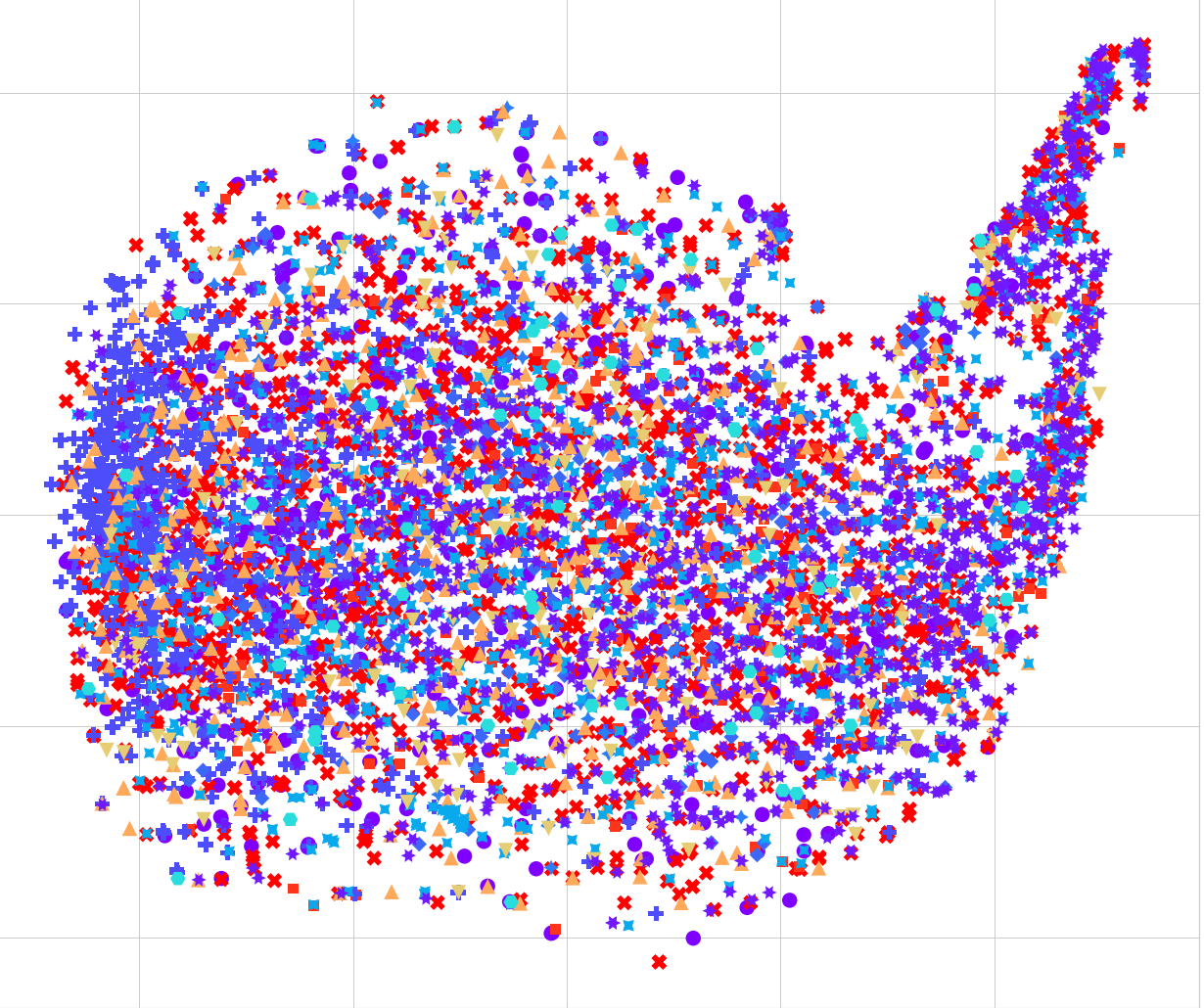}%
\label{Autoencoder_CCP_protein_subcellular}}
\hfil
\subfloat[\raggedright \scriptsize OHE (CCP-NN)]{\includegraphics[scale=0.065]{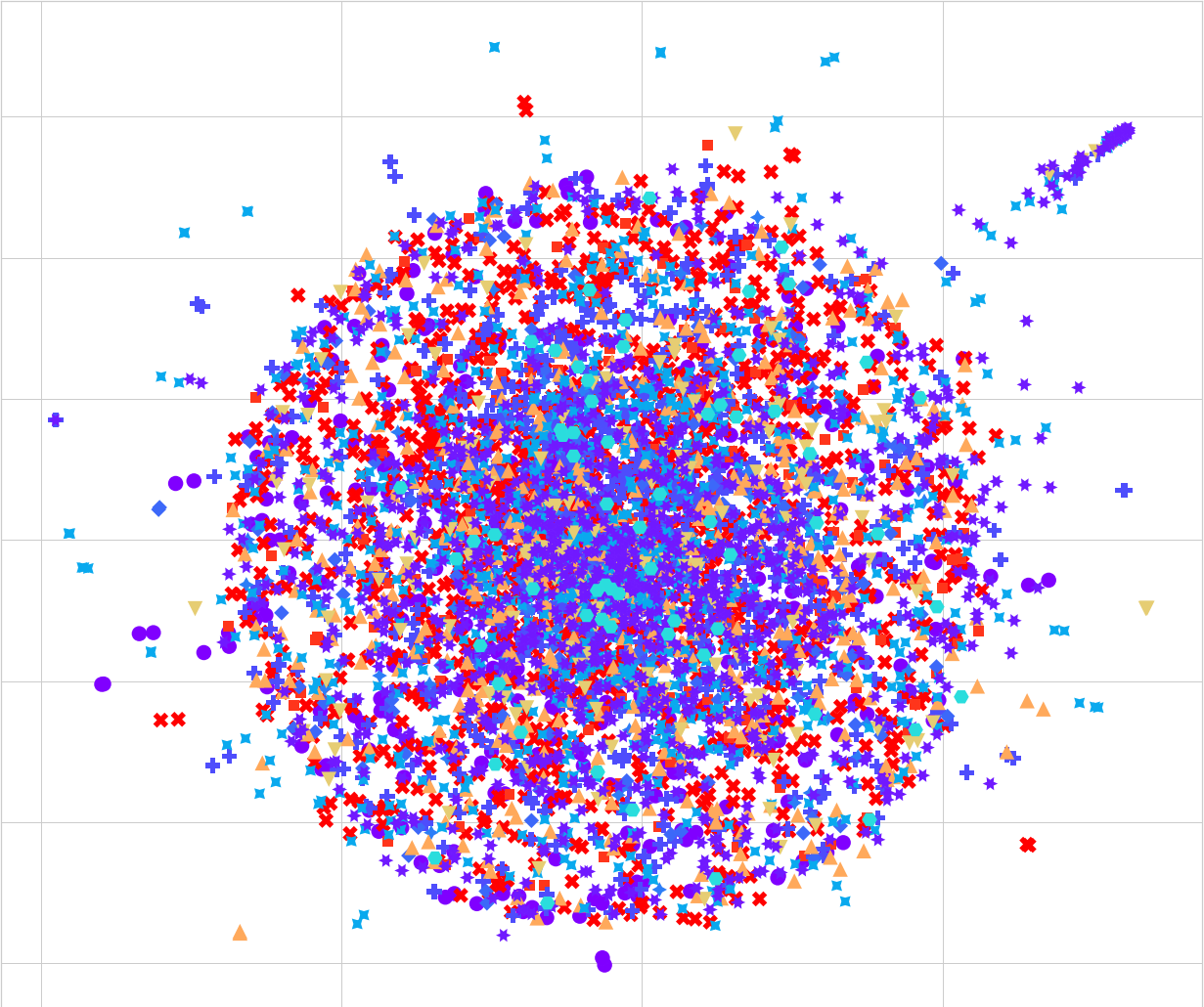}%
\label{OHE_Approx}}
\hfil
\subfloat[\raggedright \scriptsize Spike2Vec CCP-NN]{\includegraphics[scale=0.065]{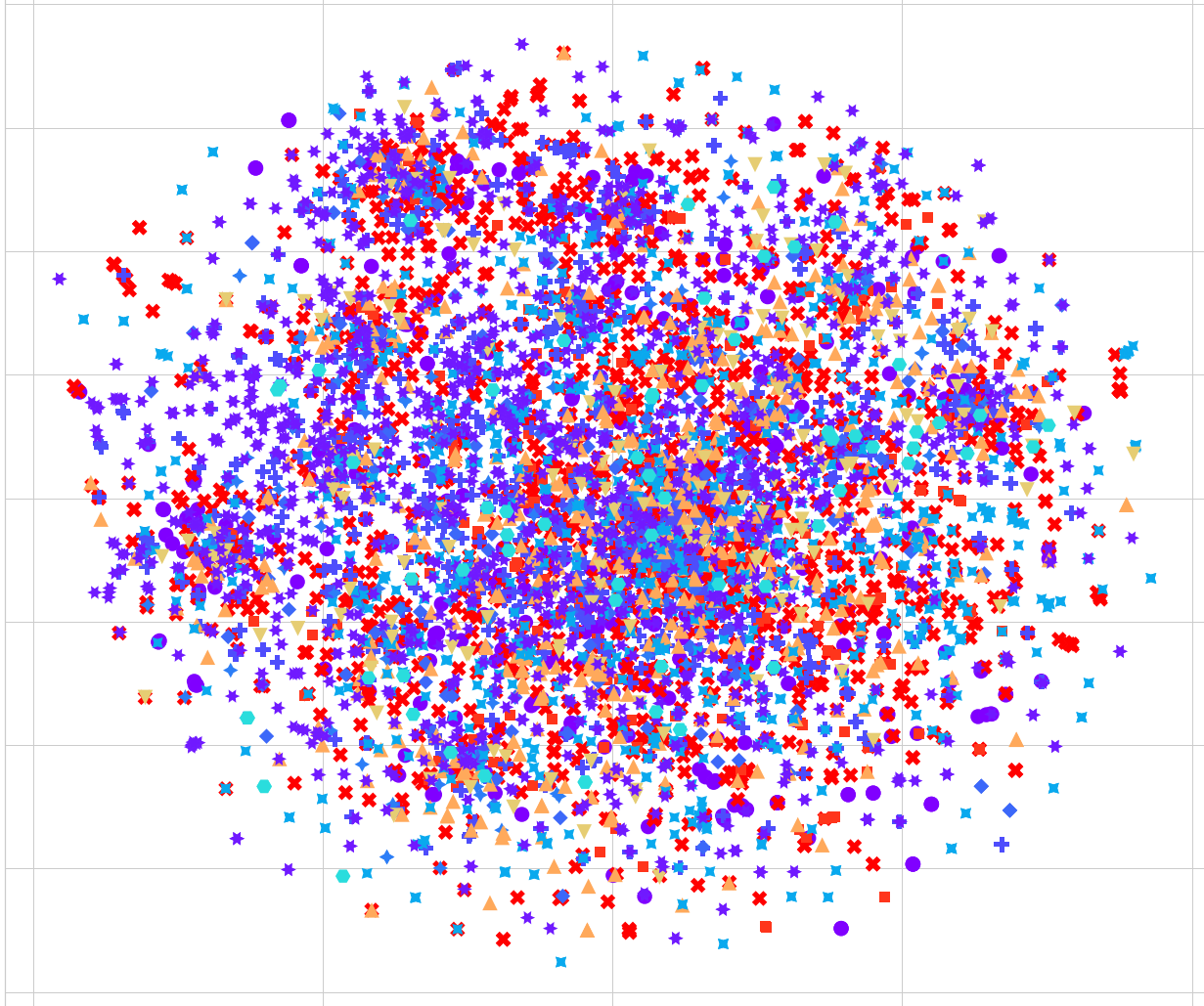}%
\label{Spike2Vec_Approx_protein_subcellular}}
\\
\subfloat[\raggedright \scriptsize PWM CCP-NN]{\includegraphics[scale=0.065]{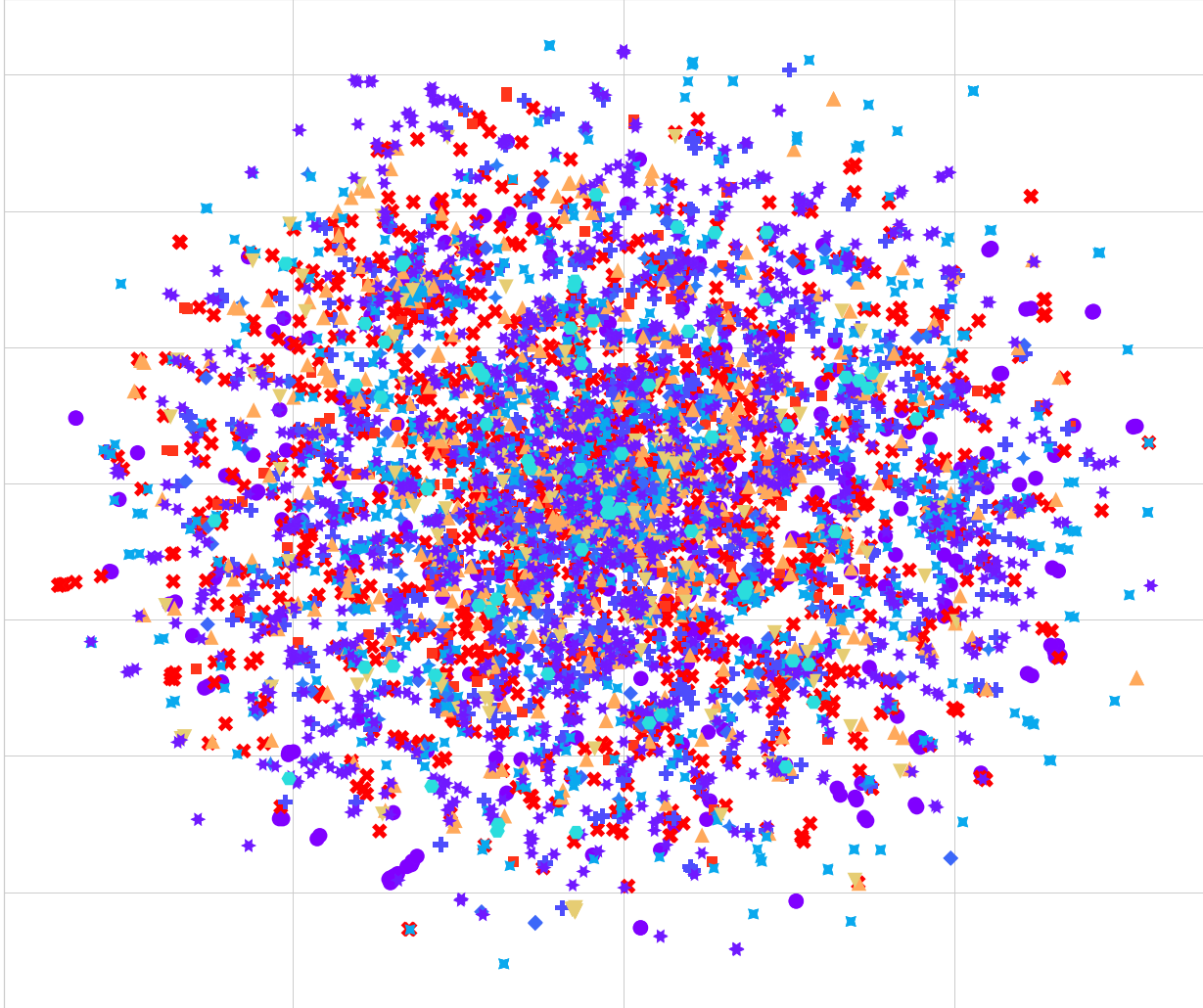}%
\label{PWM_Approx_protein_subcellular}}
\hfil
\subfloat[\raggedright \scriptsize Auto-En. CCP-NN]{\includegraphics[scale=0.065]{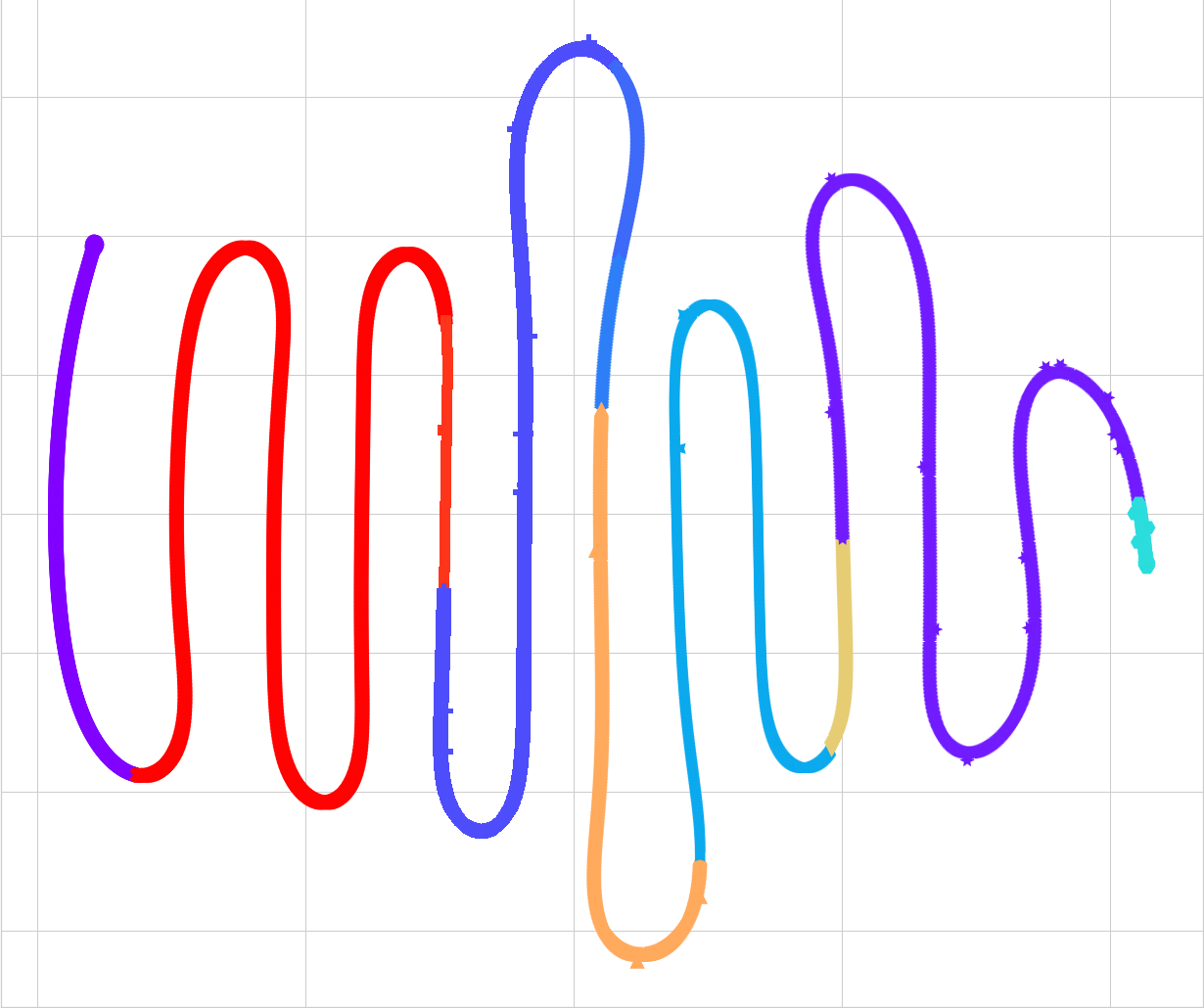}%
\label{Autoencoder_Approx_protein_subcellular}}
\\
\includegraphics[scale=0.26]{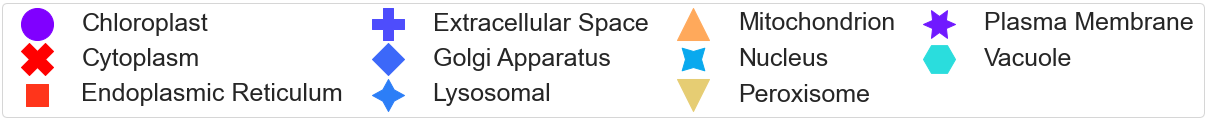}
\caption{t-SNE plots (\textbf{Protein Subcellular Data}) for different structure embeddings and Clustering and Projection methods (CCP and CCP-NN). The figure is best seen in color.}
\label{fig_tsne_protein_subcellular}
\end{figure}

\begin{figure}[!t]
\centering
\subfloat[\raggedright \scriptsize OHE (CCP)]{\includegraphics[scale=0.065]{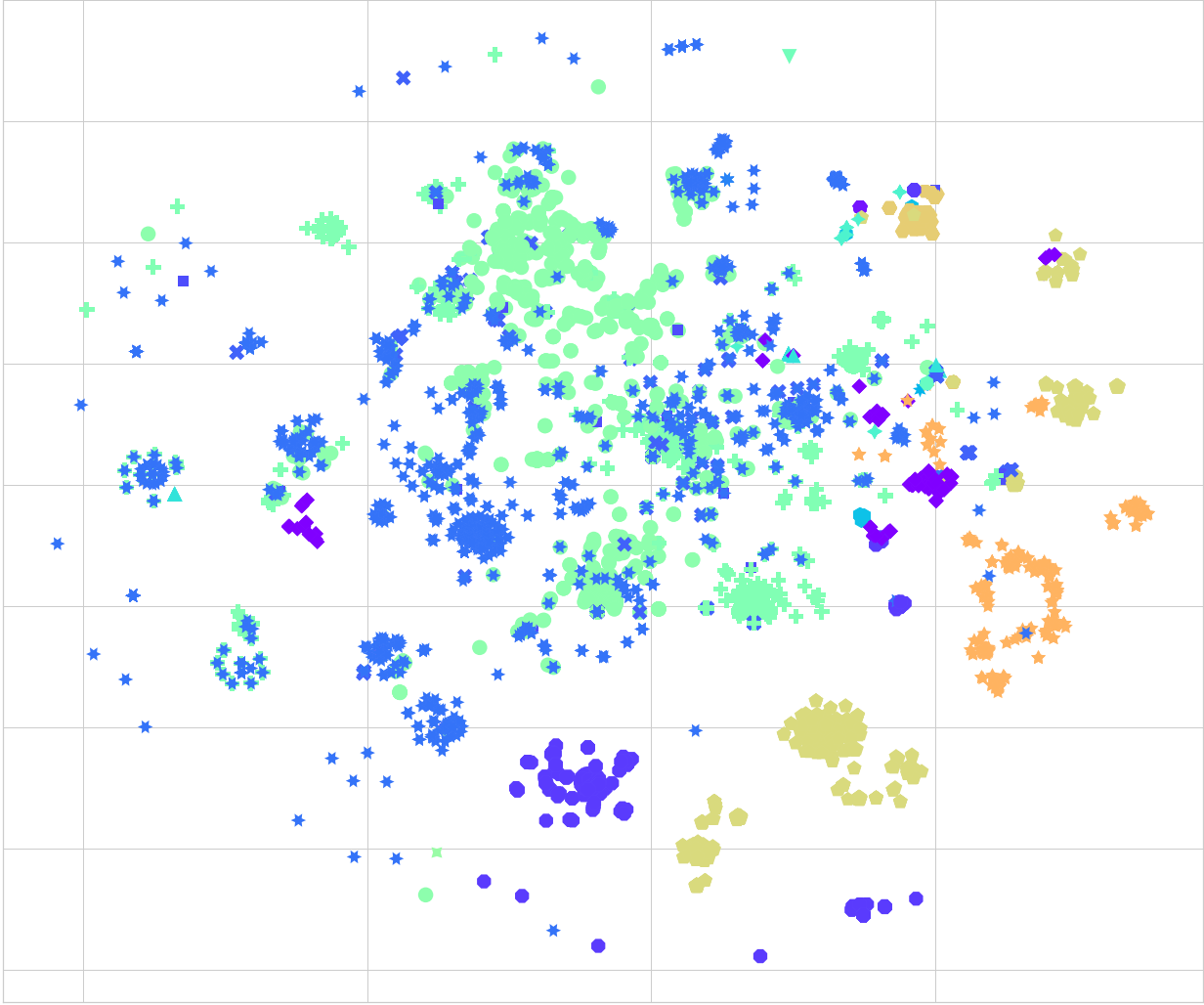}%
\label{OHE_CCP_Host}}
\hfil
\subfloat[\raggedright \scriptsize Spike2Vec (CCP)]{\includegraphics[scale=0.065]{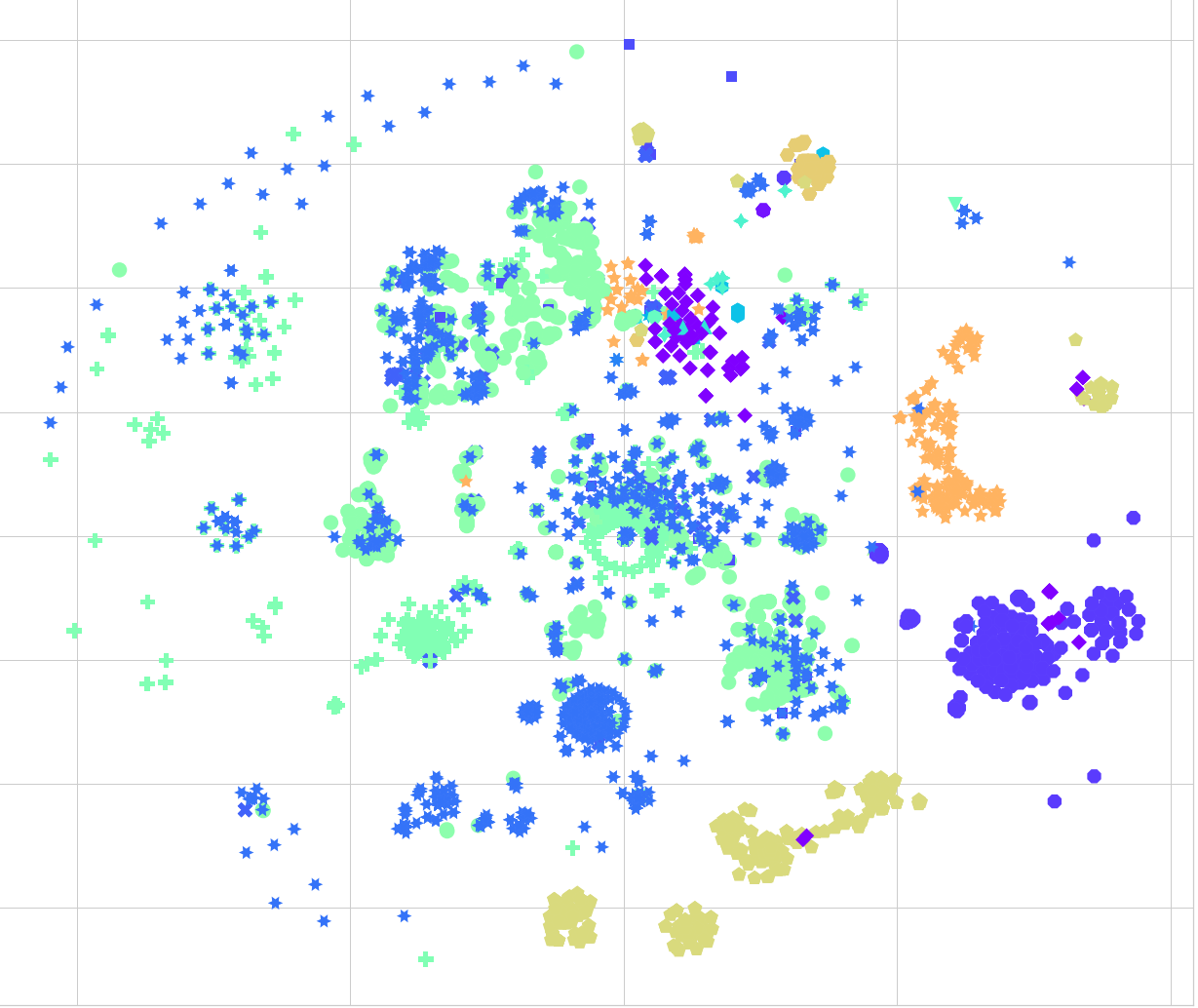}%
\label{Spike2Vec_CCP_Host}}
\hfil
\subfloat[\raggedright \scriptsize PWM2Vec (CCP)]{\includegraphics[scale=0.065]{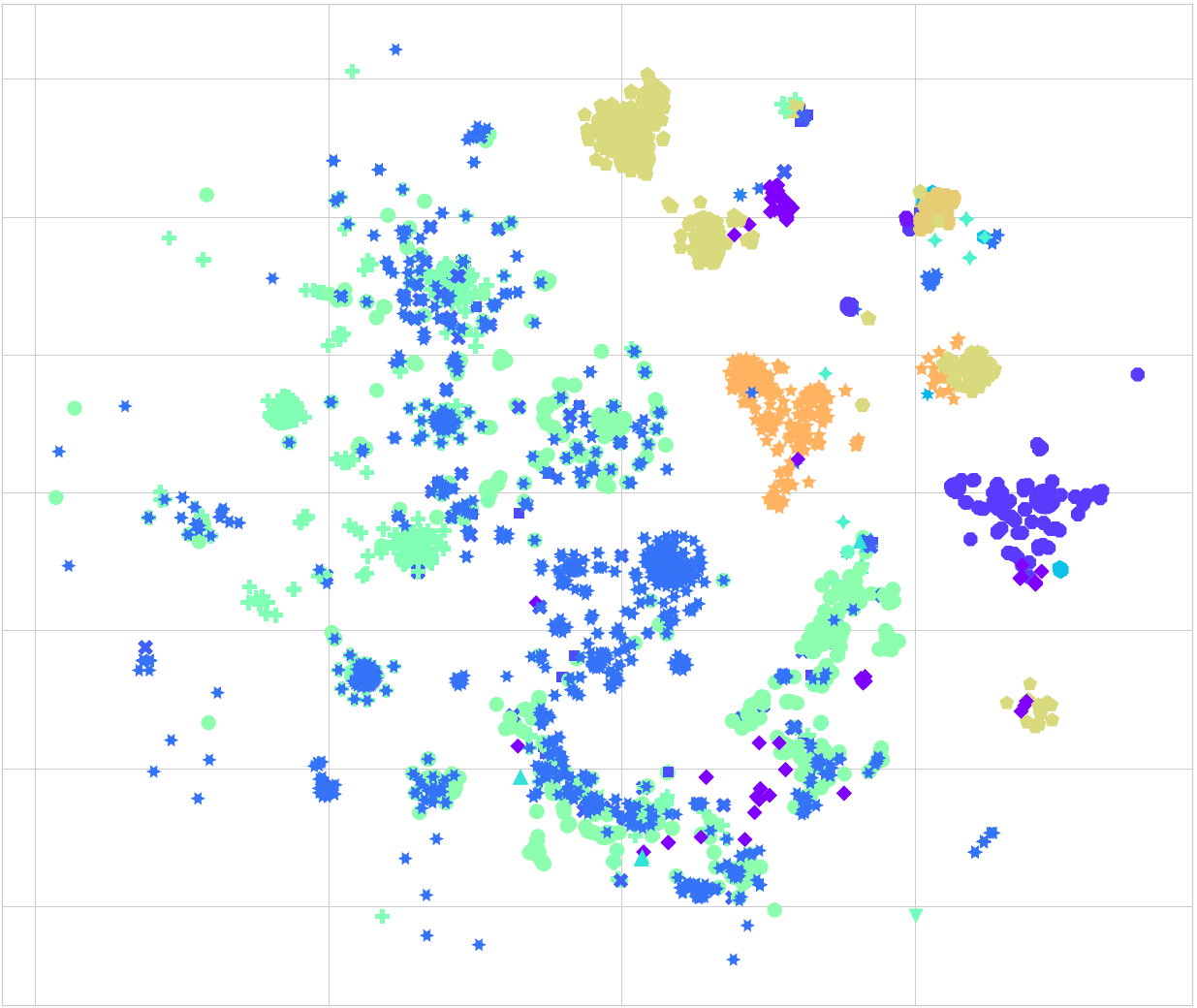}%
\label{PWM2Vec_CCP_Host}}
\\
\subfloat[\raggedright \scriptsize Autoencoder (CCP)]{\includegraphics[scale=0.065]{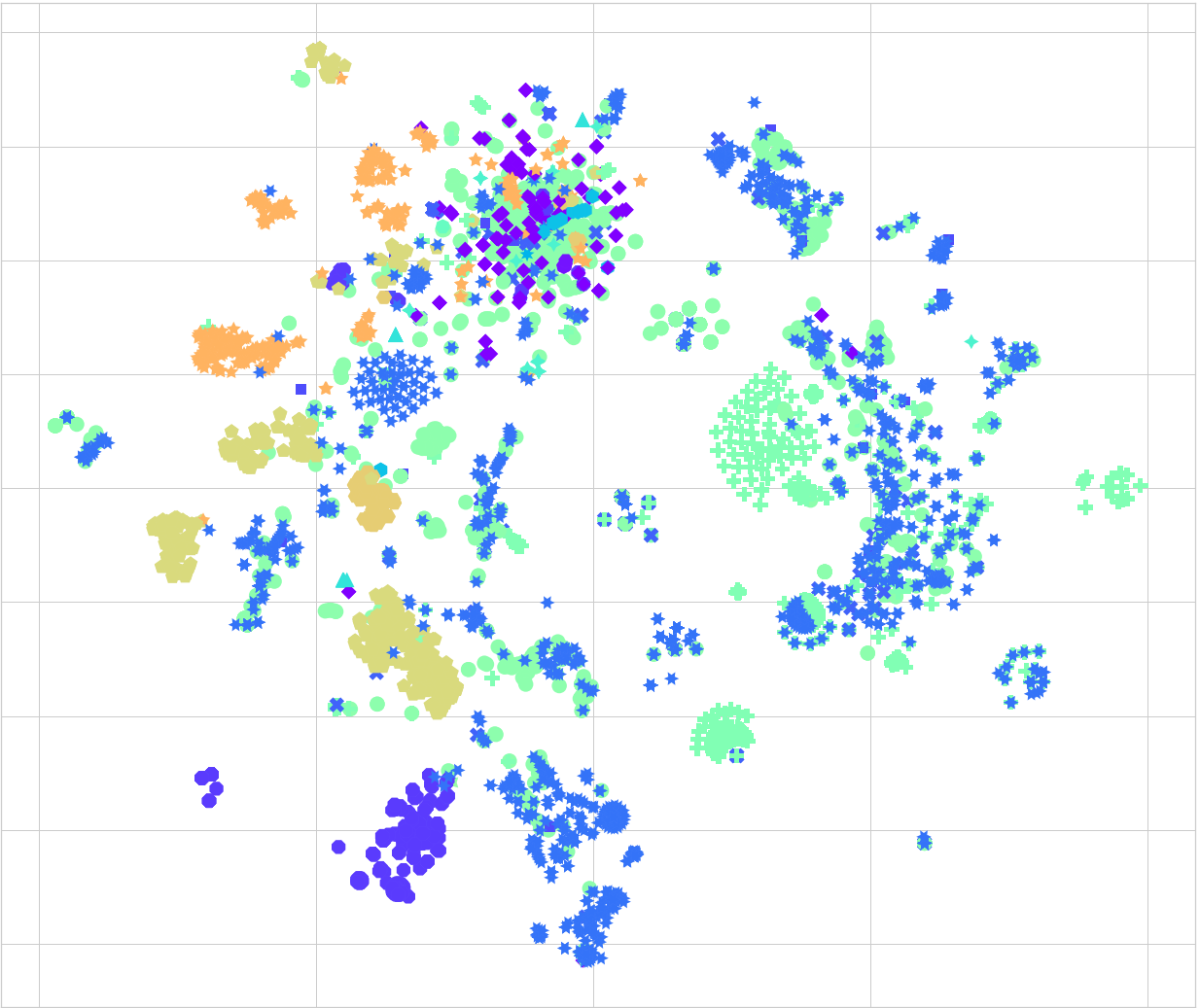}%
\label{Autoencoder_CCP_Host}}
\hfil
\subfloat[\raggedright \scriptsize OHE (CCP-NN)]{\includegraphics[scale=0.065]{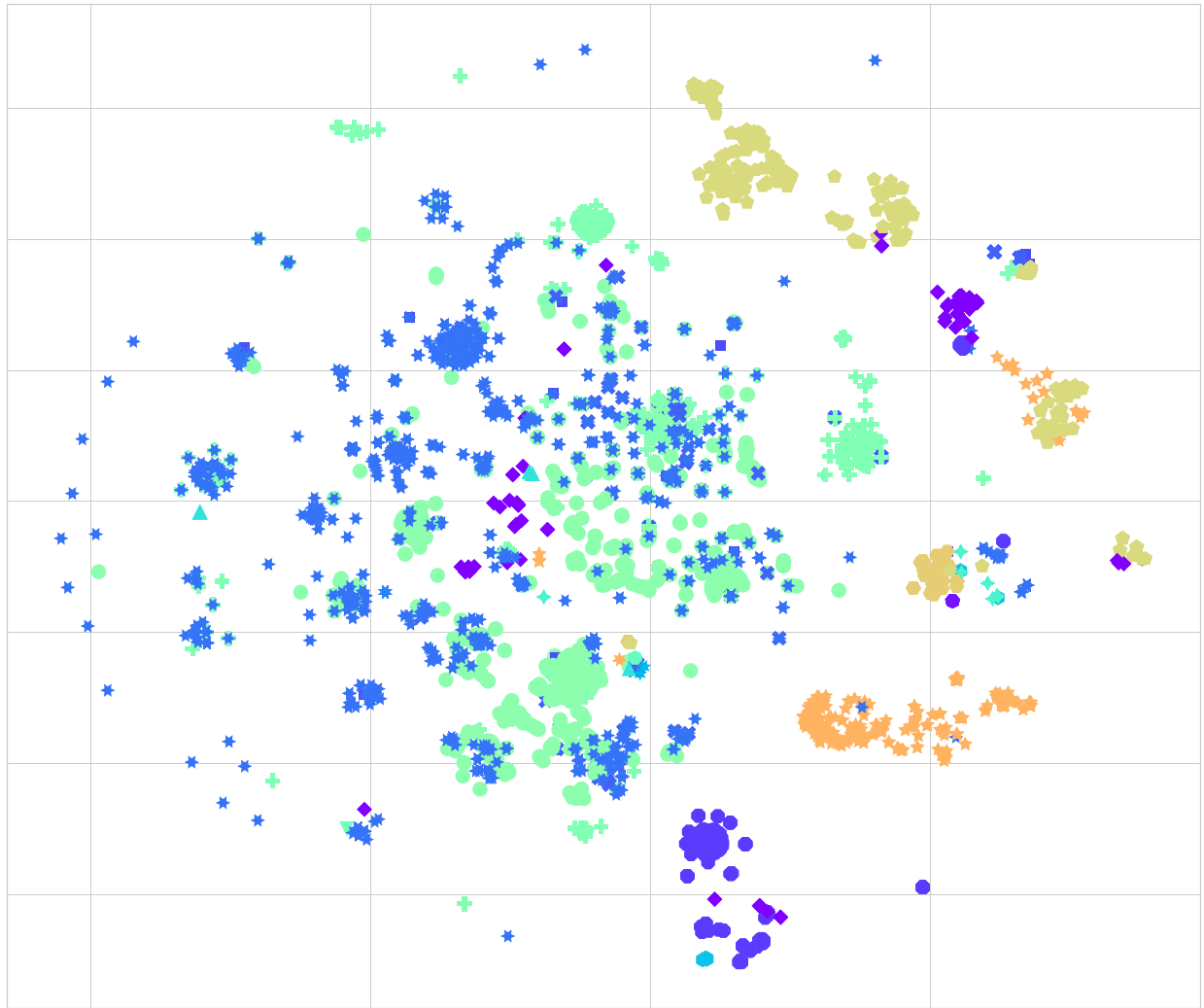}%
\label{OHE_Approx_Host}}
\hfil
\subfloat[\raggedright \scriptsize Spike2Vec CCP-NN]{\includegraphics[scale=0.065]{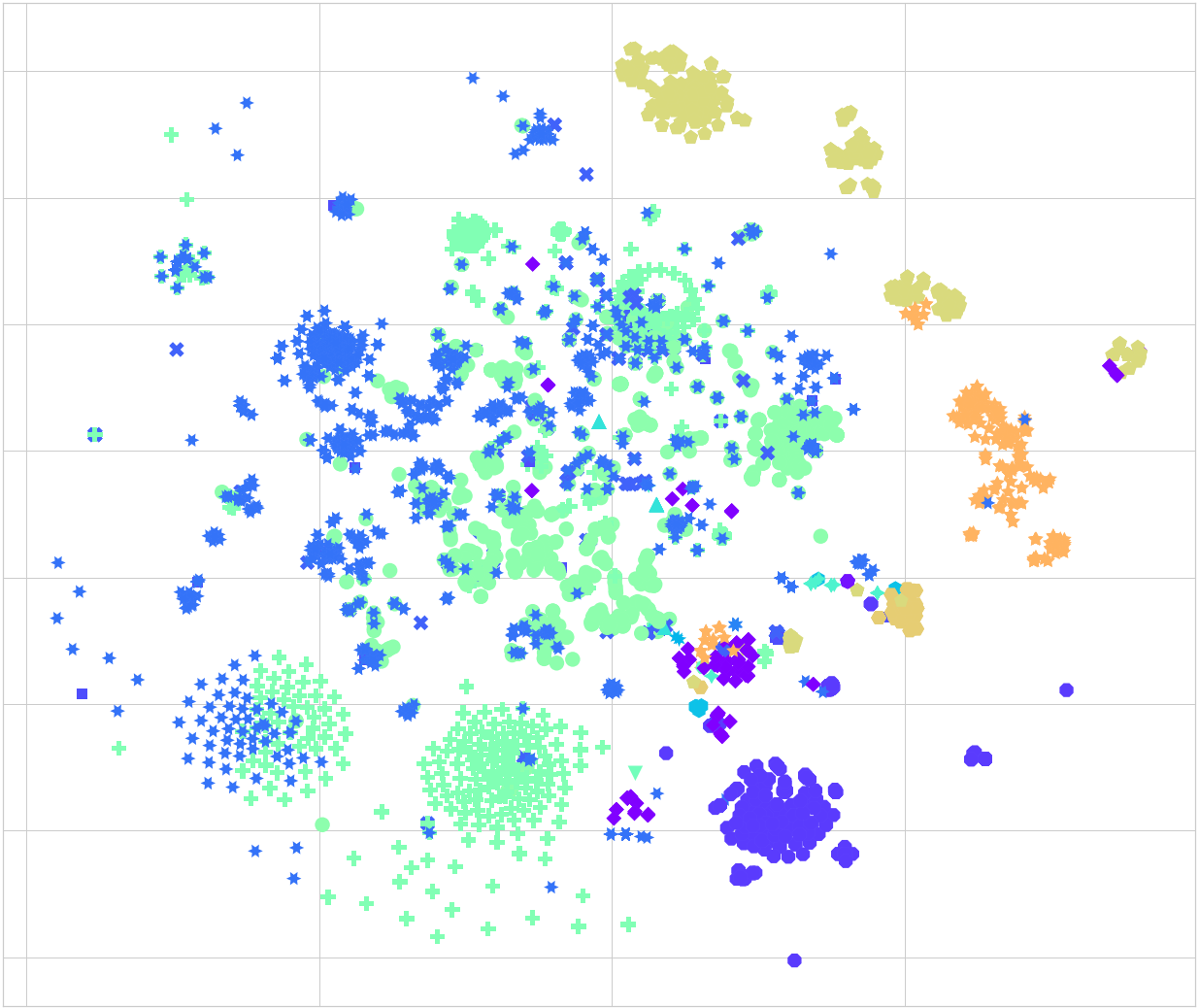}%
\label{Spike2Vec_Approx_Host}}
\\
\subfloat[\raggedright \scriptsize PWM CCP-NN]{\includegraphics[scale=0.065]{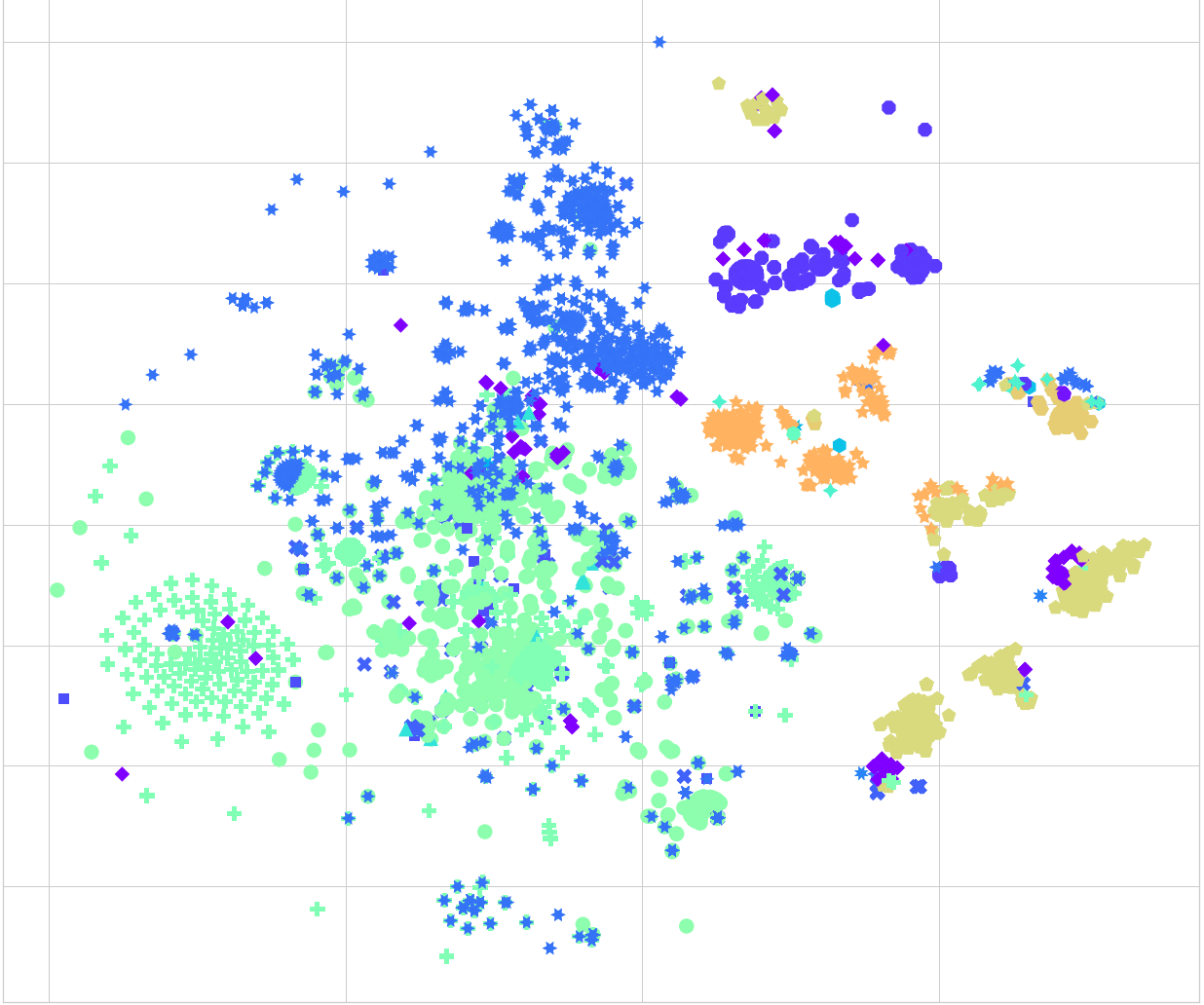}%
\label{PWM_Approx_Host}}
\hfil
\subfloat[\raggedright \scriptsize Auto-En. CCP-NN]{\includegraphics[scale=0.065]{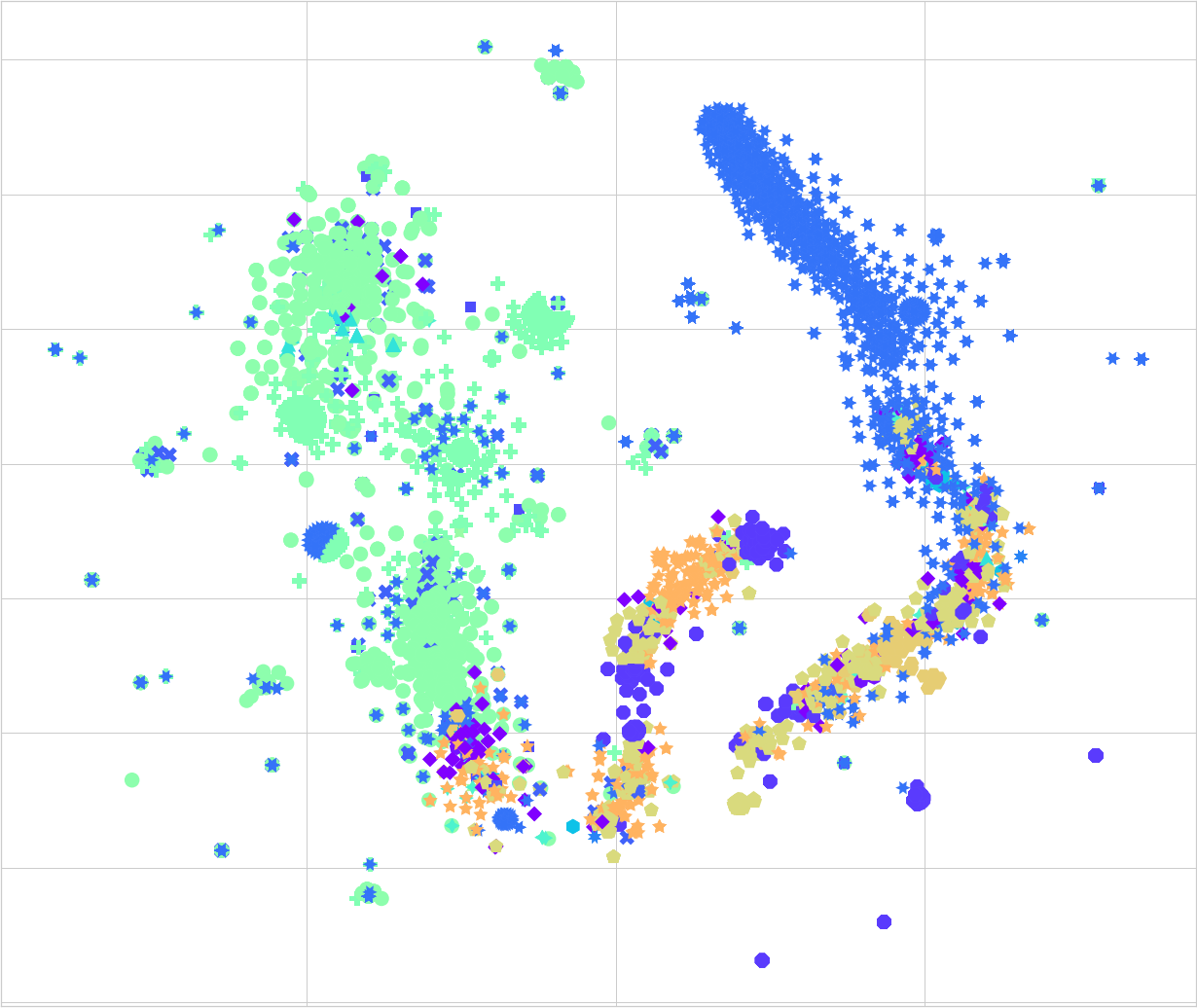}%
\label{Autoencoder_Approx_Host}}
\\
\includegraphics[scale=0.26]
{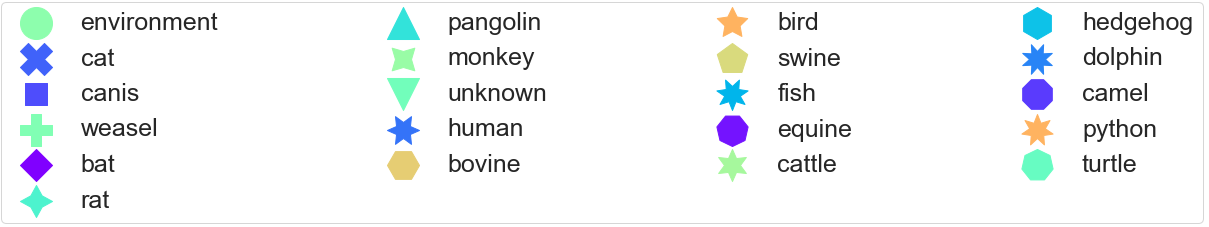}
\caption{t-SNE plots (\textbf{Coronavirus Host Data}) for different structure embeddings and Clustering and Projection methods (CCP and CCP-NN). The figure is best seen in color.}
\label{fig_tsne_host}
\end{figure}

\begin{figure}[!t]
\centering
\subfloat[\raggedright \scriptsize OHE (CCP)]{\includegraphics[scale=0.065]{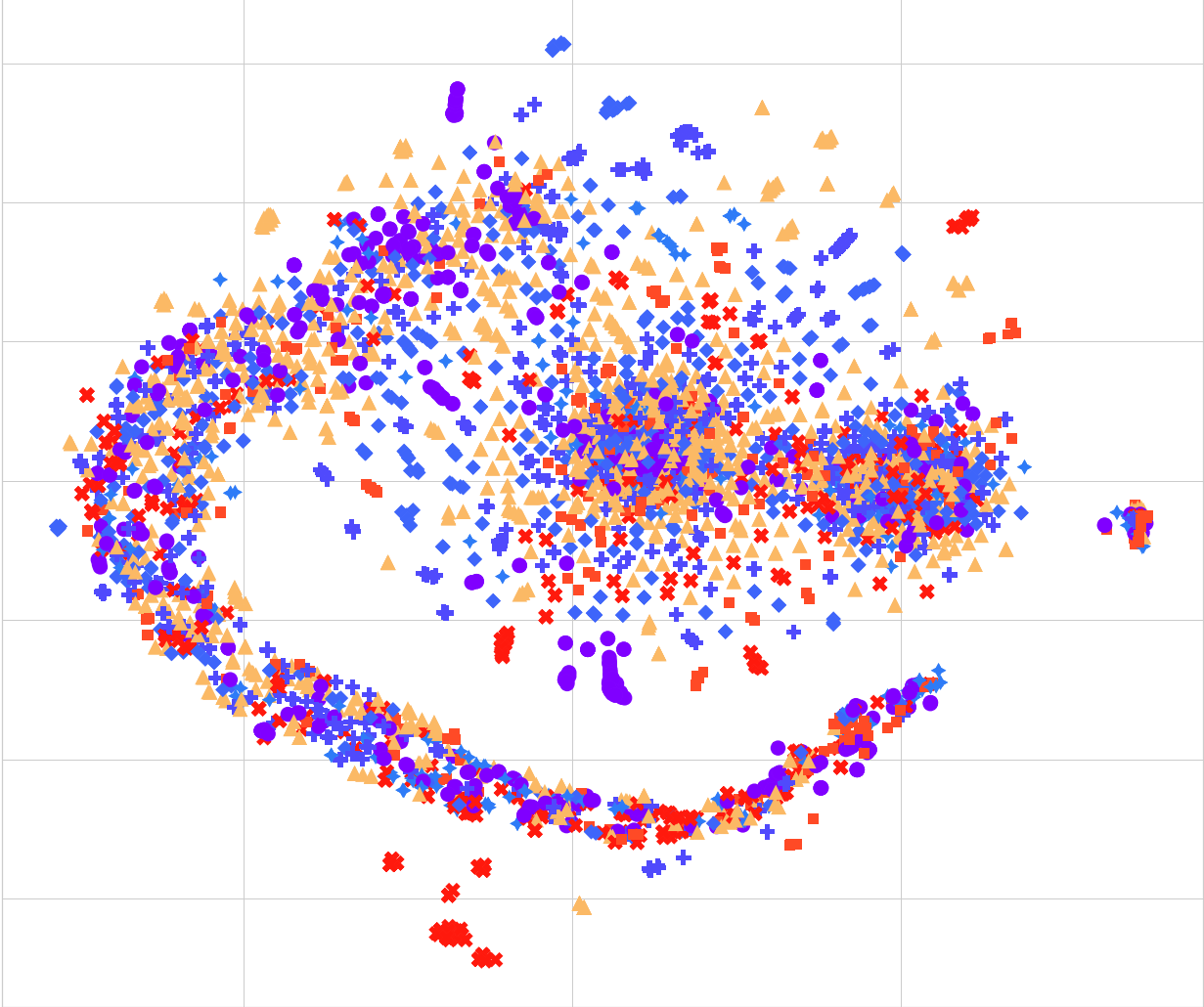}%
\label{OHE_CCP_Human_Dna}}
\hfil
\subfloat[\raggedright \scriptsize Spike2Vec (CCP)]{\includegraphics[scale=0.065]{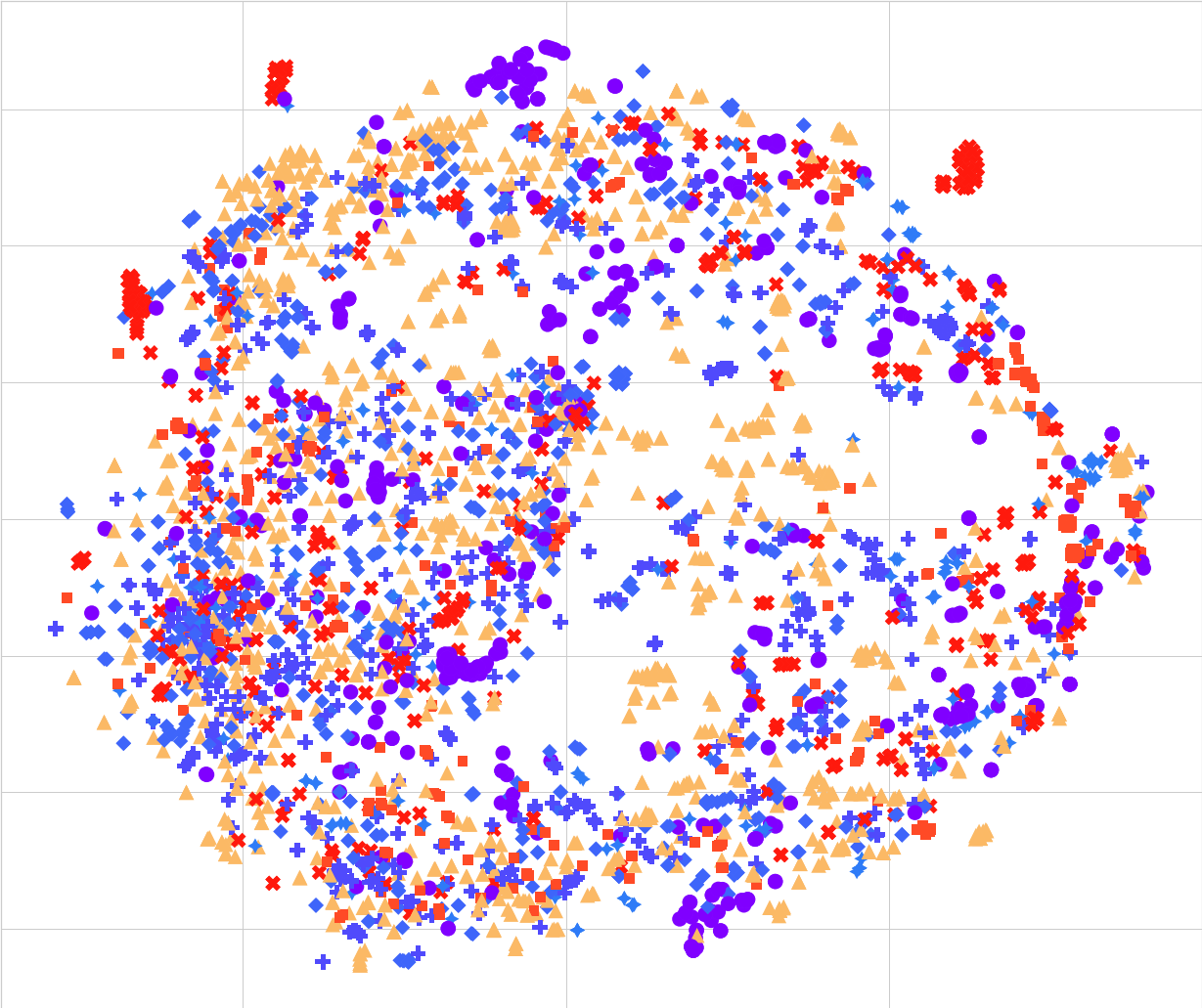}%
\label{Spike2Vec_CCP_Human_Dna}}
\hfil
\subfloat[\raggedright \scriptsize PWM2Vec (CCP)]{\includegraphics[scale=0.065]{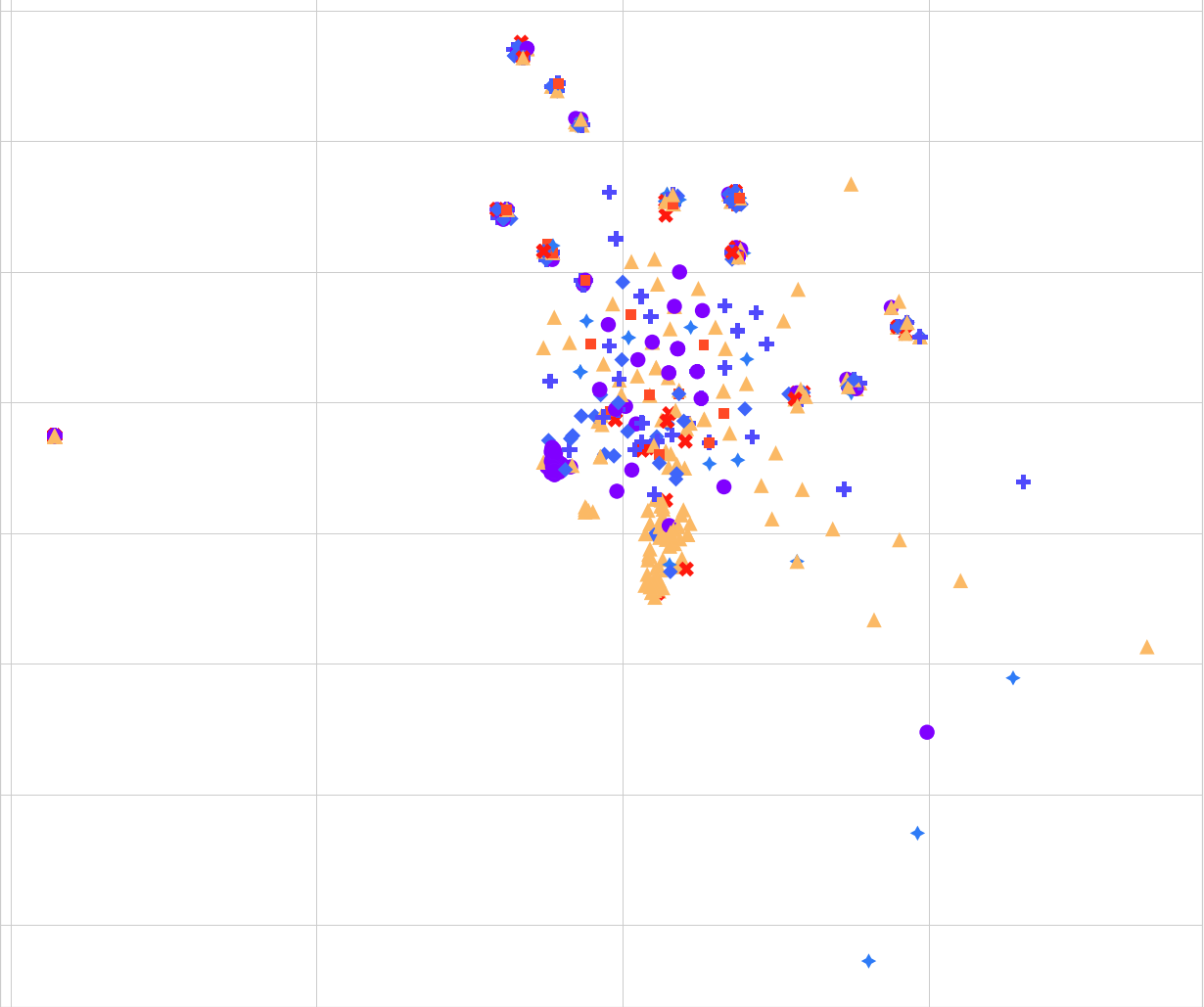}%
\label{PWM2Vec_CCP_Human_Dna}}
\\
\subfloat[\raggedright \scriptsize Autoencoder (CCP)]{\includegraphics[scale=0.065]{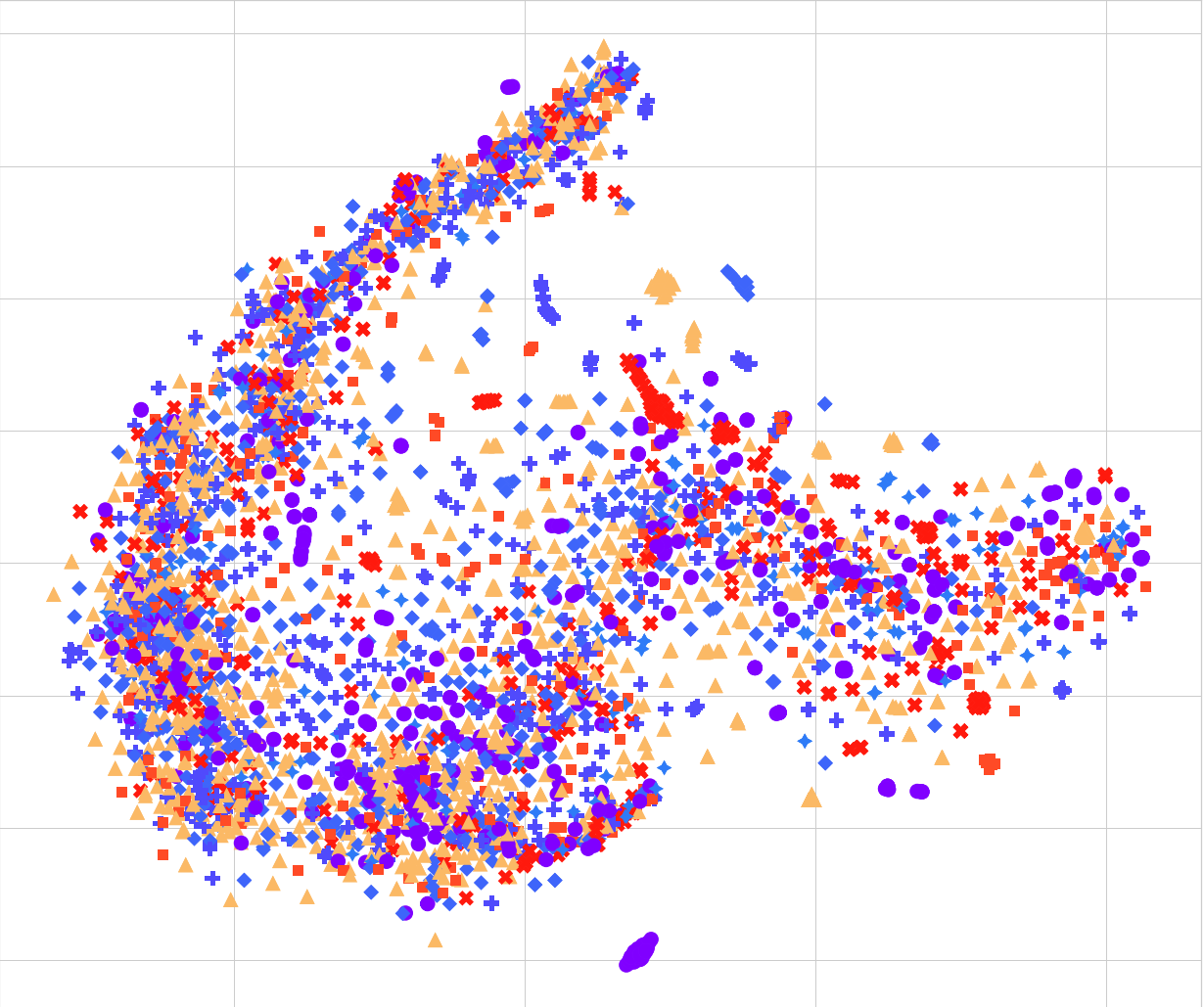}%
\label{Autoencoder_CCP_Human_Dna}}
\hfil
\subfloat[\raggedright \scriptsize OHE (CCP-NN)]{\includegraphics[scale=0.065]{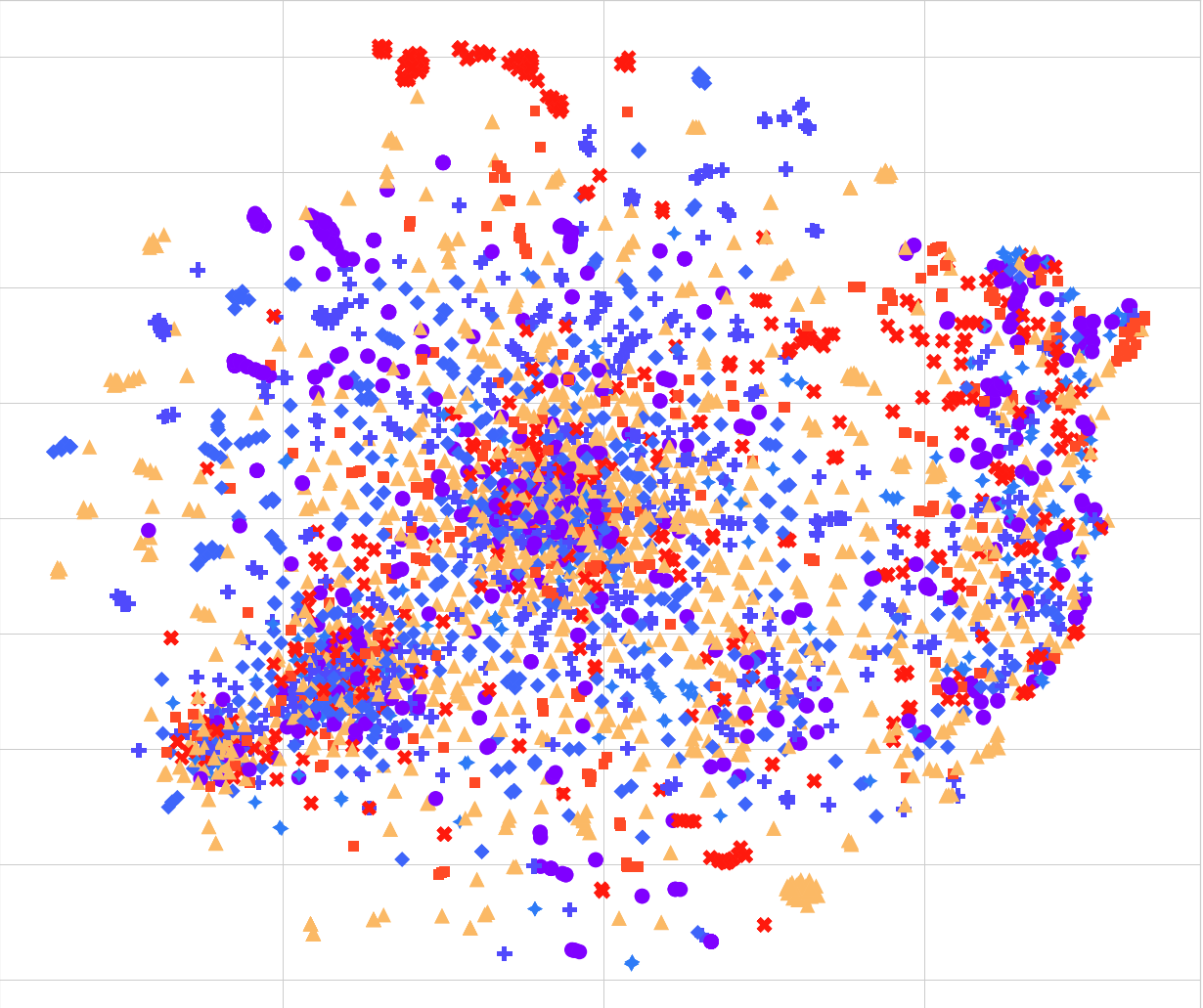}%
\label{OHE_Approx_Human_Dna}}
\hfil
\subfloat[\raggedright \scriptsize Spike2Vec CCP-NN]{\includegraphics[scale=0.065]{Figures/tSne/human_dna/OHE_Approx.png}%
\label{Spike2Vec_Approx_Human_Dna}}
\\
\subfloat[\raggedright \scriptsize PWM CCP-NN]{\includegraphics[scale=0.065]{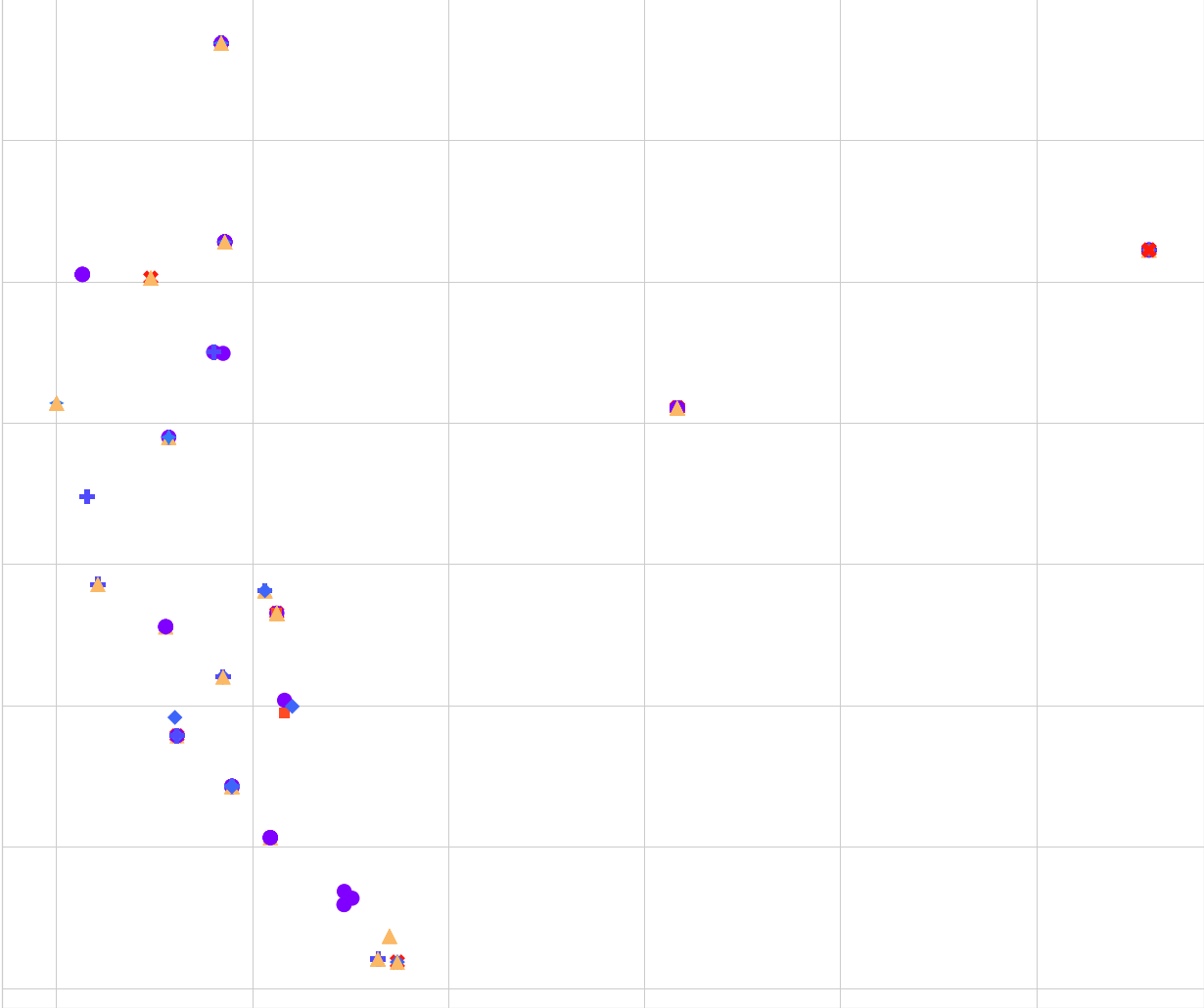}%
\label{PWM_Approx_Human_Dna}}
\hfil
\subfloat[\raggedright \scriptsize Auto-En. CCP-NN]{\includegraphics[scale=0.065]{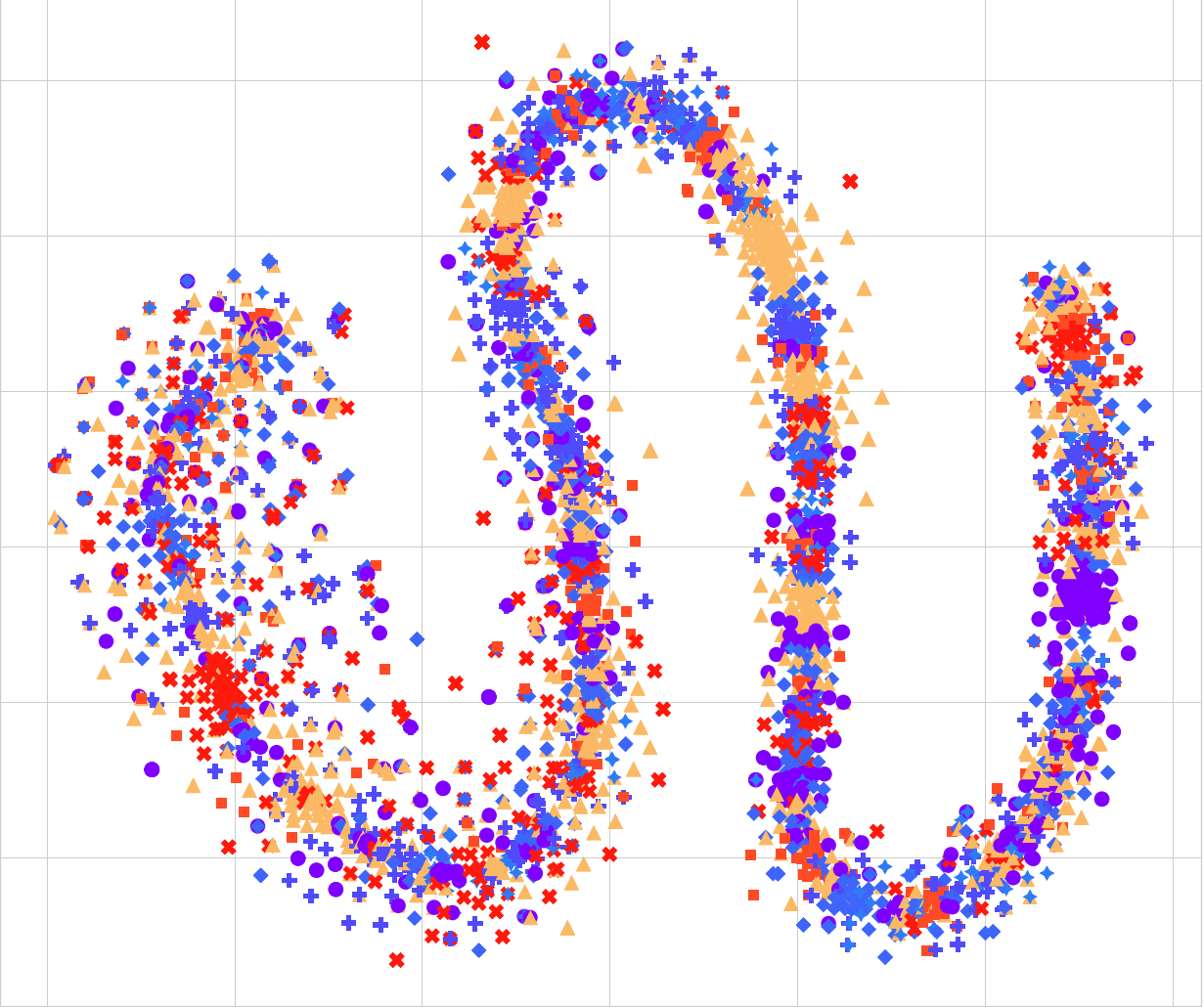}%
\label{Autoencoder_Approx_Human_Dna}}
\\
\includegraphics[scale=0.26]
{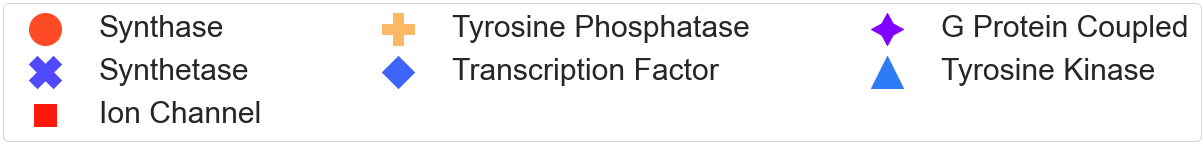}
\caption{t-SNE plots (\textbf{Coronavirus Host Data}) for different structure embeddings and Clustering and Projection methods (CCP and CCP-NN). The figure is best seen in color.}
\label{fig_tsne_human_dna}
\end{figure}

In this study, we used three datasets to assess the effectiveness of the proposed method.
We employ t-distributed stochastic neighbor embedding (t-SNE)~\cite{van2008visualizing} to examine the data for any natural (hidden) grouping. The t-SNE plots for various embedding techniques are shown in Figure~\ref{fig_tsne_protein_subcellular}, ~\ref{fig_tsne_host}, and ~\ref{fig_tsne_human_dna} for Protein Subcellular, Coronavirus Host, and Human DNA datasets, respectively. 
We can observe that t-SNE can group similar classes in the case of the Autoencoder with CCP-NN.

To classify the molecular sequences, we employed several ML models, including Support Vector Machine (SVM), Naive Bayes (NB), Multi-Layer Perceptron (MLP), K-Nearest Neighbors (KNN), Random Forest (RF), Logistic Regression (LR), and Decision Tree (DT). 
We evaluated classification performance using average accuracy, precision, recall, F1 (weighted), F1 (macro), Receiver Operator Characteristic Curve (ROC) Area Under the Curve (AUC), and training runtime. 
To preserve the original data distribution, the data for each classification task is divided into 60-10-30\% train-validation-test sets using stratified sampling. To obtain more consistent findings, we also conduct our tests by averaging the performance outcomes of $5$ runs.
We carefully considered baselines from several embedding generation categories, including feature engineering, conventional kernel matrix generation, neural networks, pre-trained language models, and pre-trained transformers for protein sequences. Table~\ref{tbl_sota_detail} contains the specifics for the baseline models. 

\begin{table}[h!]
    \caption{Baseline methods.}
    \label{tbl_sota_detail}
    \centering
    \resizebox{0.49\textwidth}{!}{
    \begin{tabular}{lp{2.2cm}p{5.5cm}c}
    \toprule
      Method & Category & Detail & Source \\
    \midrule \midrule
    \multirow{2}{*}{OHE} & \multirow{10}{2.5cm}{Feature Engineering} & The numerical vector is generated with simple one-hot encoding for each amino acid in the sequence. & ~\cite{kuzmin2020machine} \\ 
    \cmidrule{3-4}
    \multirow{2}{*}{Spike2Vec} &  & Uses the sliding window (of size k) to get $k$-mers and their count in the sequence to generate feature vectors. & ~\cite{ali2021spike2vec} \\
    \cmidrule{3-4}
    \multirow{2}{*}{PWM2Vec} &   & Uses the concept of the position-weight matrix (PWM) to generate embeddings & ~\cite{ali2022pwm2vec} \\
    \cmidrule{2-4}
    \multirow{4}{*}{String Kernel} & \multirow{4}{2.5cm}{Kernel Matrix} & Designs $n \times n$ kernel matrix that can be used with kernel classifiers or with kernel PCA to get a feature vector. & \multirow{4}{*}{~\cite{ali2022efficient}}  \\
    \cmidrule{2-4}
    \multirow{2}{*}{WDGRL} &  \multirow{4}{2.5cm}{Neural Network (NN)} &  \multirow{3}{6cm}{Take one-hot representation of biological sequence as input and design NN-based embedding method by minimizing loss} & \multirow{2}{*}{~\cite{shen2018wasserstein}}  \\
    & \\
    \multirow{2}{*}{AutoEncoder} & & & \multirow{2}{*}{~\cite{xie2016unsupervised}}  \\
    & \\
    \cmidrule{2-4}
    \multirow{3}{*}{SeqVec} & \multirow{2}{2.5cm}{Pretrained Large Language Model (LLM)} & Takes biological sequences as input and fine-tunes the weights based on a pre-trained model to get the final embedding. & \multirow{4}{*}{~\cite{heinzinger2019modeling}}  \\
    \cmidrule{2-4}
    \multirow{4}{*}{ProteinBERT} & \multirow{4}{2.5cm}{Pretrained Transformer} & A pre-trained protein sequence model to classify the given biological sequence using Transformer/Bert & \multirow{4}{*}{~\cite{brandes2022proteinbert}}  \\
     \cmidrule{2-4}
    \multirow{3}{*}{TAPE} & \multirow{3}{2.5cm}{Pretrained Transformer} & A LLM model with a self-supervised pretraining method for molecular sequence embedding generation. & \multirow{4}{*}{~\cite{rao2019evaluating}}  \\
      \bottomrule
    \end{tabular}
    }
\end{table}

\subsection{Baseline Methods}
\label{subsec_baselines}

Among the baseline methods discussed in Table~\ref{tbl_sota_detail}, we selected $4$ popular embedding generation models, including One Hot Encoding (OHE)~\cite{kuzmin2020machine}, Spike2Vec~\cite{ali2021spike2vec}, PWM2Vec~\cite{ali2022pwm2vec}, and Autoencoder~\cite{xie2016unsupervised} to be used as input to both vanilla CCP and CCP-NN for dimensionality reduction. These methods were simple to use (in terms of implementation compared to complex models like SeqVec, TAPE, and Protein Bert), easy and fast to compute (compared to WDGRL, which is computationally expensive and takes a long computational time), and generate embeddings directly, which can be used for dimensionality reduction (unlike String kernel, which generates a kernel matrix, which has to be converted to embeddings using kernel PCA, which could cause loss of information).

\subsection{Dataset Statistics}\label{sec_data_appendix}
We use $3$ datasets for our experiments. The effectiveness of the k-NN algorithm is heavily dependent on the quality and scale of the dataset, as it uses distance-based calculations. To address this, we performed data normalization during preprocessing to ensure that all features contribute equally, improving the accuracy and reliability of the model. The datasets used are listed below:

\textbf{Protein Subcellular Locations}
The Protein Subcellular Locations dataset we used consists of $5959$ unaligned protein sequences, each corresponding to a different subcellular location~\cite{ProtLoc_website_url}. The labels for the classification task are these subcellular sites, and there are $11$ unique labels in our dataset. These labels correspond to the proteins of plant cells and fungal cells, while animal cells share all localizations with them. Table~\ref{tbl_dataset_statistics_protLoc} provides the distribution of classes. 

\begin{table}[h!]
    \caption{The distribution of sequences in the \textbf{Protein Subcellular locations} data among the subcellular locations.}
    \label{tbl_dataset_statistics_protLoc}
    \centering
    \resizebox{0.49\textwidth}{!}{
    \begin{tabular}{p{2.5cm}c|p{3.2cm}c}
    \toprule
        Subcellular Locations & No. of Sequences & Subcellular Locations & No. of Sequences \\
        \midrule \midrule
        Cytoplasm &  1411 & Endoplasmic Reticulum & 198\\ 
        Plasma Membrane & 1238 & Peroxisome & 157 \\
        Extracellular Space & 843 & Golgi Apparatus & 150 \\
        Nucleus & 837 & Lysosomal & 103 \\ 
        Mitochondrion & 510 & Vacuole & 63 \\
        Chloroplast & 449 & - & - \\
        \midrule 
        - & - & \textbf{Total} & \textbf{5959} \\
        \bottomrule
    \end{tabular}
    }
    
\end{table}

\textbf{Coronavirus Host}
The NIAD Virus Pathogen Database and Analysis Resource(ViPR)~\cite{pickett2012vipr} and GISAID~\cite{gisaid_website_url} are used to retrieve Spike molecular sequences of CoVs for all hosts. Table~\ref{tbl_dataset_statistics} (in the supplementary material) comprises details about the $21$ host types with their counts of sequences that we gathered through the annotation of the total of $5558$ complete unaligned spike protein sequences. We have some hosts that have very few sequences; these sequences ensure that the dataset represents the full diversity of known hosts rather than omitting certain species entirely. In addition, a single sequence may still carry host-specific signatures that could be useful for feature analysis.

\begin{table}[h!]
    \caption{Statistics for \textbf{Coronavirus Host} dataset.}
    \label{tbl_dataset_statistics}
    \centering
    \resizebox{0.49\textwidth}{!}{
    \begin{tabular}{cc|cc|cc}
    \hline
        Host Name & \# of Sequences & Host Name & \# of Sequences & Host Name & \# of Sequences \\
        \hline \hline
        Humans & 1813 & Rats & 26 & Cats & 123 \\
        Environment &  1034 & Pangolins & 21 & Bovines & 88 \\
        Weasel & 994 & Hedgehog & 15 & Dogs & 40 \\
        Swine & 558 & Dolphin & 7 & Python & 2 \\
        Birds & 374 & Equine & 5 & Monkey & 2 \\
        Camels & 297 & Fish & 2 & Cattle & 1 \\
        Bats & 153 & Unknown & 2 & Turtle & 1\\
        \midrule
         - & - & -  & - & \textbf{Total} & \textbf{5558}\\
         \bottomrule
        \hline
    \end{tabular}
    }
   
\end{table}

\textbf{Human DNA}
The data contains $4380$ unaligned Human DNA nucleotide sequences~\cite{human_dna_website_url}. A total of $7$ unique labels comprised of a human gene family are G Protein-Coupled, Tyrosine Kinase, Tyrosine Phosphatase, Synthetase, Synthase, Ion Channel, and Transcription Factor. Table~\ref{tbl_dataset_statistics_humna_dna} (in the supplementary material) provides the statistics for the dataset.

\begin{table}[h!]
    \caption{The distribution of gene family with the count of sequences in the \textbf{Human DNA} data.}
    \label{tbl_dataset_statistics_humna_dna}
    \centering
    \resizebox{0.49\textwidth}{!}{
    \begin{tabular}{lc|cc}
    \toprule
        Gene Family & Num. of Sequences &  Gene Family & Num. of Sequences \\
        \midrule \midrule
        G Protein Coupled & 531 & Tyrosine Kinase & 534 \\
        Tyrosine Phosphatase & 349 &  Synthetase & 672 \\
        Synthase & 711 & Ion Channel & 240 \\
        Transcription Factor & 1343 & - & - \\
        \midrule
        - & - & \textbf{Total} & \textbf{4380} \\
        \bottomrule
    \end{tabular}
    }
\end{table}

\section{Results And Discussion}\label{sec_R}
In this section, we report classification and runtime results for both CCP and CCP-NN using different datasets and embedding models.

\subsection{Results For Protein Subcellular dataset}
The classification results (averaged over $5$ runs) for the proposed CCP-NN and its comparison with the CCP approach for the Protein Subcellular dataset are shown in Table~\ref{tbl_results_Part1_Protein_Subcellular}.
We can observe that the proposed CCP-NN outperforms the original CCP-based low-dimensional representation for all evaluation metrics and achieves a near-perfect predictive classification performance. The performance gain for CCP-NN (using OHE with the Decision Tree classifier and using Autoencoder with the Decision Tree classifier), compared to the best CCP-based results (Spike2Vec with the Random Forest classifier), is $50.3\%$, highlighting a significant improvement in terms of predictive accuracy.

\begin{table}[h!]
    \caption{Classification results (averaged over $5$ runs) for \textbf{Protein Subcellular} dataset using Nearest Neighbour CCP (CCP-NN) and CCP. The best value for each embedding is shown with an underline. The overall best value for each evaluation metric is shown in bold. The $\uparrow$ means a higher number is better, and the down arrow $\downarrow$ means that a lower number is better.}
    \label{tbl_results_Part1_Protein_Subcellular}
    \centering
    \resizebox{0.49\textwidth}{!}{
    \begin{tabular}{clcp{0.6cm}p{0.6cm}p{0.6cm}p{0.7cm}p{0.7cm}p{0.7cm}p{0.9cm}}
    \toprule
    & \multirow{2}{0.8cm}{Embeddings} & \multirow{2}{*}{Algo.} & \multirow{2}{*}{Acc. $\uparrow$} & \multirow{2}{*}{Prec. $\uparrow$} & \multirow{2}{*}{Recall $\uparrow$} & \multirow{2}{0.8cm}{F1 (Weig.) $\uparrow$} & \multirow{2}{0.8cm}{F1 (Macro) $\uparrow$} & \multirow{2}{0.8cm}{ROC AUC $\uparrow$} & Train Time (sec.) $\downarrow$\\
        \midrule \midrule
 \multirow{28}{1cm}{CCP ($\phi_{CCP}$)} & 
 \multirow{7}{1.2cm}{OHE}
& SVM & 0.288 & 0.228 & 0.288 & 0.196 & 0.085 & 0.517 & 17.920 \\
& & NB & 0.209 & \underline{0.445} & 0.209 & 0.215 & 0.171 & 0.569 & 0.344 \\
& & MLP & 0.321 & 0.327 & 0.321 & 0.323 & 0.199 & 0.563 & 6.457 \\
& & KNN & 0.238 & 0.212 & 0.238 & 0.197 & 0.106 & 0.515 & \underline{0.288} \\
& & RF & \underline{0.429} & 0.420 & \underline{0.429} & \underline{0.375} & 0.215 & \underline{0.580} & 3.222 \\
& & LR & 0.349 & 0.274 & 0.349 & 0.267 & 0.127 & 0.542 & 7.668 \\
& & DT & 0.306 & 0.306 & 0.306 & 0.306 & \underline{0.202} & 0.563 & 0.653  \\
\cmidrule{3-10}                                                                                     
& \multirow{7}{1.2cm}{Spike2Vec}                                                                                       
& SVM & 0.393 & 0.414 & 0.393 & 0.380 & \underline{0.247} & 0.589 & 16.626	\\
& & NB & 0.200 & 0.318 & 0.200 & 0.235 & 0.161 & 0.549 & 0.321   \\
& & MLP & 0.360 & 0.359 & 0.360 & 0.358 & 0.221 & 0.577 & 10.421 \\
& & KNN & 0.262 & 0.281 & 0.262 & 0.247 & 0.147 & 0.538 & \underline{0.279}  \\
& & RF & \underline{0.495} & \underline{0.524} & \underline{0.495} & \underline{0.440} & 0.245 & \underline{0.598} & 3.292   \\
& & LR & 0.418 & 0.412 & 0.418 & 0.387 & 0.233 & 0.582 & 10.810  \\
& & DT & 0.319 & 0.318 & 0.319 & 0.317 & 0.203 & 0.566 & 1.077   \\
\cmidrule{3-10}                                                                                       
&  \multirow{7}{1.2cm}{PWM2Vec}                                                                                 
& SVM & 0.411 & 0.424 & 0.411 & 0.412 & \underline{0.316} & \underline{0.627} & 11.482		\\
& & NB & 0.224 & 0.267 & 0.224 & 0.221 & 0.173 & 0.561 & 0.284          \\
& & MLP & 0.335 & 0.335 & 0.335 & 0.334 & 0.214 & 0.572 & 7.056        \\
& & KNN & 0.244 & 0.369 & 0.244 & 0.213 & 0.135 & 0.524 & \underline{0.229}        \\
& & RF & \underline{0.450} & \underline{0.511} & \underline{0.450} & 0.391 & 0.214 & 0.582 & 2.987          \\
& & LR & 0.445 & 0.443 & 0.445 & \underline{0.429} & 0.302 & 0.613 & 8.984         \\
& & DT & 0.281 & 0.280 & 0.281 & 0.280 & 0.183 & 0.553 & 0.924          \\
\cmidrule{3-10}                                                                                        
&  \multirow{7}{1.2cm}{Autoencoder}                                                             
 & SVM & \underline{0.300} & 0.194 & \underline{0.300} & 0.226 & 0.106 & 0.528 & 5.469	\\
& & NB & 0.204 & 0.162 & 0.204 & 0.145 & 0.089 & \underline{0.535} & \underline{0.081}     \\
& & MLP &  0.236 & 0.208 & 0.236 & 0.217 & 0.118 & 0.520 & 9.564    \\
& & KNN & 0.239 & 0.224 & 0.239 & 0.226 & \underline{0.133} & 0.529 & 0.144    \\
& & RF & 0.298 & \underline{0.229} & 0.298 & \underline{0.235} & 0.114 & 0.529 & 13.596    \\
& & LR & 0.294 & 0.216 & 0.294 & 0.231 & 0.111 & 0.529 & 2.201     \\
& & DT & 0.197 & 0.198 & 0.197 & 0.197 & 0.114 & 0.514 & 2.236     \\

\midrule
 \multirow{28}{1cm}{CCP Nearest Neighbor ($\phi_{CCP\_NN}$)} & \multirow{7}{1.2cm}{OHE}
& SVM & 0.463 & 0.489 & 0.463 & 0.472 & 0.293 & 0.619 & 14.200 \\
& & NB & 0.705 & 0.735 & 0.705 & 0.705 & 0.604 & 0.791 & 0.294 \\
& & MLP & 0.569 & 0.578 & 0.569 & 0.571 & 0.365 & 0.661 & 9.380 \\
& & KNN & 0.239 & 0.362 & 0.239 & 0.152 & 0.071 & 0.509 & 0.290 \\
& & RF & 0.939 & 0.945 & 0.939 & 0.931 & 0.814 & 0.882 & 1.976  \\
& & LR & 0.537 & 0.506 & 0.537 & 0.518 & 0.318 & 0.634 & 10.401  \\
& & DT & \textbf{\underline{0.998}} & \textbf{\underline{0.998}} & \textbf{\underline{0.998}} & \textbf{\underline{0.998}} & \textbf{\underline{0.995}} & \underline{0.997} & \underline{0.200}  \\
\cmidrule{3-10}                                                                                        
& \multirow{7}{1.2cm}{Spike2Vec}                                                                                       
& SVM & 0.587 & 0.597 & 0.587 & 0.590 & 0.410 & 0.683 & 8.628  \\
& & NB & 0.400 & 0.557 & 0.400 & 0.416 & 0.362 & 0.679 & 0.231  \\
& & MLP & 0.619 & 0.625 & 0.619 & 0.620 & 0.422 & 0.693 & 6.110  \\
& & KNN & 0.233 & 0.356 & 0.233 & 0.210 & 0.119 & 0.524 & \underline{0.222} \\
& & RF & 0.940 & 0.945 & 0.940 & 0.934 & 0.840 & 0.892 & 1.740   \\
& & LR & 0.615 & 0.594 & 0.615 & 0.591 & 0.398 & 0.670 & 7.800  \\
& & DT & \underline{0.995} & \underline{0.995} & \underline{0.995} & \underline{0.995} & \underline{0.991} & \underline{0.995} & 0.262  \\
\cmidrule{3-10}                                                                                     
&  \multirow{7}{1.2cm}{PWM2Vec}                                                                                 
& SVM & 0.520 & 0.537 & 0.520 & 0.526 & 0.368 & 0.659 & 13.341  \\
& & NB & 0.395 & 0.521 & 0.395 & 0.398 & 0.342 & 0.670 & 0.322  \\
& & MLP & 0.561 & 0.567 & 0.561 & 0.561 & 0.365 & 0.661 & 9.496 \\
& & KNN & 0.198 & 0.342 & 0.198 & 0.133 & 0.063 & 0.499 & \underline{0.261}  \\
& & RF & 0.942 & 0.946 & 0.942 & 0.936 & 0.846 & 0.895 & 3.003  \\
& & LR & 0.571 & 0.551 & 0.571 & 0.552 & 0.369 & 0.655 & 11.213  \\
& & DT & \underline{0.995} & \underline{0.995} & \underline{0.995} & \underline{0.995} & \underline{0.993} & \underline{0.996} & 0.452 \\
\cmidrule{3-10}                                                                                                       
&  \multirow{7}{1.2cm}{Autoencoder}                                                             
 & SVM & 0.814 & 0.671 & 0.814 & 0.734 & 0.411 & 0.716 & 0.612  \\
& & NB & 0.950 & 0.962 & 0.950 & 0.952 & 0.907 & 0.981 & \textbf{\underline{0.065}}  \\
& & MLP & 0.933 & 0.911 & 0.933 & 0.918 & 0.861 & 0.933 & 2.851  \\
& & KNN & 0.997 & 0.997 & 0.997 & 0.997 & 0.991 & 0.996 & 0.085  \\
& & RF & 0.998 & 0.998 & 0.998 & 0.998 & 0.994 & 0.998 & 2.290  \\
& & LR & 0.439 & 0.202 & 0.439 & 0.274 & 0.114 & 0.558 & 0.815  \\
& & DT & \textbf{\underline{0.998}} & \textbf{\underline{0.998}} & \textbf{\underline{0.998}} & \textbf{\underline{0.998}} & \textbf{\underline{0.995}} & \textbf{\underline{0.998}} & 0.427  \\ 
         \bottomrule
         \end{tabular}
    }

\end{table}

The comparison of the best performing proposed method from Table~\ref{tbl_results_Part1_Protein_Subcellular} (i.e., CCP-NN with Autoencoder) with the existing baseline models (without CCP or CPP-NN) is shown in Table~\ref{tbl_results_Protein_Subcellular_org}. We can observe that the proposed method significantly outperforms all baselines for all evaluation metrics other than the training runtime. Specifically, in terms of average accuracy, the proposed method with Autoencoder embedding achieves $28\%$ improvement compared to the second best (i.e., Protein Bert, a pre-trained transformer-based model) and achieves a near-perfect average accuracy score in the case of the Protein Subcellular dataset.

\begin{table}[h!]
    \caption{Classification result comparisons (averaged over $5$ runs) for the best performing proposed method (i.e., CCP-NN with Autoencoder) with baselines on \textbf{Protein Subcellular} dataset. The best value for each embedding is shown in underlined. The overall best value for each evaluation metric is shown in bold.
    The $\uparrow$ means a higher number is better, and the down arrow $\downarrow$ means that a lower number is better.
    }
    \label{tbl_results_Protein_Subcellular_org}
\centering
\resizebox{0.45\textwidth}{!}{
 \begin{tabular}{p{1.2cm}lp{0.5cm}p{0.5cm}p{0.5cm}p{0.5cm}p{0.5cm}p{0.5cm}p{0.8cm}}
    \toprule
        \multirow{2}{1.2cm}{Embeddings} & \multirow{2}{*}{Algo.} & \multirow{2}{*}{Acc. $\uparrow$} & \multirow{2}{*}{Prec. $\uparrow$} & \multirow{2}{0.8cm}{Recall $\uparrow$} & \multirow{2}{0.8cm}{F1 (Weig.) $\uparrow$} & \multirow{2}{0.8cm}{F1 (Macro) $\uparrow$} & \multirow{2}{0.8cm}{ROC AUC $\uparrow$} & Train Time (sec.) $\downarrow$\\
        \midrule \midrule
         \multirow{7}{1cm}{OHE}
        & SVM & \underline{0.530} & \underline{0.516} & \underline{0.530} & \underline{0.509} & \underline{0.355} & \underline{0.647} & 706.196	\\
        & NB & 0.131 & 0.201 & 0.131 & 0.137 & 0.091 & 0.514 & 16.515   \\
        & MLP & 0.390 & 0.401 & 0.390 & 0.390 & 0.255 & 0.595 & 187.259 \\
        & KNN & 0.242 & 0.114 & 0.242 & 0.095 & 0.036 & 0.500 & 8.465   \\
        & RF & 0.404 & 0.410 & 0.404 & 0.329 & 0.171 & 0.563 & 32.347   \\
        & LR & 0.515 & 0.498 & 0.515 & 0.492 & 0.335 & 0.637 & \underline{6.140}    \\
        & DT & 0.307 & 0.300 & 0.307 & 0.303 & 0.193 & 0.560 & 25.282   \\
       \cmidrule{2-9}
        \multirow{7}{1cm}{Spike2Vec}
        & SVM & \underline{0.575} & \underline{0.579} & \underline{0.575} & \underline{0.571} & \underline{0.483} & \underline{0.706} & 111.398	\\
        & NB & 0.253 & 0.368 & 0.253 & 0.253 & 0.182 & 0.578 & 3.095      \\
        & MLP & 0.478 & 0.489 & 0.478 & 0.481 & 0.345 & 0.645 & 36.700    \\
        & KNN & 0.279 & 0.416 & 0.279 & 0.215 & 0.136 & 0.531 & \underline{1.889}     \\
        & RF & 0.480 & 0.528 & 0.480 & 0.429 & 0.237 & 0.592 & 7.353      \\
        & LR & 0.564 & 0.566 & 0.564 & 0.555 & 0.453 & 0.687 & 8.075      \\
        & DT & 0.293 & 0.289 & 0.293 & 0.290 & 0.183 & 0.554 & 2.890      \\
       \cmidrule{2-9}
        \multirow{7}{1cm}{PWM2Vec}
        & SVM & 0.423 & 0.444 & 0.423 & 0.426 & 0.339 & 0.640 & 79.182 \\
         & NB & 0.293 & 0.312 & 0.293 & 0.241 & 0.206 & 0.581 & \underline{0.810} \\
         & MLP & 0.309 & 0.315 & 0.309 & 0.310 & 0.206 & 0.568 & 111.598 \\
         & KNN & 0.285 & 0.461 & 0.285 & 0.247 & 0.192 & 0.549 & 1.964 \\
         & RF & 0.436 & \underline{0.496} & 0.436 & 0.379 & 0.210 & 0.577 & 84.261 \\
         & LR & \underline{0.470} & 0.476 & \underline{0.470} & \underline{0.470} & \underline{0.351} & \underline{0.645} & 96.467 \\
         & DT & 0.306 & 0.316 & 0.306 & 0.310 & 0.196 & 0.561 & 34.803 \\

        \cmidrule{2-9} 
        \multirow{7}{1cm}{Autoencoder}
        & SVM & 0.431 & 0.447 & 0.431 & 0.435 & 0.315 & 0.632 & 95.840 \\
        & NB & 0.228 & 0.305 & 0.228 & 0.205 & 0.161 & 0.569 & \underline{0.316} \\
        & MLP & 0.412 & 0.389 & 0.412 & 0.399 & 0.253 & 0.598 & 126.795 \\
        & KNN & 0.275 & 0.292 & 0.275 & 0.219 & 0.127 & 0.529 & 1.970 \\
        & RF & 0.381 & 0.347 & 0.381 & 0.306 & 0.163 & 0.558 & 30.260 \\
        & LR & \underline{0.464} & \underline{0.452} & \underline{0.464} & \underline{0.455} & \underline{0.332} & \underline{0.639} & 138.959 \\
        & DT & 0.228 & 0.232 & 0.228 & 0.229 & 0.150 & 0.533 & 15.367 \\
        
        \cmidrule{2-9}
        \multirow{7}{1cm}{String Kernel}
        & SVM  & 0.496 & 0.510 & 0.496 & 0.501 & 0.395 & 0.674 & 5.277 \\   
         & NB   & 0.301 & 0.322 & 0.301 & 0.265 & 0.243 & 0.593 & 0.136 \\
         & MLP  & 0.389 & 0.390 & 0.389 & 0.388 & 0.246 & 0.591 & 7.263 \\
         & KNN  & 0.372 & 0.475 & 0.372 & 0.370 & 0.272 & 0.586 & \underline{0.395} \\
         & RF   & 0.473 & 0.497 & 0.473 & 0.411 & 0.218 & 0.585 & 7.170 \\
         & LR   & \underline{0.528} & \underline{0.525} & \underline{0.528} & \underline{0.525} & \underline{0.415} & \underline{0.678} & 8.194 \\
         & DT   & 0.328 & 0.335 & 0.328 & 0.331 & 0.207 & 0.568 & 2.250 \\
         \cmidrule{2-9}        
           \multirow{7}{1cm}{WDGRL}  
           & SVM & \underline{0.229} & 0.098 & \underline{0.229} & 0.137 & 0.057 & \underline{0.503} & 1.752 \\
           & NB & 0.206 & 0.154 & 0.206 & 0.158 & 0.073 & 0.501 & \textbf{\underline{0.008}} \\
           & MLP & 0.218 & 0.136 & 0.218 & 0.151 & 0.067 & 0.502 & 11.287 \\
           & KNN & 0.170 & 0.154 & 0.170 & 0.158 & \underline{0.086} & 0.500 & 0.273 \\
           & RF & 0.211 & \underline{0.167} & 0.211 & \underline{0.163} & 0.079 & \underline{0.503} & 2.097 \\
           & LR & \underline{0.229} & 0.098 & \underline{0.229} & 0.137 & 0.057 & \underline{0.503} & 0.112 \\
           & DT & 0.152 & 0.154 & 0.152 & 0.153 & \underline{0.086} & 0.498 & 0.082 \\
             
          \cmidrule{2-9}
        \multirow{7}{1cm}{SeqVec}
         & SVM & 0.412 & 0.425 & 0.412 & \underline{0.421} & 0.306 & 0.611 & 10.241 \\
         & NB & 0.205 & 0.297 & 0.205 & 0.196 & 0.154 & 0.542 & \underline{0.125} \\
         & MLP & 0.403 & 0.377 & 0.404 & 0.384 & 0.231 & 0.574 & 21.495 \\
         & KNN & 0.244 & 0.271 & 0.245 & 0.201 & 0.114 & 0.511 & 1.141 \\
         & RF & 0.362 & 0.323 & 0.362 & 0.295 & 0.155 & 0.541 & 5.137 \\
         & LR & \underline{0.451} & \underline{0.444} & \underline{0.451} & \underline{0.421} & \underline{0.323} & \underline{0.624} & 4.427 \\
         & DT & 0.213 & 0.221 & 0.213 & 0.224 & 0.149 & 0.517 & 7.752 \\
         \cmidrule{2-9}  
        \multirow{1}{1cm}{Protein Bert}
         & \_ &  0.718 & 0.715 & 0.718 & 0.706 & 0.572 & 0.765 & 16341.85 \\

        \cmidrule{2-9} 
          \multirow{7}{1cm}{TAPE} 
         & SVM & 0.637 & 0.640 & 0.637 & 0.636 & 0.552 & \underline{0.760} & 8.553 \\
         & NB & 0.377 & 0.508 & 0.377 & 0.375 & 0.300 & 0.662 & \underline{0.311} \\
         & MLP & 0.590 & 0.590 & 0.590 & 0.589 & 0.432 & 0.695 & 5.296 \\
         & KNN & 0.595 & 0.600 & 0.595 & 0.589 & 0.468 & 0.710 & 0.160 \\
         & RF & 0.600 & 0.622 & 0.600 & 0.572 & 0.405 & 0.666 & 28.819 \\ 
         & LR & \underline{0.671} & \underline{0.664} & \underline{0.671} & \underline{0.660} & \underline{0.553} & 0.746 & 15.968 \\
         & DT & 0.417 & 0.424 & 0.417 & 0.420 & 0.300 & 0.620 & 11.233 \\

         \cmidrule{2-9} 
           \multirow{7}{1cm}{$\phi_{CCP\_NN}$ (ours) - Autoencoder} 
        & SVM & 0.814 & 0.671 & 0.814 & 0.734 & 0.411 & 0.716 & 0.612 \\
        & NB & 0.950 & 0.962 & 0.950 & 0.952 & 0.907 & 0.981 & \underline{0.065}   \\
        & MLP & 0.933 & 0.911 & 0.933 & 0.918 & 0.861 & 0.933 & 2.851 \\
        & KNN & 0.997 & 0.997 & 0.997 & 0.997 & 0.991 & 0.996 & 0.085  \\
        & RF & 0.998 & 0.998 & 0.998 & 0.998 & 0.994 & 0.998 & 2.290 \\
        & LR & 0.439 & 0.202 & 0.439 & 0.274 & 0.114 & 0.558 & 0.815  \\
        & DT & \textbf{\underline{0.998}} & \textbf{\underline{0.998}} & \textbf{\underline{0.998}} & \textbf{\underline{0.998}} & \textbf{\underline{0.995}} & \textbf{\underline{0.998}} & 0.427 \\        
         \bottomrule
         \end{tabular}
}
 
\end{table}

The standard deviation (SD) results (for $5$ runs) for the baselines and the proposed method are shown in Table~\ref{tbl_results_protein_subcellular_std} for the Protein Subcellular dataset. We can observe that in the majority of the cases, the SD values are towards the lower end (i.e. $<0.02$), which shows that there is not much variation in the results for different experimental runs having a random train-test split. 

\begin{table}[h!]
    \caption{Classification results (standard deviation values over $5$ runs) on  \textbf{Protein Subcellular} datasets for different evaluation metrics.}
    \label{tbl_results_protein_subcellular_std}
\centering
\resizebox{0.49\textwidth}{!}{
 \begin{tabular}{p{1.2cm}lp{0.9cm}p{0.9cm}p{0.9cm}p{0.9cm}p{0.9cm}p{0.9cm}p{1.4cm}}
    \toprule
        \multirow{2}{1.2cm}{Embeddings} & \multirow{2}{*}{Algo.} & \multirow{2}{*}{Acc.} & \multirow{2}{*}{Prec.} & \multirow{2}{*}{Recall} & \multirow{2}{0.9cm}{F1 (Weig.)} & \multirow{2}{0.9cm}{F1 (Macro)} & \multirow{2}{0.9cm}{ROC AUC} & Train Time (sec.)\\
        \midrule \midrule
        \multirow{7}{*}{OHE}
 & SVM & 0.005339 & 0.005547 & 0.005339 & 0.004908 & 0.00894 & 0.003092 & 3.916014 \\
 & NB & 0.165977 & 0.047923 & 0.165977 & 0.124314 & 0.028412 & 0.021885 & 0.234008  \\
 & MLP & 0.010331 & 0.011387 & 0.010331 & 0.010000 & 0.024026 & 0.013581 & 5.913928  \\
 & KNN & 0.015266 & 0.00962 & 0.015266 & 0.014351 & 0.016323 & 0.006342 & 2.590869 \\
 & RF & 0.007615 & 0.010725 & 0.007615 & 0.008411 & 0.016425 & 0.007389 & 0.433674 \\
 & LR & 0.006048 & 0.005895 & 0.006048 & 0.006385 & 0.01425 & 0.005388 & 2.752432 \\
 & DT & 0.004937 & 0.005132 & 0.004937 & 0.004463 & 0.003758 & 0.002974 & 0.437479 \\
        \cmidrule{2-9} 
        \multirow{7}{*}{Spike2Vec}
 & SVM & 0.01981 & 0.02825 & 0.01981 & 0.02270 & 0.01555 & 0.00959 & 0.47610 \\
 & NB & 0.00864 & 0.03391 & 0.00864 & 0.01524 & 0.01116 & 0.00755 & 0.00741 \\
 & MLP & 0.00267 & 0.00347 & 0.00267 & 0.00356 & 0.00022 & 0.00257 & 5.75725 \\
 & KNN & 0.01714 & 0.02418 & 0.01714 & 0.01927 & 0.01990 & 0.01220 & 0.00993 \\
 & RF & 0.00814 & 0.00652 & 0.00814 & 0.00645 & 0.00206 & 0.00159 & 0.03782 \\
 & LR & 0.00726 & 0.01165 & 0.00726 & 0.01068 & 0.00633 & 0.00329 & 0.03185 \\
 & DT & 0.01457 & 0.01302 & 0.01457 & 0.01389 & 0.01716 & 0.00889 & 0.00797 \\
        \cmidrule{2-9} 
        \multirow{7}{*}{PWM2Vec}
& SVM & 0.01386 & 0.01648 & 0.01386 & 0.01651 & 0.01519 & 0.00998 & 0.39967 \\
 & NB & 0.01718 & 0.02313 & 0.01718 & 0.02055 & 0.01788 & 0.00860 & 0.00820 \\
 & MLP & 0.01803 & 0.01902 & 0.01803 & 0.01891 & 0.01228 & 0.00753 & 7.02375 \\
 & KNN & 0.01043 & 0.01275 & 0.01043 & 0.01042 & 0.00859 & 0.00419 & 0.03936 \\
 & RF & 0.02213 & 0.01162 & 0.02213 & 0.02231 & 0.01838 & 0.01264 & 0.04062 \\
 & LR & 0.01802 & 0.01861 & 0.01802 & 0.02047 & 0.01716 & 0.01037 & 0.00798 \\
 & DT & 0.00925 & 0.01388 & 0.00925 & 0.01112 & 0.00742 & 0.00530 & 0.01723 \\
         \cmidrule{2-9}
        \multirow{7}{*}{String Kernel}
 & SVM & 0.00952 & 0.00581 & 0.00952 & 0.00786 & 0.00160 & 0.01255 & 0.08849 \\
 & NB & 0.03677 & 0.05084 & 0.03677 & 0.03489 & 0.02923 & 0.01730 & 0.00058 \\
 & MLP & 0.02344 & 0.06729 & 0.02344 & 0.03073 & 0.03126 & 0.01631 & 4.14838 \\
 & KNN & 0.01650 & 0.01933 & 0.01650 & 0.01870 & 0.01455 & 0.00640 & 0.00274 \\
 & RF & 0.02286 & 0.03968 & 0.02286 & 0.02447 & 0.02789 & 0.01317 & 0.07689 \\
 & LR & 0.00959 & 0.03215 & 0.00959 & 0.01878 & 0.02042 & 0.00588 & 0.00183 \\
 & DT & 0.03106 & 0.03357 & 0.03106 & 0.03238 & 0.03658 & 0.02079 & 0.00885 \\
          \cmidrule{2-9} 
           \multirow{7}{*}{WDGRL}  
     & SVM & 0.005067 & 0.00313 & 0.005067 & 0.004219 & 0.000845 & 0.007464 & 0.036333 \\
 & NB & 0.018709 & 0.028927 & 0.018709 & 0.012416 & 0.009551 & 0.007475 & 0.000338 \\
 & MLP & 0.00703 & 0.037114 & 0.00703 & 0.01059 & 0.011773 & 0.006923 & 0.932333 \\
 & KNN & 0.00703 & 0.005878 & 0.00703 & 0.006132 & 0.003794 & 0.002335 & 0.001434 \\
 & RF & 0.008417 & 0.011911 & 0.008417 & 0.008514 & 0.006176 & 0.003087 & 0.044674 \\
 & LR & 0.005067 & 0.016623 & 0.005067 & 0.011562 & 0.013997 & 0.004144 & 0.000185 \\
 & DT & 0.017918 & 0.01796 & 0.017918 & 0.018219 & 0.016001 & 0.009088 & 0.002927 \\
  \cmidrule{2-9}
 \multirow{7}{1.9cm}{Autoencoder}  
  & SVM & 0.00920 & 0.01191 & 0.00920 & 0.01139 & 0.00167 & 0.00084 & 0.41194 \\
 & NB & 0.12364 & 0.07618 & 0.12364 & 0.09485 & 0.03558 & 0.01111 & 0.04840 \\
 & MLP & 0.01107 & 0.01326 & 0.01107 & 0.01200 & 0.01462 & 0.00857 & 0.74852 \\
 & KNN & 0.01132 & 0.01190 & 0.01132 & 0.01157 & 0.01692 & 0.01040 & 0.02262 \\
 & RF & 0.00792 & 0.01075 & 0.00792 & 0.00992 & 0.01423 & 0.00511 & 0.36076 \\
 & LR & 0.00877 & 0.01212 & 0.00877 & 0.01124 & 0.00138 & 0.00081 & 0.30021 \\
 & DT & 0.00890 & 0.01560 & 0.00890 & 0.01162 & 0.02010 & 0.00778 & 0.21743 \\
  \cmidrule{2-9}
\multirow{7}{1.9cm}{SeqVec}
 & SVM & 0.00921 & 0.00812 & 0.00921 & 0.00927 & 0.00116 & 0.00061 & 0.25855 \\
 & NB & 0.10878 & 0.04510 & 0.10878 & 0.07567 & 0.03016 & 0.01570 & 0.04303 \\
 & MLP & 0.00943 & 0.01155 & 0.00943 & 0.01044 & 0.02354 & 0.01569 & 0.77613 \\
 & KNN & 0.01221 & 0.01064 & 0.01221 & 0.00991 & 0.01766 & 0.00902 & 0.02768 \\
 & RF & 0.00694 & 0.01022 & 0.00694 & 0.00697 & 0.01370 & 0.00901 & 0.33069 \\
 & LR & 0.01052 & 0.00901 & 0.01052 & 0.01049 & 0.00148 & 0.00071 & 0.10309 \\
 & DT & 0.00974 & 0.00994 & 0.00974 & 0.00919 & 0.02002 & 0.01160 & 0.07449 \\
 \cmidrule{2-9} 
\multirow{1}{1.9cm}{Protein Bert}
 & \_ & 0.008258 & 0.005042 & 0.008258 & 0.006823 & 0.001385 & 0.010888 & 0.076791 \\
 \cmidrule{2-9} 
 \multirow{7}{1.9cm}{CCP}
      & SVM & 0.00648 & 0.00599 & 0.00498 & 0.00745 & 0.01347 & 0.00793 & 0.32149 \\
 & NB & 0.01322 & 0.01478 & 0.01395 & 0.01544 & 0.01815 & 0.01133 & 0.53631 \\
 & MLP & 0.00698 & 0.00894 & 0.00643 & 0.00845 & 0.01322 & 0.00813 & 3.13462 \\
 & KNN & 0.01231 & 0.01056 & 0.01487 & 0.01254 & 0.01998 & 0.01676 & 0.76152 \\
 & RF & 0.00542 & 0.00854 & 0.00322 & 0.00233 & 0.01543 & 0.00953 & 0.91258 \\
 & LR & 0.00435 & 0.00743 & 0.00434 & 0.00643 & 0.00734 & 0.00532 & 0.26145 \\
 & DT & 0.00543 & 0.00512 & 0.00743 & 0.00832 & 0.01843 & 0.00743 & 0.29754 \\
 \cmidrule{2-9} 
\multirow{7}{1.9cm}{CCP-NN}
      & SVM & 0.00499 & 0.00674 & 0.00612 & 0.00687 & 0.01611 & 0.00974 & 0.22764\\
 & NB & 0.01527 & 0.01412 & 0.01357 & 0.01314 & 0.01809 & 0.01167 & 0.51931\\
 & MLP & 0.00925 & 0.00814 & 0.00853 & 0.00847 & 0.01363 & 0.00874 & 1.49651\\
 & KNN & 0.01156 & 0.01013 & 0.01432 & 0.01216 & 0.03311 & 0.01356 & 0.54301\\
 & RF & 0.00567 & 0.00854 & 0.00565 & 0.00542 & 0.01653 & 0.00756 & 0.78123\\
 & LR & 0.00654 & 0.00432 & 0.00632 & 0.00425 & 0.00753 & 0.00594 & 0.34325\\
 & DT & 0.00712 & 0.00578 & 0.00713 & 0.00835 & 0.01831 & 0.00845 & 0.29543\\
         \bottomrule
         \end{tabular}
}
\end{table}

Detailed results along with their discussion of the Coronavirus Host and Human DNA datasets are reported in Section~\ref{sec_host_results_appendix} and Section~\ref{sec_HumanDNA_results_appendix}, respectively.
The observed improvement in classification results when using the proposed method over the original CCP method can be attributed to several technical and logical factors discussed in detail in Section~\ref{sec_discuss_appendix}.

\subsection{Results For Coronavirus Host Data}\label{sec_host_results_appendix}
The classification results (averaged over $5$ runs) for the proposed CCP-NN and its comparison with the CCP approach for the Coronavirus Host dataset are shown in Table~\ref{tbl_results_part1_host_data}.
The best values for each embedding method are underlined, while the overall best values among all methods are shown in bold. We can observe that the proposed CCP-NN outperforms the original CCP-based low-dimensional representation for all evaluation metrics (other than the classifier training runtime) and achieves near-perfect predictive classification performance. Although the performance gain for CCP-NN compared to the best CCP-based results is not significant, it still outperforms CCP for all evaluation metrics other than training runtime.

The comparison of the best performing proposed method from Table~\ref{tbl_results_part1_host_data}, i.e., CCP-NN with Spike2Vec, with the existing baseline models is shown in Table~\ref{tbl_results_Host_org} for the Coronavirus Host dataset. We can observe that the proposed method significantly outperforms all baselines for all evaluation metrics other than the training runtime. Specifically, in terms of average accuracy, the CCP-NN with Spike2Vec embedding achieves a $1.7\%$ improvement compared to the second best results (i.e., original Spike2Vec with Random Forest and Logistic Regression classifiers) and achieves a higher average accuracy score in the case of the Coronavirus Host dataset.

\begin{table}[h!]
\caption{Classification results (averaged over $5$ runs) on \textbf{Coronavirus Host} dataset for different evaluation metrics using Nearest Neighbour CCP (CCP-NN) and CCP. The best value for each embedding is shown with an underline. The overall best value for each evaluation metric is shown in bold.
The $\uparrow$ means a higher number is better, and the down arrow $\downarrow$ means that a lower number is better.
}
    \label{tbl_results_part1_host_data}
    \centering
  \resizebox{0.49\textwidth}{!}{
    \begin{tabular}{clcp{0.5cm}p{0.5cm}p{0.5cm}p{0.5cm}p{0.5cm}p{0.5cm}p{0.8cm}}
    \toprule
    & \multirow{2}{1.2cm}{Embeddings} & \multirow{2}{*}{Algo.} & \multirow{2}{*}{Acc. $\uparrow$} & \multirow{2}{*}{Prec. $\uparrow$} & \multirow{2}{*}{Recall $\uparrow$} & \multirow{2}{0.8cm}{F1 (Weig.) $\uparrow$} & \multirow{2}{0.8cm}{F1 (Macro) $\uparrow$} & \multirow{2}{0.8cm}{ROC AUC $\uparrow$} & Train Time (sec.) $\downarrow$\\
        \midrule \midrule
 \multirow{28}{0.8cm}{CCP ($\phi_{CCP}$)} & 
 \multirow{7}{1cm}{OHE}
& SVM & 0.819 & 0.817 & 0.819 & 0.815 & 0.651 & 0.820 & 45.784 	\\
& & NB & 0.616 & 0.738 & 0.616 & 0.620 & 0.442 & 0.753 & \underline{0.312}		\\
& & MLP & 0.830 & 0.823 & 0.830 & 0.815 & 0.587 & 0.798 & 29.640    \\
& & KNN & 0.814 & 0.815 & 0.814 & 0.805 & 0.621 & 0.805 & 0.503      \\
& & RF & \underline{0.846} & 0.843 & \underline{0.846} & 0.835 & 0.667 & 0.828 & 1.185 \\
& & LR & 0.833 & 0.813 & 0.833 & 0.807 & 0.600 & 0.804 & 20.693     \\
& & DT & 0.845 & \underline{0.847} & 0.845 & \underline{0.837} & \underline{0.677} & \underline{0.835} & 0.283       \\

\cmidrule{3-10}                                                                                      
& \multirow{7}{1cm}{Spike2Vec}                               
& SVM & 0.794 & 0.811 & 0.794 & 0.781 & 0.666 & 0.827 & 22.888	\\
& & NB & 0.606 & 0.742 & 0.606 & 0.557 & 0.442 & 0.729 & \underline{1.609}       \\
& & MLP & 0.829 & 0.840 & 0.829 & 0.822 & 0.649 & 0.835 & 416.592  \\
& & KNN & 0.805 & 0.817 & 0.805 & 0.802 & 0.626 & 0.818 & 2.226      \\
& & RF & \underline{0.852} & \underline{0.853} & \underline{0.852} & \underline{0.846} & \underline{0.712} & \underline{0.840} & 8.844       \\
& & LR & 0.769 & 0.796 & 0.769 & 0.757 & 0.630 & 0.796 & 60.627     \\
& & DT & 0.826 & 0.827 & 0.826 & 0.821 & 0.602 & 0.813 & 2.207       \\
\cmidrule{3-10}
 
&  \multirow{7}{1cm}{PWM2Vec}                                                                                 
& SVM & 0.820 & 0.816 & 0.820 & 0.811 & 0.643 & 0.846 & 4.292		\\
& & NB & 0.434 & 0.434 & 0.434 & 0.358 & 0.358 & 0.714 & 0.412       \\
& & MLP & 0.804 & 0.804 & 0.804 & 0.795 & 0.589 & 0.797 & 7.458      \\
& & KNN & 0.798 & 0.800 & 0.798 & 0.792 & 0.613 & 0.811 & \underline{0.234}      \\
& & RF & \underline{0.833} & \underline{0.829} & \underline{0.833} & \underline{0.825} & \underline{0.671} & \underline{0.845} & 6.868       \\
& & LR & 0.810 & 0.808 & 0.810 & 0.799 & 0.579 & 0.787 & 16.907     \\
& & DT & 0.793 & 0.790 & 0.793 & 0.789 & 0.616 & 0.808 & 2.642       \\
\cmidrule{3-10}                                                                                        
&  \multirow{7}{1cm}{Autoencoder}                                                    & SVM & 0.630 & 0.620 & 0.630 & 0.605 & 0.234 & 0.615 & 3.189	\\
& & NB & 0.440 & 0.489 & 0.440 & 0.382 & 0.354 & 0.695 & \underline{\textbf{0.138}}   \\
& & MLP & 0.706 & 0.680 & 0.706 & 0.689 & 0.325 & 0.655 & 9.281  \\
& & KNN & 0.730 & 0.749 & 0.730 & 0.732 & 0.482 & 0.753 & 0.186  \\
& & RF & \underline{0.818} & \underline{0.819} & \underline{0.818} & \underline{0.810} & \underline{0.646} & \underline{0.799} & 7.642   \\
& & LR & 0.625 & 0.625 & 0.625 & 0.598 & 0.229 & 0.605 & 3.425   \\
& & DT & 0.755 & 0.753 & 0.755 & 0.750 & 0.516 & 0.764 & 1.218   \\
\midrule
 \multirow{28}{1cm}{CCP NN ($\phi_{CCP\_NN}$)} & 
 \multirow{7}{1cm}{OHE}
& SVM & 0.815 & 0.812 & 0.815 & 0.814 & 0.640 & 0.812 & 39.413 \\
& & NB & 0.596 & 0.711 & 0.596 & 0.597 & 0.354 & 0.704 & 0.295  \\
& & MLP & 0.830 & 0.806 & 0.830 & 0.808 & 0.575 & 0.796 & 31.610  \\
& & KNN & 0.813 & 0.814 & 0.813 & 0.802 & 0.592 & 0.785 & 0.508  \\
& & RF & \underline{0.843} & \underline{0.832} & \underline{0.843} & \underline{0.825} & \underline{0.641} & \underline{0.819} & 1.134 \\
& & LR & 0.778 & 0.763 & 0.778 & 0.750 & 0.393 & 0.700 & 20.612 \\
& & DT & 0.835 & 0.828 & 0.835 & 0.820 & 0.633 & 0.818 & \underline{0.179} \\
\cmidrule{3-10}                                                                                      
& \multirow{7}{1cm}{Spike2Vec}                               
& SVM & 0.827 & 0.832 & 0.827 & 0.818 & 0.758 & 0.871 & 17.080  \\
& & NB & 0.583 & 0.724 & 0.583 & 0.524 & 0.454 & 0.729 & 1.523   \\
& & MLP & 0.837 & 0.844 & 0.837 & 0.831 & 0.710 & 0.866 & 334.025   \\
& & KNN & 0.810 & 0.812 & 0.810 & 0.806 & 0.677 & 0.848 & 1.489 \\
& & RF & \underline{\textbf{0.855}} & \underline{\textbf{0.855}} & \underline{\textbf{0.855}} & \underline{\textbf{0.849}} & \underline{\textbf{0.770}} & \underline{\textbf{0.872}} & 5.547  \\
& & LR & 0.800 & 0.813 & 0.800 & 0.792 & 0.738 & 0.861 & 45.502  \\
& & DT & 0.845 & 0.846 & 0.845 & 0.841 & 0.703 & 0.864 & \underline{1.343} \\
\cmidrule{3-10}  
 
&  \multirow{7}{1cm}{PWM2Vec}                    
& SVM & 0.839 & 0.836 & 0.839 & 0.831 & 0.612 & 0.797 & 4.404 \\
& & NB & 0.648 & 0.700 & 0.648 & 0.633 & 0.476 & 0.746 & 0.390  \\
& & MLP & 0.825 & 0.826 & 0.825 & 0.818 & 0.584 & 0.788 & 6.728  \\
& & KNN & 0.781 & 0.787 & 0.781 & 0.777 & 0.591 & 0.781 & \underline{0.237}  \\
& & RF & \underline{0.854} & \underline{0.856} & \underline{0.854} & \underline{0.849} & \underline{0.682} & \underline{0.835} & 6.202  \\
& & LR & 0.824 & 0.815 & 0.824 & 0.811 & 0.545 & 0.755 & 15.188 \\
& & DT & 0.824 & 0.827 & 0.824 & 0.820 & 0.594 & 0.795 & 2.739  \\
\cmidrule{3-10}                                                                                        
&  \multirow{7}{1cm}{Autoencoder}                                                    
 & SVM & 0.494 & 0.347 & 0.494 & 0.404 & 0.110 & 0.547 & 3.865 \\
& & NB & 0.468 & 0.578 & 0.468 & 0.490 & 0.153 & 0.647 & \underline{0.141} \\
& & MLP & 0.696 & 0.671 & 0.696 & 0.671 & 0.283 & 0.631 & 9.500 \\
& & KNN & 0.729 & 0.733 & 0.729 & 0.721 & 0.447 & 0.715 & 0.168  \\
& & RF & \underline{0.834} & \underline{0.836} & \underline{0.834} & \underline{0.830} & \underline{0.582} & \underline{0.799} & 6.296  \\
& & LR & 0.464 & 0.332 & 0.464 & 0.372 & 0.099 & 0.538 & 2.518 \\
& & DT & 0.805 & 0.809 & 0.805 & 0.802 & 0.532 & 0.783 & 1.104  \\
         \bottomrule
         \end{tabular}
    }
\end{table}

\begin{table}[h!]
     \caption{Classification result comparisons (averaged over $5$ runs) for the best performing proposed method (i.e., CCP-NN with Spike2Vec) with baselines on \textbf{Coronavirus Host} dataset for different evaluation metrics. The best values are in bold. 
     The $\uparrow$ means a higher number is better, and the down arrow $\downarrow$ means that a lower number is better.
     }
    \label{tbl_results_Host_org}
\centering
\resizebox{0.45\textwidth}{!}{
 \begin{tabular}{p{1.2cm}lp{0.8cm}p{0.8cm}p{0.8cm}p{0.8cm}p{0.8cm}p{0.8cm}p{1cm}}
    \toprule
        \multirow{2}{1.2cm}{Embeddings} & \multirow{2}{*}{Algo.} & \multirow{2}{*}{Acc. $\uparrow$} & \multirow{2}{*}{Prec. $\uparrow$} & \multirow{2}{*}{Recall $\uparrow$} & \multirow{2}{1.2cm}{F1 (Weig.) $\uparrow$} & \multirow{2}{1.2cm}{F1 (Macro) $\uparrow$} & \multirow{2}{1.2cm}{ROC AUC $\uparrow$} & Train Time (sec.) $\downarrow$\\
        \midrule \midrule
         \multirow{7}{1.2cm}{OHE}
        & SVM & 0.822 & 0.834 & 0.822 & 0.823 & 0.728 & 0.839 & 389.128 \\
        & NB & 0.677 & 0.808 & 0.677 & 0.654 &  0.517 & 0.815 & 56.741 \\
        & MLP & 0.779 & 0.761 & 0.779 & 0.757 & 0.622 & 0.715 & 390.289 \\
        & KNN & 0.805 & 0.794 & 0.805 & 0.792 & 0.674 & 0.781 & 16.211 \\
        & RF & \underline{0.836} & 0.831 & \underline{0.836} & 0.822 &  0.709 & \underline{0.832} & 151.911 \\
        & LR & 0.835 & \underline{0.849} & 0.835 & \underline{0.824} &  \underline{0.734} & \underline{0.832} & 48.786 \\
        & DT & 0.824 & 0.833 & 0.824 & 0.811 &  0.679 & 0.810 & \underline{21.581} \\
       \cmidrule{2-9}
        \multirow{7}{1.2cm}{Spike2Vec}
        & SVM & 0.848 & \underline{0.852} & 0.848 & 0.842 & 0.739 & \underline{0.883} & 191.066 	\\
        & NB & 0.661 & 0.768 & 0.661 & 0.661 &  0.522 & 0.764 & 10.220     \\
        & MLP & 0.815 & 0.837 & 0.815 & 0.814 & 0.640 & 0.835 & 46.624    \\
        & KNN & 0.782 & 0.794 & 0.782 & 0.781 & 0.686 & 0.832 & 82.112    \\
        & RF & \underline{0.853} & 0.848 & \underline{0.853} & 0.845 & 0.717 & 0.864 & 15.915     \\
        & LR & \underline{0.853} & \underline{0.852} & \underline{0.853} & \underline{0.846} & \underline{0.757} & 0.879 & 60.620     \\
        & DT & 0.829 & 0.827 & 0.829 & 0.825 & 0.696 & 0.855 & \underline{4.261}      \\
       \cmidrule{2-9}
        \multirow{7}{1.2cm}{PWM2Vec}
        & SVM & 0.799 & 0.806 & 0.799 & 0.801 & 0.648 & 0.859 & 44.793	\\
        & NB & 0.381 & 0.584 & 0.381 & 0.358 & 0.400 & 0.683 & \underline{2.494}    \\
        & MLP & 0.782 & 0.792 & 0.782 & 0.778 & 0.693 & 0.848 & 21.191  \\
        & KNN & 0.786 & 0.782 & 0.786 & 0.779 & 0.679 & 0.838 & 12.933  \\
        & RF & \underline{0.836} & \underline{0.839} & \underline{0.836} & \underline{0.828} & \underline{0.739} & \underline{0.862} & 7.690    \\
        & LR & 0.809 & 0.815 & 0.809 & 0.800 & 0.728 & 0.852 & 274.91  \\
        & DT & 0.801 & 0.802 & 0.801 & 0.797 & 0.633 & 0.829 & 4.537    \\

        \cmidrule{2-9} 
        \multirow{7}{1.2cm}{Autoencoder}
        & SVM & 0.602 & 0.588 & 0.602 & 0.590 &  0.519 & 0.759 & 2575.9 \\
         & NB & 0.261 & 0.520 & 0.261 & 0.303 & 0.294 & 0.673 & 21.74 \\
         & MLP & 0.486 & 0.459 & 0.486 & 0.458 & 0.216 &  0.594 & 29.93 \\
         & KNN & 0.763 & 0.764 & 0.763 & 0.755 & 0.547 & 0.784 & \underline{18.51} \\
         & RF &  \underline{0.800} & \underline{0.796} & \underline{0.800} & \underline{0.791} & \underline{0.648} & \underline{0.815} &  57.90\\
         & LR & 0.717 & 0.750 & 0.717 & 0.702 & 0.564 & 0.812 & 11072.6 \\
         & DT & 0.772 & 0.767 & 0.772 & 0.765 & 0.571 & 0.808 & 121.36 \\ 
        
        \cmidrule{2-9}
        \multirow{7}{1.2cm}{String Kernel}
        & SVM & 0.601 & 0.673 & 0.601 & 0.602 & 0.325 & 0.624 & 5.198 \\
             & NB & 0.230 & 0.665 & 0.230 & 0.295 & 0.162 & 0.625 & \underline{0.131} \\
             & MLP & 0.647 & 0.696 & 0.647 & 0.641 & 0.302 & 0.628 & 42.322 \\
             & KNN & 0.613 & 0.623 & 0.613 & 0.612 & 0.310 & 0.629 & 0.434 \\
             & RF & \underline{0.668} & 0.692 & \underline{0.668} & \underline{0.663} & \underline{0.360} & \underline{0.658} & 4.541 \\
             & LR & 0.554 & \underline{0.724} & 0.554 & 0.505 & 0.193 & 0.568 & 5.096 \\
             & DT &  0.646 & 0.674 & 0.646 & 0.643 & 0.345 & 0.653 & 1.561 \\
         \cmidrule{2-9}        
           \multirow{7}{1.2cm}{WDGRL}  
        & SVM & 0.329 & 0.108 & 0.329 & 0.163 & 0.029 & \underline{0.500} & 2.859 \\
         & NB & 0.004 & 0.095 & 0.004 & 0.007 & 0.002 &  0.496 & \textbf{\underline{0.008}}  \\
         & MLP & 0.328 & 0.136 & 0.328 & 0.170 & 0.032 & 0.499 & 5.905  \\
         & KNN & 0.235 & 0.198 & 0.235 & 0.211 & \underline{0.058} & 0.499 & 0.081  \\
         & RF & 0.261 & 0.196 & 0.261 & \underline{0.216} & 0.051 & 0.499 &  1.288 \\
         & LR & \underline{0.332} & 0.149 & \underline{0.332} & 0.177 & 0.034 & \underline{0.500} & 0.365   \\
         & DT & 0.237 & \underline{0.202} & 0.237 & 0.211 & 0.054 & 0.498 & 0.026  \\
             
          \cmidrule{2-9}
        \multirow{7}{1.2cm}{SeqVec}
        & SVM & 0.711 & 0.745 & 0.711 & 0.698 & 0.497 & 0.747 & 0.751 \\
         & NB & 0.503 & 0.636 & 0.503 & 0.554 & 0.413 & 0.648 & \underline{0.012} \\
         & MLP & 0.718 & 0.748 & 0.718 & 0.708 & 0.407 & 0.706 & 10.191  \\
         & KNN & 0.815 & 0.806 & 0.815 & 0.809 & 0.588 & 0.800 & 0.418  \\
         & RF & \underline{0.833} & \underline{0.824} & \underline{0.833} & \underline{0.828} & \underline{0.678} & \underline{0.839} &  1.753   \\
         & LR & 0.673 & 0.683 & 0.673 & 0.654 & 0.332 & 0.660 & 1.177   \\
         & DT & 0.778 & 0.786 & 0.778 & 0.781 & 0.618 & 0.825 & 0.160  \\ 
         \cmidrule{2-9}  
        \multirow{1}{1.9cm}{Protein Bert}
         & \_ & 0.799 &  0.806 & 0.799 & 0.789 & 0.715 & 0.841 & 15742.9 \\


\cmidrule{2-9}
  \multirow{7}{1.6cm}{TAPE}
  & SVM & 0.818 & 0.823 & 0.818 & 0.811 & 0.711 & \underline{0.854} & 3.201 \\
 & NB & 0.482 & 0.587 & 0.482 & 0.442 & 0.400 & 0.712 & 0.494 \\
 & MLP & 0.812 & 0.819 & 0.812 & 0.802 & 0.665 & 0.828 & 3.737 \\
 & KNN & 0.793 & 0.797 & 0.793 & 0.789 & 0.633 & 0.818 & \underline{0.150} \\
 & RF & \underline{0.830} & \underline{0.834} & \underline{0.830} & \underline{0.823} & \underline{0.725} & 0.846 & 13.656 \\
 & LR & 0.779 & 0.797 & 0.779 & 0.764 & 0.628 & 0.794 & 11.325 \\
 & DT & 0.785 & 0.786 & 0.785 & 0.782 & 0.578 & 0.798 & 4.675 \\
 
        \cmidrule{2-9} 
           \multirow{7}{1.2cm}{$\phi_{CCP\_NN}$ (ours) - Spike2Vec} 
        & SVM & 0.309 & 0.216 & 0.309 & 0.152 & 0.078 & 0.503 & 7.596  \\
        & NB & 0.172 & 0.392 & 0.172 & 0.107 & 0.116 & 0.523 & \underline{0.174}  \\
        & MLP & 0.341 & 0.324 & 0.341 & 0.272 & 0.201 & 0.543 & 218.782  \\
        & KNN & 0.419 & 0.412 & 0.419 & 0.412 & 0.361 & 0.629 & 1.140  \\
        & RF & \underline{\textbf{0.870}} & \underline{\textbf{0.870}} & \underline{\textbf{0.870}} & \underline{\textbf{0.870}} & \underline{\textbf{0.864}} & \underline{\textbf{0.921}} & 19.982  \\
        & LR & 0.309 & 0.280 & 0.309 & 0.152 & 0.077 & 0.503 & 1.911  \\
        & DT & 0.808 & 0.808 & 0.808 & 0.807 & 0.796 & 0.880 & 3.766  \\

        
         \bottomrule
         \end{tabular}
}
\end{table}

Table~\ref{tbl_std_org_host} displays the results of the standard deviation (SD) (averaged over 5 runs) for both the baseline methods and the approach proposed in the Coronavirus Host dataset. The results indicate that in most cases the SD values are relatively low, usually below 0.02. This observation suggests that the classification results remain consistent between different experimental runs with random train-test splits. The low variability in the SD values reflects the stability of the classification performance reported for both the proposed and the baseline models.

\begin{table}[h!]
  \caption{Standard Deviation values of 5 runs for Classification results on the proposed and baseline methods for \textbf{Coronavirus Host} dataset.
  }
  \label{tbl_std_org_host}
\centering
\resizebox{0.49\textwidth}{!}{
 \begin{tabular}{p{1.2cm}lp{.8cm}p{.8cm}p{.8cm}p{.8cm}p{.8cm}p{.8cm}p{1.2cm}}
    \toprule
        \multirow{2}{1.2cm}{Embeddings} & \multirow{2}{*}{Algo.} & \multirow{2}{*}{Acc.} & \multirow{2}{*}{Prec.} & \multirow{2}{*}{Recall} & \multirow{2}{0.8cm}{F1 (Weig.) } & \multirow{2}{0.8cm}{F1 (Macro) } & \multirow{2}{0.8cm}{ROC AUC} & Train Time (sec.) \\
        \midrule \midrule
    \multirow{7}{*}{OHE}  
      & SVM & 0.010697 & 0.010309 & 0.010697 & 0.009847 & 0.006786 & 0.004762 & 1.818067 \\
 & NB & 0.014762 & 0.011664 & 0.014762 & 0.012998 & 0.008811 & 0.00592 & 0.047871 \\
 & MLP & 0.01903 & 0.027088 & 0.01903 & 0.022536 & 0.017811 & 0.006836 & 1.241431 \\
 & KNN & 0.005715 & 0.007481 & 0.005715 & 0.004432 & 0.005243 & 0.002815 & 0.405597 \\
 & RF & 0.011174 & 0.010344 & 0.011174 & 0.011743 & 0.013327 & 0.006616 & 0.201381 \\
 & LR & 0.059575 & 0.039483 & 0.059575 & 0.057232 & 0.060847 & 0.036954 & 96.708299 \\
 & DT & 0.010625 & 0.010962 & 0.010625 & 0.010682 & 0.011695 & 0.005174 & 0.157993 \\
 \cmidrule{2-9}	
    \multirow{7}{*}{Spike2Vec}  
     & SVM & 0.02187 & 0.03118 & 0.02187 & 0.02506 & 0.01717 & 0.01059 & 0.52562 \\
 & NB & 0.00954 & 0.03743 & 0.00954 & 0.01682 & 0.01232 & 0.00833 & 0.00818 \\
 & MLP & 0.00295 & 0.00383 & 0.00295 & 0.00393 & 0.00025 & 0.00284 & 6.35600 \\
 & KNN & 0.01892 & 0.02670 & 0.01892 & 0.02128 & 0.02197 & 0.01347 & 0.01097 \\
 & RF & 0.00898 & 0.00720 & 0.00898 & 0.00712 & 0.00227 & 0.00176 & 0.04175 \\
 & LR & 0.00802 & 0.01287 & 0.00802 & 0.01179 & 0.00698 & 0.00363 & 0.03516 \\
 & DT & 0.01608 & 0.01437 & 0.01608 & 0.01533 & 0.01894 & 0.00981 & 0.00880 \\
    \cmidrule{2-9}	
    \multirow{7}{*}{PWM2Vec}  
     & SVM & 0.01459 & 0.01735 & 0.01459 & 0.01737 & 0.01599 & 0.01051 & 0.42070 \\
 & NB & 0.01808 & 0.02434 & 0.01808 & 0.02163 & 0.01882 & 0.00905 & 0.00863 \\
 & MLP & 0.01898 & 0.02002 & 0.01898 & 0.01990 & 0.01293 & 0.00793 & 7.39342 \\
 & KNN & 0.01098 & 0.01342 & 0.01098 & 0.01097 & 0.00904 & 0.00441 & 0.04144 \\
 & RF & 0.02330 & 0.01223 & 0.02330 & 0.02348 & 0.01934 & 0.01331 & 0.04276 \\
 & LR & 0.01896 & 0.01959 & 0.01896 & 0.02155 & 0.01807 & 0.01092 & 0.00840 \\
 & DT & 0.00974 & 0.01461 & 0.00974 & 0.01171 & 0.00781 & 0.00558 & 0.01814 \\
    \cmidrule{2-9}	
    \multirow{7}{*}{String Kernel}  
 & SVM & 0.00892 & 0.00545 & 0.00892 & 0.00737 & 0.00150 & 0.01176 & 0.08293 \\
 & NB & 0.03446 & 0.04765 & 0.03446 & 0.03270 & 0.02739 & 0.01621 & 0.00054 \\
 & MLP & 0.02197 & 0.06306 & 0.02197 & 0.02880 & 0.02930 & 0.01528 & 3.28975 \\
 & KNN & 0.01546 & 0.01811 & 0.01546 & 0.01752 & 0.01364 & 0.00600 & 0.00257 \\
 & RF & 0.02143 & 0.03719 & 0.02143 & 0.02293 & 0.02613 & 0.01234 & 0.07206 \\
 & LR & 0.00898 & 0.03013 & 0.00898 & 0.01760 & 0.01914 & 0.00551 & 0.00171 \\
 & DT & 0.02911 & 0.03146 & 0.02911 & 0.03035 & 0.03428 & 0.01948 & 0.00829 \\
     \cmidrule{2-9}	
    \multirow{7}{*}{WDGRL}  
    & SVM & 0.008378 & 0.005078 & 0.008378 & 0.006888 & 0.00141 & 0.002417 & 0.034498 \\
 & NB & 0.008720 & 0.055455 & 0.00872 & 0.022475 & 0.022506 & 0.007747 & 0.000352 \\
 & MLP & 0.016103 & 0.010655 & 0.016103 & 0.018511 & 0.014246 & 0.007622 & 2.437331 \\
 & KNN & 0.010047 & 0.009635 & 0.010047 & 0.010275 & 0.011106 & 0.006126 & 0.002751 \\
 & RF & 0.013395 & 0.019497 & 0.013395 & 0.015266 & 0.018384 & 0.008316 & 0.035652 \\
 & LR & 0.007675 & 0.094971 & 0.007675 & 0.008903 & 0.005274 & 0.00175 & 0.001188 \\
 & DT & 0.009280 & 0.008941 & 0.00928 & 0.009266 & 0.007579 & 0.004472 & 0.004392 \\
\cmidrule{2-9}	
    \multirow{7}{1.9cm}{Autoencoder}  
  & SVM & 0.00956 & 0.00974 & 0.00956 & 0.01059 & 0.00127 & 0.00094 & 0.35010 \\
 & NB & 0.05871 & 0.04630 & 0.05871 & 0.05414 & 0.02578 & 0.01233 & 0.03049 \\
 & MLP & 0.00846 & 0.01142 & 0.00846 & 0.00776 & 0.01468 & 0.00882 & 1.00276 \\
 & KNN & 0.00631 & 0.00803 & 0.00631 & 0.00807 & 0.00699 & 0.00493 & 0.01380 \\
 & RF & 0.00338 & 0.00548 & 0.00338 & 0.00381 & 0.01029 & 0.00786 & 0.72100 \\
 & LR & 0.00982 & 0.00982 & 0.00982 & 0.01072 & 0.00128 & 0.00093 & 0.19975 \\
 & DT & 0.01025 & 0.00968 & 0.01025 & 0.00968 & 0.02163 & 0.00821 & 0.09998 \\

\cmidrule{2-9}	
    \multirow{7}{1.9cm}{SeqVec}  
  & SVM & 0.00729 & 0.00924 & 0.00729 & 0.00924 & 0.00102 & 0.00031 & 0.29267 \\
 & NB & 0.14408 & 0.06203 & 0.14408 & 0.11723 & 0.02721 & 0.01650 & 0.01298 \\
 & MLP & 0.01185 & 0.01247 & 0.01185 & 0.01129 & 0.02103 & 0.00999 & 0.68591 \\
 & KNN & 0.01281 & 0.01485 & 0.01281 & 0.01456 & 0.02329 & 0.01131 & 0.05557 \\
 & RF & 0.01050 & 0.01599 & 0.01050 & 0.01294 & 0.01490 & 0.00771 & 0.36725 \\
 & LR & 0.00795 & 0.00984 & 0.00795 & 0.01007 & 0.00126 & 0.00053 & 0.22741 \\
 & DT & 0.01119 & 0.01183 & 0.01119 & 0.01277 & 0.02537 & 0.00989 & 0.12363 \\
 \cmidrule{2-9}
 \multirow{1}{1.9cm}{Protein Bert}
         & \_ & 0.02965 & 0.03204 & 0.02965 & 0.03091 & 0.03492 & 0.01984 & 0.00845 \\
     \cmidrule{2-9}	
    \multirow{7}{1.2cm}{$\phi_{CCP\_NN}$ (ours) - Spike2Vec}  
       & SVM & 0.011316 & 0.020199 & 0.011316 & 0.017867 & 0.026025 & 0.014217 & 0.279999 \\
 & NB & 0.012632 & 0.018465 & 0.012632 & 0.012929 & 0.013448 & 0.006475 & 0.009835 \\
 & MLP & 0.007226 & 0.009827 & 0.007226 & 0.007801 & 0.014985 & 0.009328 & 4.298934 \\
 & KNN & 0.010121 & 0.012496 & 0.010121 & 0.0117 & 0.024522 & 0.012687 & 1.301837 \\
 & RF & 0.007323 & 0.011286 & 0.007323 & 0.00848 & 0.006726 & 0.001826 & 0.159344 \\
 & LR & 0.014064 & 0.02447 & 0.014064 & 0.016745 & 0.010772 & 0.005139 & 0.430449 \\
 & DT & 0.006303 & 0.011474 & 0.006303 & 0.007942 & 0.012572 & 0.006275 & 0.016553 \\
    \bottomrule
  \end{tabular}
  }
\end{table}

\subsection{Results For Human DNA Data}\label{sec_HumanDNA_results_appendix}
Table~\ref{tbl_results_part1_human_DNA_data} presents the classification results for both the proposed CCP-NN and the conventional CCP approach on the Human DNA dataset. The results are averaged over 5 runs. The best values for each embedding method are underlined, and the overall best values among all methods are shown in bold. It is evident that the proposed CCP-NN consistently outperforms the original CCP-based low-dimensional representation for all evaluation metrics, except for the classifier training runtime. The classification accuracy achieved by CCP-NN is notably higher, with the best-performing results using CCP-NN (with Autoencoder and Random Forest Classifier) showing a significant improvement of $10.8\%$ compared to the best results obtained from the original CCP (using One-Hot Encoding with Random Forest classifier).

\begin{table}[h!]
    \caption{Classification results (averaged over $5$ runs) on \textbf{Human DNA} dataset for different evaluation metrics using Nearest Neighbour CCP (CCP-NN) and CCP. The best value for each embedding is shown with an underline. The overall best value for each evaluation metric is shown in bold.
   The $\uparrow$ means a higher number is better, and the down arrow $\downarrow$ means that a lower number is better.
    }
    \label{tbl_results_part1_human_DNA_data}
    \centering
   \resizebox{0.49\textwidth}{!}{
    \begin{tabular}{clcp{0.5cm}p{0.5cm}p{0.5cm}p{0.5cm}p{0.5cm}p{0.5cm}p{0.8cm}}
    \toprule
    & \multirow{2}{1.2cm}{Embeddings} & \multirow{2}{*}{Algo.} & \multirow{2}{*}{Acc. $\uparrow$} & \multirow{2}{*}{Prec. $\uparrow$} & \multirow{2}{0.8cm}{Recall $\uparrow$} & \multirow{2}{0.8cm}{F1 (Weig.) $\uparrow$} & \multirow{2}{0.8cm}{F1 (Macro) $\uparrow$} & \multirow{2}{0.8cm}{ROC AUC $\uparrow$} & Train Time (sec.) $\downarrow$\\
        \midrule \midrule
\multirow{28}{1cm}{CCP ($\phi_{CCP}$)} & 
 \multirow{7}{1cm}{OHE}
& SVM & 0.370 & 0.466 & 0.370 & 0.260 & 0.213 & 0.553 & 6.585	\\
& & NB &  0.375 & 0.598 & 0.375 & 0.371 & 0.369 & 0.634 & \underline{0.138}   \\
& & MLP & 0.667 & 0.672 & 0.667 & 0.669 & 0.644 & 0.795 & 5.791  \\
& & KNN & 0.581 & 0.595 & 0.581 & 0.580 & 0.549 & 0.732 & 0.189  \\
& & RF & \underline{0.762} & \underline{0.813} & \underline{0.762} & \underline{0.763} & \underline{0.763} & \underline{0.829} & 1.789   \\
& & LR & 0.378 & 0.541 & 0.378 & 0.270 & 0.228 & 0.559 & 4.224   \\
& & DT & 0.661 & 0.663 & 0.661 & 0.661 & 0.638 & 0.791 & 0.554   \\

\cmidrule{3-10}                                                                                    
& \multirow{7}{1cm}{Spike2Vec}                        
& SVM & 0.447 & 0.417 & 0.447 & 0.385 & 0.303 & 0.602 & 1.212		\\
& & NB & 0.215 & 0.313 & 0.215 & 0.180 & 0.148 & 0.543 & \underline{\textbf{0.012}}       \\
& & MLP & 0.526 & 0.523 & 0.526 & 0.516 & 0.467 & 0.687 & 10.664    \\
& & KNN & 0.601 & 0.612 & 0.601 & 0.602 & 0.565 & 0.750 & 0.137      \\
& & RF & \underline{0.725} & \underline{0.753} & \underline{0.725} & \underline{0.721} & \underline{0.709} & \underline{0.806} & 2.693       \\
& & LR & 0.437 & 0.443 & 0.437 & 0.385 & 0.312 & 0.605 & 0.184       \\
& & DT & 0.593 & 0.597 & 0.593 & 0.594 & 0.561 & 0.746 & 0.142       \\                \cmidrule{3-10}                                 
&  \multirow{7}{1cm}{PWM2Vec}                   
& SVM & 0.312 & 0.302 & 0.312 & 0.162 & 0.085 & 0.505 & 2.884		\\
& & NB & 0.095 & 0.324 & 0.095 & 0.058 & 0.051 & 0.508 & \underline{0.017}         \\
& & MLP & 0.311 & 0.312 & 0.311 & 0.165 & 0.088 & 0.505 & 318.449    \\
& & KNN & 0.194 & 0.320 & 0.194 & 0.113 & 0.080 & 0.509 & 1.537        \\
& & RF & \underline{0.315} & \underline{0.346} & \underline{0.315} & \underline{0.178} & \underline{0.104} & \underline{0.509} & 1.952         \\
& & LR &  0.313 & 0.277 & 0.313 & 0.166 & 0.089 & 0.506 & 0.826         \\
& & DT & 0.310 & 0.293 & 0.310 & 0.174 & 0.100 & 0.507 & 0.062         \\
\cmidrule{3-10}                                                                                         
&  \multirow{7}{1cm}{Autoencoder}                                                    
 & SVM & 0.440 & 0.504 & 0.440 & 0.394 & 0.341 & 0.606 & 3.620		\\
& & NB & 0.192 & 0.312 & 0.192 & 0.147 & 0.147 & 0.539 & \underline{0.076}       \\
& & MLP & 0.485 & 0.483 & 0.485 & 0.481 & 0.440 & 0.672 & 33.924   \\
& & KNN & 0.496 & 0.498 & 0.496 & 0.494 & 0.462 & 0.685 & 0.262      \\
& & RF & \underline{0.593} & \underline{0.715} & \underline{0.593} & \underline{0.585} & \underline{0.576} & \underline{0.713} & 14.667     \\
& & LR & 0.414 & 0.452 & 0.414 & 0.370 & 0.308 & 0.593 & 2.159       \\
& & DT & 0.476 & 0.479 & 0.476 & 0.476 & 0.447 & 0.679 & 3.027       \\
\midrule
 \multirow{28}{1cm}{CCP Nearest Neighbor ($\phi_{CCP\_NN}$)} & 
 \multirow{7}{1cm}{OHE}
& SVM & 0.618 & 0.621 & 0.618 & 0.614 & 0.588 & 0.749 & 5.475 \\
& & NB & 0.357 & 0.573 & 0.357 & 0.344 & 0.343 & 0.624 & \underline{0.141} \\
& & MLP & 0.667 & 0.669 & 0.667 & 0.666 & 0.639 & 0.791 & 5.556 \\
& & KNN & 0.569 & 0.581 & 0.569 & 0.568 & 0.538 & 0.729 & 0.178  \\
& & RF & \underline{0.768} & \underline{0.828} & \underline{0.768} & \underline{0.771} & \underline{0.776} & \underline{0.834} & 1.727  \\
& & LR & 0.579 & 0.595 & 0.579 & 0.569 & 0.536 & 0.713 & 5.182 \\
& & DT & 0.675 & 0.677 & 0.675 & 0.675 & 0.653 & 0.801 & 0.498  \\
\cmidrule{3-10}                        
& \multirow{7}{1cm}{Spike2Vec}                           
& SVM & 0.309 & 0.226 & 0.309 & 0.153 & 0.087 & 0.506 & 1.467  \\
& & NB & 0.204 & 0.280 & 0.204 & 0.163 & 0.169 & 0.533 & \underline{0.013}  \\
& & MLP & 0.307 & 0.296 & 0.307 & 0.283 & 0.236 & 0.554 & 30.027 \\
& & KNN & 0.361 & 0.360 & 0.361 & 0.357 & 0.314 & 0.604 & 0.130  \\
& & RF & \underline{0.558} & \underline{0.690} & \underline{0.558} & \underline{0.545} & \underline{0.530} & \underline{0.688} & 3.561  \\
& & LR & 0.322 & 0.388 & 0.322 & 0.193 & 0.133 & 0.519 & 0.216  \\
& & DT & 0.532 & 0.532 & 0.532 & 0.531 & 0.502 & 0.709 & 0.234  \\                     \cmidrule{3-10}                                  
&  \multirow{7}{1cm}{PWM2Vec}            
& SVM & \underline{0.316} & 0.161 & \underline{0.316} & 0.160 & 0.080 & 0.503 & 2.380 \\
& & NB & 0.069 & 0.301 & 0.069 & 0.032 & 0.038 & \underline{0.506} & 0.015  \\
& & MLP & 0.314 & 0.147 & 0.314 & 0.158 & 0.078 & 0.502 & 131.941 \\
& & KNN & 0.222 & \underline{0.303} & 0.222 & 0.106 & 0.067 & 0.504 & 1.198 \\
& & RF & 0.315 & 0.190 & 0.315 & 0.161 & 0.081 & 0.503 & 0.941 \\
& & LR & 0.314 & 0.140 & 0.314 & 0.159 & 0.079 & 0.502 & 0.494  \\
& & DT & 0.315 & 0.192 & 0.315 & \underline{0.163} & \underline{0.083} & 0.503 & \underline{0.005}  \\
\cmidrule{3-10}                  
&  \multirow{7}{1cm}{Autoencoder}                                                    
 & SVM & 0.309 & 0.216 & 0.309 & 0.152 & 0.078 & 0.503 & 7.596  \\
& & NB & 0.172 & 0.392 & 0.172 & 0.107 & 0.116 & 0.523 & \underline{0.174}  \\
& & MLP & 0.341 & 0.324 & 0.341 & 0.272 & 0.201 & 0.543 & 218.782  \\
& & KNN & 0.419 & 0.412 & 0.419 & 0.412 & 0.361 & 0.629 & 1.140  \\
& & RF & \underline{\textbf{0.870}} & \underline{\textbf{0.870}} & \underline{\textbf{0.870}} & \underline{\textbf{0.870}} & \underline{\textbf{0.864}} & \underline{\textbf{0.921}} & 19.982  \\
& & LR & 0.309 & 0.280 & 0.309 & 0.152 & 0.077 & 0.503 & 1.911  \\
& & DT & 0.808 & 0.808 & 0.808 & 0.807 & 0.796 & 0.880 & 3.766  \\
         \bottomrule
         \end{tabular}
    }
\end{table}

\begin{table}[h!]
    \caption{Classification result comparisons (averaged over $5$ runs) for the best performing proposed method (i.e., CCP-NN with Autoencoder) with baselines on \textbf{Human DNA} dataset for different evaluation metrics. The best values are in bold. 
    The $\uparrow$ means a higher number is better, and the down arrow $\downarrow$ means that a lower number is better.}
    \label{tbl_results_Human_DNA_org}
\centering
\resizebox{0.45\textwidth}{!}{
 \begin{tabular}{p{1.2cm}lp{0.5cm}p{0.5cm}p{0.5cm}p{0.5cm}p{0.5cm}p{0.5cm}p{0.8cm}}
    \toprule
        \multirow{2}{1.2cm}{Embeddings} & \multirow{2}{*}{Algo.} & \multirow{2}{*}{Acc. $\uparrow$} & \multirow{2}{*}{Prec. $\uparrow$} & \multirow{2}{0.8cm}{Recall $\uparrow$} & \multirow{2}{0.8cm}{F1 (Weig.) $\uparrow$} & \multirow{2}{0.8cm}{F1 (Macro) $\uparrow$} & \multirow{2}{0.8cm}{ROC AUC $\uparrow$} & Train Time (sec.) $\downarrow$\\
        \midrule \midrule
         \multirow{7}{2cm}{OHE}
        & SVM & 0.579 & 0.599 & 0.579 & 0.576 & 0.561 & 0.721 & 10.475 \\
         & NB & 0.165 & 0.142 & 0.165 & 0.101 & 0.125 & 0.529 & \underline{0.746} \\
         & MLP & 0.600 & 0.611 & 0.600 & 0.612 & 0.564 & 0.723 & 45.785 \\
         & KNN & 0.638 & 0.649 & 0.638 & 0.640 & 0.598 & 0.754 & 1.574 \\
         & RF & \underline{0.722} & \underline{0.768} & \underline{0.722} & \underline{0.741} & \underline{0.729} & \underline{0.811} & 5.749 \\
         & LR & 0.566 & 0.568 & 0.566 & 0.574 & 0.521 & 0.698 & 9.781 \\
         & DT & 0.611 & 0.615 & 0.611 & 0.619 & 0.590 & 0.747 & 0.749 \\
       \cmidrule{2-9}
        \multirow{7}{2cm}{Spike2Vec}
        & SVM & 0.597 & 0.602 & 0.597 & 0.589 & 0.563 & 0.733 & 4.612 \\
         & NB & 0.175 & 0.143 & 0.175 & 0.106 & 0.128 & 0.532 & \underline{0.039} \\
         & MLP & 0.618 & 0.618 & 0.618 & 0.613 & 0.573 & 0.747 & 22.292 \\
         & KNN & 0.640 & 0.653 & 0.640 & 0.642 & 0.608 & 0.772 & 0.561 \\
         & RF & \underline{0.752} & \underline{0.773} & \underline{0.752} & \underline{0.749} & \underline{0.736} & \underline{0.824} & 2.558 \\
         & LR & 0.569 & 0.570 & 0.569 & 0.555 & 0.525 & 0.710 & 2.074 \\
         & DT & 0.621 & 0.624 & 0.621 & 0.621 & 0.594 & 0.765 & 0.275 \\
       \cmidrule{2-9}
        \multirow{7}{2cm}{PWM2Vec}
        & SVM &  0.302 & 0.241 & 0.302 & 0.165 & 0.091 & 0.505 & 10011.3 \\
         & NB &  0.084 & \underline{0.442} & 0.084 & 0.063 & 0.066 & \underline{0.511} & 4.565 \\
         & MLP &  \underline{0.310} & 0.350 & \underline{0.310} & 0.175 & 0.107 & 0.510 & 320.555 \\
         & KNN &  0.121 & 0.337 & 0.121 & 0.093 & 0.077 & 0.509 & 2.193 \\
         & RF &  0.309 & 0.332 & 0.309 & \underline{0.181} & 0.110 & 0.510 & 65.250 \\
         & LR & 0.304 & 0.257 & 0.304 & 0.167 & 0.094 & 0.506 & 23.651 \\
         & DT &  0.306 & 0.284 & 0.306 & \underline{0.181} & \underline{0.111} & 0.509 & \underline{1.861} \\
        \cmidrule{2-9} 
        \multirow{7}{2cm}{Autoencoder}
        & SVM & 0.621 & 0.638 & 0.621 & 0.624 & 0.593 & 0.769 & 22.230 \\
             & NB &  0.260 & 0.426 & 0.260 & 0.247 & 0.268 & 0.583 & 0.287 \\
             & MLP & 0.621 & 0.624 & 0.621 & 0.620 & 0.578 & 0.756 & 111.809 \\
             & KNN & 0.565 & 0.577 & 0.565 & 0.568 & 0.547 & 0.732 & \underline{1.208} \\
             & RF & 0.689 & 0.738 & 0.689 & 0.683 & 0.668 & 0.774 & 20.131 \\
             & LR & \underline{0.692} & \underline{0.700} & \underline{0.692} & \underline{0.693} & \underline{0.672} & \underline{0.799} & 58.369 \\
             & DT & 0.543 & 0.546 & 0.543 & 0.543 & 0.515 & 0.718 & 10.616 \\
        
        \cmidrule{2-9}
        \multirow{7}{2cm}{String Kernel}
        & SVM  &  0.618 & 0.617 & 0.618 & 0.613 & 0.588 & 0.753 & 39.791 \\
         & NB   &  0.338 & 0.452 & 0.338 & 0.347 & 0.333 & 0.617 & \underline{0.276} \\
         & MLP  & 0.597 & 0.595 & 0.597 & 0.593 & 0.549 & 0.737 & 331.068 \\
         & KNN  &  0.645 & 0.657 & 0.645 & 0.646 & 0.612 & 0.774 & 1.274 \\
         & RF   &  \underline{0.731} & \underline{0.776} & \underline{0.731} & \underline{0.729} & \underline{0.723} & \underline{0.808} & 12.673 \\
         & LR   &  0.571 & 0.570 & 0.571 & 0.558 & 0.532 & 0.716 & 2.995 \\
         & DT   & 0.630 & 0.631 & 0.630 & 0.630 & 0.598 & 0.767 & 2.682 \\
         \cmidrule{2-9}        
           \multirow{7}{2cm}{WDGRL}  
        & SVM & 0.318 & 0.101 & 0.318 & 0.154 & 0.069 & 0.500 & 0.751 \\
             & NB & 0.232 & 0.214 & 0.232 & 0.196 & 0.138 & 0.517 & \textbf{\underline{0.004}} \\
             & MLP &  0.326 & 0.286 & 0.326 & 0.263 & 0.186 & 0.535 & 8.613 \\
             & KNN & 0.317 & 0.317 & 0.317 & 0.315 & 0.266 & 0.574 & 0.092 \\
             & RF & \underline{0.453} & \underline{0.501} & \underline{0.453} & \underline{0.430} & \underline{0.389} & \underline{0.625} & 1.124 \\
             & LR & 0.323 & 0.279 & 0.323 & 0.177 & 0.095 & 0.507 & 0.041 \\
             & DT & 0.368 & 0.372 & 0.368 & 0.369 & 0.328 & 0.610 & 0.047 \\
             
          \cmidrule{2-9}
        \multirow{7}{2cm}{SeqVec}
        & SVM & 0.656 & 0.661 & 0.656 & 0.652 & 0.611 & \underline{0.791} & 0.891 \\
         & NB & 0.324 & 0.445 & 0.312 & 0.295 & 0.282 & 0.624 & \underline{0.036} \\
         & MLP & 0.657 & 0.633 & 0.653 & 0.646 & 0.616 & 0.783 & 12.432 \\
         & KNN & 0.592 & 0.606 & 0.592 & 0.591 & 0.552 & 0.717 & 0.571 \\
         & RF & 0.713 & \underline{0.724} & 0.701 & 0.702 & \underline{0.693} & 0.752 & 2.164 \\
         & LR & \underline{0.725} & 0.715 & \underline{0.726} & \underline{0.725} & 0.685 & 0.784 & 1.209 \\
         & DT & 0.586 & 0.553 & 0.585 & 0.577 & 0.557 & 0.736 & 0.24 \\
         \cmidrule{2-9}  
        \multirow{1}{1.9cm}{Protein Bert}
         & \_ & 0.542 & 0.580 & 0.542 & 0.514 & 0.447 & 0.675 & 58681.57 \\


        \cmidrule{2-9} 
           \multirow{7}{2cm}{TAPE} 
         & SVM & 0.601 & 0.600 & 0.601 & 0.598 & 0.567 & 0.748 & 18.323 \\
 & NB & 0.231 & 0.334 & 0.231 & 0.224 & 0.213 & 0.558 & 0.340 \\
 & MLP & 0.538 & 0.533 & 0.538 & 0.526 & 0.462 & 0.689 & 7.882 \\
 & KNN & 0.532 & 0.539 & 0.532 & 0.533 & 0.503 & 0.712 & 0.173 \\
 & RF & 0.689 & 0.705 & 0.689 & 0.686 & 0.672 & 0.791 & 42.389 \\
 & LR & 0.544 & 0.548 & 0.544 & 0.519 & 0.456 & 0.683 & 19.215 \\
 & DT & 0.582 & 0.585 & 0.582 & 0.582 & 0.553 & 0.743 & 15.769 \\
 \cmidrule{2-9} 
           \multirow{7}{2cm}{$\phi_{CCP\_NN}$ (ours) - Autoencoder} 
        & SVM & 0.309 & 0.216 & 0.309 & 0.152 & 0.078 & 0.503 & 7.596  \\
        & NB & 0.172 & 0.392 & 0.172 & 0.107 & 0.116 & 0.523 & \underline{0.174}  \\
        & MLP & 0.341 & 0.324 & 0.341 & 0.272 & 0.201 & 0.543 & 218.782  \\
        & KNN & 0.419 & 0.412 & 0.419 & 0.412 & 0.361 & 0.629 & 1.140  \\
        & RF & \underline{\textbf{0.870}} & \underline{\textbf{0.870}} & \underline{\textbf{0.870}} & \underline{\textbf{0.870}} & \underline{\textbf{0.864}} & \underline{\textbf{0.921}} & 19.982  \\
        & LR & 0.309 & 0.280 & 0.309 & 0.152 & 0.077 & 0.503 & 1.911  \\
        & DT & 0.808 & 0.808 & 0.808 & 0.807 & 0.796 & 0.880 & 3.766  \\
        
         \bottomrule
         \end{tabular}
}
\end{table}

The comparison of the best performing proposed method from Table~\ref{tbl_results_part1_human_DNA_data}, i.e., CCP-NN with Autoencoder, with the existing baseline models is shown in Table~\ref{tbl_results_Human_DNA_org} for the Human DNA dataset. We can observe that the proposed method significantly outperforms all baselines for all evaluation metrics other than the training runtime. Specifically, in terms of average accuracy, the CCP-NN with Autoencoder embedding achieves $11.8\%$ improvement compared to the second-best results (i.e., original Spike2Vec with Random Forest classifier).

It is noteworthy that the pre-trained Protein Bert exhibited significantly poorer performance on the Human DNA dataset compared to its performance on the Protein Subcellular and Coronavirus Host datasets. The underlying reason for this discrepancy lies in the fact that the Protein Bert model is designed and trained specifically on molecular sequence data. Consequently, when faced with nucleotide sequences of Human DNA, the model struggles to generalize effectively, leading to its subpar performance. In contrast, the proposed method demonstrated the highest performance among all approaches, outperforming the baseline methods on the Human DNA dataset. This indicates the robustness and efficacy of our proposed method in handling diverse biological sequence data.

Table~\ref{tbl_std_org_human_dna} presents the standard deviation (SD) results (averaged over 5 runs) for the baseline methods and our proposed approach to the Human DNA dataset. The findings reveal that the majority of the standard deviations. values are relatively low, generally below 0.02. This observation indicates that the classification results exhibit consistency across various experimental runs with random train-test splits. The low variability in the SD values highlights the stability of the reported classification performance for both the proposed and baseline models.

\begin{table}[h!]
 \caption{Standard Deviation values of 5 runs for Classification results on the proposed and baseline methods for the \textbf{Human DNA} dataset.}
  \label{tbl_std_org_human_dna}
\centering
\resizebox{0.49\textwidth}{!}{
 \begin{tabular}{p{1.2cm}lp{.8cm}p{.8cm}p{.8cm}p{.8cm}p{.8cm}p{.8cm}p{1.2cm}}
    \toprule
        \multirow{2}{1.2cm}{Embeddings} & \multirow{2}{*}{Algo.} & \multirow{2}{*}{Acc.} & \multirow{2}{*}{Prec.} & \multirow{2}{*}{Recall} & \multirow{2}{0.8cm}{F1 (Weig.)} & \multirow{2}{0.8cm}{F1 (Macro)} & \multirow{2}{0.8cm}{ROC AUC} & Train Time (sec.) \\
        \midrule \midrule
        \multirow{7}{*}{OHE}  
         & SVM & 0.009508 & 0.011373 & 0.009508 & 0.012585 & 0.011098 & 0.004223 & 0.669254 \\
 & NB & 0.137802 & 0.030246 & 0.137802 & 0.121778 & 0.019086 & 0.010169 & 0.023381 \\
 & MLP & 0.013534 & 0.017288 & 0.013534 & 0.01404 & 0.014754 & 0.00558 & 1.455186 \\
 & KNN & 0.00772 & 0.008617 & 0.00772 & 0.008645 & 0.024282 & 0.010772 & 0.215671 \\
 & RF & 0.005861 & 0.014158 & 0.005861 & 0.007086 & 0.012801 & 0.005187 & 0.316404 \\
 & LR & 0.009137 & 0.005479 & 0.009137 & 0.011615 & 0.003251 & 0.001162 & 0.125465 \\
 & DT & 0.008326 & 0.008945 & 0.008326 & 0.008091 & 0.007818 & 0.004461 & 0.100935 \\
  \cmidrule{2-9}	
    \multirow{7}{*}{Spike2Vec}  
     & SVM & 0.02377 & 0.03390 & 0.02377 & 0.02724 & 0.01866 & 0.01151 & 0.57132 \\
 & NB & 0.01037 & 0.04069 & 0.01037 & 0.01829 & 0.01339 & 0.00906 & 0.00889 \\
 & MLP & 0.00321 & 0.00416 & 0.00321 & 0.00427 & 0.00027 & 0.00308 & 6.90869 \\
 & KNN & 0.02056 & 0.02902 & 0.02056 & 0.02313 & 0.02388 & 0.01464 & 0.01192 \\
 & RF & 0.00977 & 0.00783 & 0.00977 & 0.00774 & 0.00247 & 0.00191 & 0.04538 \\
 & LR & 0.00871 & 0.01398 & 0.00871 & 0.01282 & 0.00759 & 0.00395 & 0.03822 \\
 & DT & 0.01748 & 0.01562 & 0.01748 & 0.01666 & 0.02059 & 0.01067 & 0.00957 \\
    \cmidrule{2-9}	
    \multirow{7}{*}{PWM2Vec}  
    & SVM & 0.01751 & 0.02082 & 0.01751 & 0.02085 & 0.01919 & 0.01261 & 0.50484 \\
 & NB & 0.02170 & 0.02921 & 0.02170 & 0.02596 & 0.02258 & 0.01086 & 0.01035 \\
 & MLP & 0.02278 & 0.02402 & 0.02278 & 0.02388 & 0.01552 & 0.00951 & 8.87211 \\
 & KNN & 0.01318 & 0.01611 & 0.01318 & 0.01316 & 0.01085 & 0.00529 & 0.04972 \\
 & RF & 0.02796 & 0.01468 & 0.02796 & 0.02818 & 0.02321 & 0.01597 & 0.05131 \\
 & LR & 0.02276 & 0.02351 & 0.02276 & 0.02586 & 0.02168 & 0.01310 & 0.01008 \\
 & DT & 0.01169 & 0.01753 & 0.01169 & 0.01405 & 0.00937 & 0.00669 & 0.02177 \\
    \cmidrule{2-9}	
    \multirow{7}{*}{Autoencoder}  
     & SVM & 0.007347 & 0.031924 & 0.007347 & 0.007065 & 0.028499 & 0.013436 & 0.434526 \\
 & NB & 0.01508 & 0.019538 & 0.01508 & 0.015753 & 0.024613 & 0.009869 & 0.017177 \\
 & MLP & 0.009726 & 0.013115 & 0.009726 & 0.010498 & 0.021136 & 0.009565 & 1.394414 \\
 & KNN & 0.005735 & 0.005619 & 0.005735 & 0.006138 & 0.037517 & 0.016833 & 0.094046 \\
 & RF & 0.003454 & 0.004077 & 0.003454 & 0.003639 & 0.028805 & 0.01733 & 0.185307 \\
 & LR & 0.011574 & 0.032636 & 0.011574 & 0.013635 & 0.02211 & 0.008987 & 0.27776 \\
 & DT & 0.00641 & 0.008259 & 0.00641 & 0.006691 & 0.036235 & 0.008376 & 0.108608 \\
    \cmidrule{2-9}	
    \multirow{7}{*}{String Kernel}  
 & SVM & 0.00981 & 0.00599 & 0.00981 & 0.00811 & 0.00165 & 0.01293 & 0.09123 \\
 & NB & 0.03791 & 0.05241 & 0.03791 & 0.03597 & 0.03013 & 0.01783 & 0.00059 \\
 & MLP & 0.02417 & 0.06937 & 0.02417 & 0.03168 & 0.03223 & 0.01681 & 4.27668 \\
 & KNN & 0.01701 & 0.01992 & 0.01701 & 0.01928 & 0.01500 & 0.00659 & 0.00283 \\
 & RF & 0.02357 & 0.04091 & 0.02357 & 0.02523 & 0.02875 & 0.01358 & 0.07927 \\
 & LR & 0.00988 & 0.03315 & 0.00988 & 0.01936 & 0.02105 & 0.00606 & 0.00188 \\
 & DT & 0.03202 & 0.03461 & 0.03202 & 0.03338 & 0.03771 & 0.02143 & 0.00912 \\
     \cmidrule{2-9}	
    \multirow{7}{*}{WDGRL}  
     & SVM & 0.003191 & 0.001912 & 0.003191 & 0.002604 & 0.00054 & 0.003424 & 0.040458 \\
 & NB & 0.013202 & 0.015189 & 0.013202 & 0.017859 & 0.015814 & 0.007535 & 0.000162 \\
 & MLP & 0.013313 & 0.021275 & 0.013313 & 0.01608 & 0.015356 & 0.007229 & 2.113733 \\
 & KNN & 0.007286 & 0.010892 & 0.007286 & 0.010094 & 0.008831 & 0.003216 & 0.000944 \\
 & RF & 0.011424 & 0.022522 & 0.011424 & 0.012721 & 0.018023 & 0.008342 & 0.022051 \\
 & LR & 0.003251 & 0.011279 & 0.003251 & 0.004737 & 0.003721 & 0.000956 & 0.0014 \\
 & DT & 0.009034 & 0.011169 & 0.009034 & 0.00988 & 0.015741 & 0.00895 & 0.004752 \\
     \cmidrule{2-9}	
    \multirow{7}{*}{SeqVec}  
      & SVM & 0.004323 & 0.0053 & 0.004323 & 0.002444 & 0.00354 & 0.002263 & 0.249421 \\
 & NB & 0.003381 & 0.007385 & 0.003381 & 0.003252 & 0.002496 & 0.002215 & 0.004853 \\
 & MLP & 0.006487 & 0.00829 & 0.006487 & 0.005008 & 0.003721 & 0.001173 & 1.397002 \\
 & KNN & 0.005323 & 0.005685 & 0.005323 & 0.004541 & 0.002293 & 0.001867 & 0.082383 \\
 & RF & 0.008888 & 0.009345 & 0.008888 & 0.007638 & 0.005948 & 0.003357 & 0.592595 \\
 & LR & 0.00786 & 0.003496 & 0.00786 & 0.006579 & 0.002844 & 0.002397 & 3.962773 \\
 & DT & 0.004853 & 0.006327 & 0.004853 & 0.005272 & 0.005007 & 0.002373 & 0.324555 \\
\cmidrule{2-9}	
    \multirow{1}{*}{Protein Bert}  
 & \_ & 0.02685 & 0.02874 & 0.02547 & 0.02874 & 0.03024 & 0.01774 & 0.00788 \\
     \cmidrule{2-9}	
     \multirow{7}{1.2cm}{TAPE} 
     & SVM & 0.00877 & 0.006212 & 0.00877 & 0.00792 & 0.005215 & 0.002805 & 0.514809 \\
 & NB & 0.020059 & 0.026968 & 0.020059 & 0.022348 & 0.018922 & 0.011747 & 0.0111 \\
 & MLP & 0.017366 & 0.020384 & 0.017366 & 0.018008 & 0.022814 & 0.010171 & 2.903999 \\
 & KNN & 0.006358 & 0.008292 & 0.006358 & 0.007089 & 0.009437 & 0.004215 & 0.031404 \\
 & RF & 0.00615 & 0.006748 & 0.00615 & 0.005848 & 0.009419 & 0.00505 & 5.711691 \\
 & LR & 0.009299 & 0.018944 & 0.009299 & 0.009405 & 0.010728 & 0.006394 & 1.191378 \\
 & DT & 0.01841 & 0.018301 & 0.01841 & 0.018729 & 0.016768 & 0.009301 & 1.218549 \\
 \cmidrule{2-9} 
    \multirow{7}{1.2cm}{$\phi_{CCP\_NN}$ (ours) - Autoencoder} 
      & SVM & 0.009508 & 0.011373 & 0.009508 & 0.012585 & 0.011098 & 0.004223 & 0.669254 \\
 & NB & 0.137802 & 0.030246 & 0.137802 & 0.121778 & 0.019086 & 0.010169 & 0.023381 \\
 & MLP & 0.013534 & 0.017288 & 0.013534 & 0.01404 & 0.014754 & 0.00558 & 1.455186 \\
 & KNN & 0.00772 & 0.008617 & 0.00772 & 0.008645 & 0.024282 & 0.010772 & 0.215671 \\
 & RF & 0.005861 & 0.014158 & 0.005861 & 0.007086 & 0.012801 & 0.005187 & 0.316404 \\
 & LR & 0.009137 & 0.005479 & 0.009137 & 0.011615 & 0.003251 & 0.001162 & 0.125465 \\
 & DT & 0.008326 & 0.008945 & 0.008326 & 0.008091 & 0.007818 & 0.004461 & 0.100935 \\
    \bottomrule
  \end{tabular}
  }
\end{table}

\subsection{Runtime Evaluation}

For all datasets, we additionally report $\%$ improvement for running the $\phi_{CCP\_NN}$ compared to $\phi_{CCP}$ in terms of runtime. For computing the runtime performance gain, we use the following expression: 
\begin{equation}
    \text{\% improvement} \;\;=\;\; \dfrac{\text{$R_{\phi_{CCP}}$ $-$ $R_{\phi_{CCP\_NN}}$}}{\text{$R_{\phi_{CCP}}$}} \times 100
\end{equation}
where $R_{\phi_{CCP}}$ represents the runtime of the original CCP method while $R_{\phi_{CCP\_NN}}$ corresponds to the runtime for our NN-based CCP computation. 

The computational runtime for $R_{\phi_{CCP\_NN}}$ and $R_{\phi_{CCP}}$ along with the performance gain is reported in Tables~\ref{tbl_runtime_Subcellular}, ~\ref{tbl_runtime_host}, and ~\ref{tbl_runtime_Human_DNA} for the Protein Subcellular, Human DNA and Coronavirus Host datasets, respectively.
For protein subcellular data, we can observe that for all $4$ embedding methods as input to CCP and CCP-NN, the performance gain (i.e., percentage improvement) for our CCP-NN is $23.32\%$, $72.34\%$, $55.93\%$, and $92.88\%$, for OHE, Spike2Vec, PWM2Vec, and Autoencoder, respectively.
For the Coronavirus Host dataset, 
We can again observe that in terms of computation runtime performance gain, the proposed CCP-NN significantly outperforms the original CCP by $39.429\%$, $55.950\%$, $97.865\%$, and $93.989\%$ for OHE, Spike2Vec, PWM2Vec, and Autoencoder, respectively.
Similarly, for the Human DNA dataset,
We can observe a $94.385 \%$ (OHE), $90.243\%$ (for Spike2Vec), $91.456 \%$ (PWM2Vec), and $85.425\%$ (for Autoencoder) improvement in computational runtime for CCP-NN compared to CCP.

\begin{figure}[!t]
\centering
\subfloat[Protein Subcellular]{\includegraphics[scale=0.5]{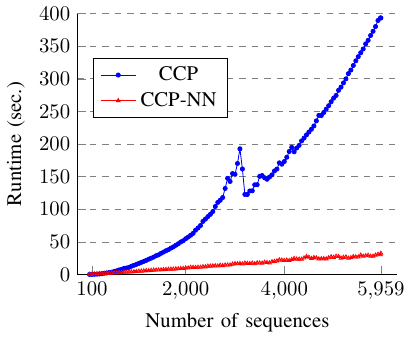}%
\label{fig_embedding_runtime_protein_subcellular}}
\hfil
\subfloat[Coronavirus Host]{\includegraphics[scale=0.5]{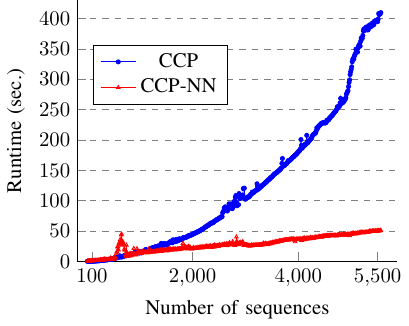}%
\label{fig_embedding_runtime_host}}
\hfil
\subfloat[Human DNA]{\includegraphics[scale=0.5]{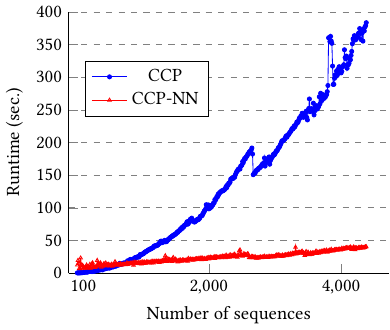}%
\label{fig_embedding_runtime_human_dna}
}
\caption{Runtime for embedding generation of Autoencoder with an increasing number of data points for different datasets. The figure is best seen in color.}
\label{fig_embedding_runtime}
\end{figure}

To observe the increase in runtime with the increasing number of data points (i.e., embeddings), we select the overall best-performing embedding method with CCP and CCP-NN, i.e., Autoencoder, and compute the runtime with an increasing number of embeddings. The runtime results are reported in Figure~\ref{fig_embedding_runtime}. We can observe that the vanilla CCP's runtime increases very quickly as we increase the number of embeddings. On the other hand, the runtime increase for the CCP-NN is very slow, showing its better scalability property.

\begin{table}[h!]
  \caption{Runtime comparison for CCP and CCP-NN using different embedding for \textbf{Protein Subcellular Data}.}
  \label{tbl_runtime_Subcellular}
    \centering
    \resizebox{0.49\textwidth}{!}{
    \begin{tabular}{lp{1.5cm}p{1.5cm}p{2cm}}
    \toprule
        \multirow{3}{*}{Embedding} & $\phi_{CCP\_NN}$ \newline in seconds & $\phi_{CPP}$ \newline in seconds & \% Improvement for $\phi_{CCP\_NN}$ over $\phi_{CPP}$ \\
        \midrule \midrule
        OHE & 1358.408 & 1772.998 & 23.32\% \\ 
        Spike2Vec & 392.002 & 1417.402 & 72.34\% \\
        PWM2Vec & 664.225 & 1507.067 & 55.93\% \\
        Autoencoder & 74.364 & 1058.623 & 92.88\% \\ 
        \bottomrule
    \end{tabular}
  }

\end{table}

\begin{table}[h!]
  \caption{The runtime comparison for CCP and CCP-NN using different embedding for \textbf{Coronavirus Host Data}.}
  \label{tbl_runtime_host}
    \centering
    \resizebox{0.49\textwidth}{!}{
     \begin{tabular}{lp{1.5cm}p{1.6cm}p{2cm}}
    \toprule
        \multirow{3}{*}{Embedding} & $\phi_{CCP\_NN}$ \newline in seconds & $\phi_{CPP}$ \newline in seconds & \% Improvement for $\phi_{CCP\_NN}$ over $\phi_{CPP}$ \\
        \midrule \midrule
        OHE & 2875.668 & 4747.570 & 39.429\% \\ 
        Spike2Vec & 650.596 & 1476.935 & 55.950\% \\ 
        PWM2Vec & 131.138 & 6141.813 & 97.865\% \\ 
        Autoencoder & 44.404 & 738.684 & 93.989\% \\
        \bottomrule
    \end{tabular}
  }
\end{table}

\begin{table}[h!]
  \caption{The runtime comparison for CCP and CCP-NN using different embedding for \textbf{Human DNA Data}.}
  \label{tbl_runtime_Human_DNA}
    \centering
    \resizebox{0.49\textwidth}{!}{
    \begin{tabular}{lp{1.5cm}p{1.6cm}p{2cm}}
    \toprule
        \multirow{3}{*}{Embedding} & $\phi_{CCP\_NN}$ \newline in seconds & $\phi_{CPP}$ \newline in seconds & \% Improvement for $\phi_{CCP\_NN}$ over $\phi_{CPP}$ \\
        \midrule \midrule
        OHE & 13.640 & 242.940 & 94.385 \% \\ 
        Spike2Vec & 27.544 & 282.286 & 90.243\% \\ 
        PWM2Vec & 7.093 & 83.026 & 91.456 \% \\ 
        Autoencoder & 166.852 & 1144.768 & 85.425\% \\
        \bottomrule
    \end{tabular}
  }
\end{table}

\subsection{Statistical Significance}
We used the Student's t-test and calculated the $p$-values using the average and standard deviations (SD) of $5$ runs to determine whether the computed classification results are statistically significant.
Given that SD values are extremely low for all datasets, i.e., $<0.002$ in most cases, we noticed that $p$-values were $< 0.05$ in the majority of cases and for all embedding methods, hence validating the statistical significance of the findings.

\subsection{Discussion}\label{sec_discuss_appendix}
The observed improvement in classification results when using the proposed method over the original CCP method can be attributed to several technical and logical factors.

\paragraph{\textbf{Efficient Nearest Neighbor Search}} The proposed method employs the nearest neighbor (NN) search technique using the AnnoyIndex. Using NN, the algorithm reduces the computational complexity of pairwise distance calculations. In high-dimensional spaces such as biological sequencing data, where feature dimensions can be large, the NN-based approach significantly speeds up the computation, allowing better handling of the complexity and potentially better capturing of correlations between features.

\paragraph{\textbf{Handling High-Dimensional Data}} In sequence classification, feature spaces can often be high-dimensional due to the representation of various attributes. The proposed method can better preserve the feature information in the low-dimensional space than the original CCP due to efficient nearest-neighbor computation, leading to more meaningful density estimations. The nearest neighbor search focuses on capturing local patterns in the data that are particularly relevant for correlations between features. Using the density estimation from an NN-based neighbor search, the method effectively identifies the relationships between relevant features, thus better representing the underlying patterns.

\paragraph{\textbf{Robustness to Noisy Data}} Data may contain noise and outliers, which can negatively impact the accuracy of the correlation estimation. The NN-based method can be more robust to noisy data, as it focuses on local patterns rather than global distances. It implicitly handles noise using neighbor relationships, leading to more reliable density estimates and better classification results.

\paragraph{\textbf{Optimal Feature Clustering}} The proposed method partitions the features into clusters based on their correlation patterns. Using NN-based density estimation, the method identifies more optimal feature clusters, which in turn can enhance the separation of target classes. The ability to detect relevant feature subsets for classification can contribute to improved accuracy. 

In general, the superior classification results obtained with the proposed method can be attributed to its efficient handling of high-dimensional data, the use of NN for nearest neighbor search, enhanced correlation estimation, robustness to noise, and optimal feature clustering. These advantages collectively enable the method to capture the underlying patterns and provide more discriminative representations for improved classification performance, as demonstrated in a variety of datasets, including protein subcellular localization data, Coronavirus host data, and Human DNA in our experiments.

\subsection{Results Comparisons on Gene Expression Datasets}
To evaluate the performance of the proposed $CCP\_NN$ method compared to the original CCP method without relying on any specific type of embeddings, we conducted experiments using two gene expression datasets: Leukemia~\cite{ding2005minimum} and ALL-AML~\cite{golub1999molecular}. 
The Leukemia dataset includes microarray gene expression values from leukemia $72$ patients, where each patient has $7070$ gene expression values. The binary classification task differentiates between ``myeloid" and ``lymphoblastic" leukemia. The ALL-AML dataset consists of $7129$ gene expression profiles of $72$ patients with acute lymphoblastic leukemia (ALL) and acute myeloid leukemia (AML), hence the binary cancer classification task is performed. 

The results are summarized in Tables~\ref{tbl_results_lukemia} and \ref{tbl_results_ALL_AML} for the Leukemia and ALL-AML datasets, respectively. We can observe that the $CCP\_NN$ method demonstrates competitive performance compared to the original CCP method for both datasets. This behavior shows that even without relying on a specific embedding generation method, the proposed method exhibits substantial improvements over the original CCP method, demonstrating its potential in gene expression analysis and classification tasks. The detailed comparisons and results are provided in the tables, where the best values for each method are underlined, and the overall best values are highlighted in bold.

\begin{table}[h!]
    \caption{Classification result comparisons (averaged over $5$ runs) for $\phi_{CCP}$ and $\phi_{CCP\_NN}$  on \textbf{Lukemia Gene Expression} dataset for different evaluation metrics. The best values for each method are underlined, while the overall best values are in bold. 
    The $\uparrow$ means a higher number is better, and the down arrow $\downarrow$ means that a lower number is better.
 }
    \label{tbl_results_lukemia}
\centering
\resizebox{0.49\textwidth}{!}{
 \begin{tabular}{p{1.2cm}lp{0.8cm}p{0.8cm}p{0.8cm}p{0.8cm}p{0.8cm}p{0.8cm}p{1cm}}
    \toprule
        \multirow{2}{1.2cm}{Embeddings} & \multirow{2}{*}{Algo.} & \multirow{2}{*}{Acc. $\uparrow$} & \multirow{2}{*}{Prec. $\uparrow$} & \multirow{2}{*}{Recall $\uparrow$} & \multirow{2}{0.8cm}{F1 (Weig.) $\uparrow$} & \multirow{2}{0.8cm}{F1 (Macro) $\uparrow$} & \multirow{2}{0.8cm}{ROC AUC $\uparrow$} & Train Time (sec.) $\downarrow$\\
        \midrule \midrule
\multirow{7}{1.2cm}{$\phi_{CCP}$} 
 & SVM & \underline{0.909} & 0.923 & \underline{0.909} & \underline{0.910} & \underline{0.902} & 0.911 & \underline{0.002} \\
 & NB & 0.900 & \underline{0.926} & 0.900 & 0.901 & 0.896 & 0.910 & \underline{0.002} \\
 & MLP & 0.882 & 0.907 & 0.882 & 0.882 & 0.874 & 0.889 & 0.117 \\
 & KNN & 0.864 & 0.879 & 0.864 & 0.865 & 0.855 & 0.866 & 0.003 \\
 & RF & 0.900 & 0.925 & 0.900 & 0.901 & 0.895 & \underline{0.912} & 0.188 \\
 & LR & 0.900 & 0.914 & 0.900 & 0.901 & 0.892 & 0.899 & 0.006 \\
 & DT & 0.745 & 0.781 & 0.745 & 0.731 & 0.711 & 0.718 & 0.015 \\
 \cmidrule{2-9} 
           \multirow{7}{1.2cm}{$\phi_{CCP\_NN}$ (ours)} 
 & SVM & 0.691 & 0.508 & 0.691 & 0.582 & 0.408 & 0.489 & 0.003 \\
 & NB & 0.618 & 0.728 & 0.618 & 0.629 & 0.596 & 0.677 & \textbf{\underline{0.001}} \\
 & MLP & 0.809 & 0.851 & 0.809 & 0.798 & 0.737 & 0.738 & 0.152 \\
 & KNN & 0.827 & 0.844 & 0.827 & 0.823 & 0.772 & 0.770 & 0.003 \\
 & RF & 0.936 & 0.947 & 0.936 & 0.938 & 0.919 & 0.944 & 0.210 \\
 & LR & 0.691 & 0.508 & 0.691 & 0.582 & 0.408 & 0.489 & 0.005 \\
 & DT & \textbf{\underline{0.945}} & \textbf{\underline{0.954}} & \textbf{\underline{0.945}} & \textbf{\underline{0.947}} & \textbf{\underline{0.926}} & \textbf{\underline{0.948}} & 0.017 \\
        
         \bottomrule
         \end{tabular}
}
\end{table}

\begin{table}[h!]
    \caption{Classification result comparisons (averaged over $5$ runs) for $\phi_{CCP}$ and $\phi_{CCP\_NN}$  on \textbf{ALL-AML} dataset for different evaluation metrics. The best values for each method are underlined, while the overall best values are in bold.
    The $\uparrow$ means a higher number is better, and the down arrow $\downarrow$ means that a lower number is better.
 }
    \label{tbl_results_ALL_AML}
\centering
\resizebox{0.49\textwidth}{!}{
 \begin{tabular}{p{1.2cm}lp{0.8cm}p{0.8cm}p{0.8cm}p{0.8cm}p{0.8cm}p{0.8cm}p{1cm}}
    \toprule
        \multirow{2}{1.2cm}{Embeddings} & \multirow{2}{*}{Algo.} & \multirow{2}{*}{Acc. $\uparrow$} & \multirow{2}{*}{Prec. $\uparrow$} & \multirow{2}{*}{Recall $\uparrow$} & \multirow{2}{0.8cm}{F1 (Weig.) $\uparrow$} & \multirow{2}{0.8cm}{F1 (Macro) $\uparrow$} & \multirow{2}{0.8cm}{ROC AUC $\uparrow$} & Train Time (sec.) $\downarrow$\\
        \midrule \midrule
\multirow{7}{1.2cm}{$\phi_{CCP}$} 
 & SVM & 0.955 & 0.959 & 0.955 & 0.954 & 0.952 & 0.950 & 0.004 \\
 & NB & 0.936 & 0.940 & 0.936 & 0.936 & 0.931 & 0.932 & \underline{0.003} \\
 & MLP & 0.936 & 0.937 & 0.936 & 0.936 & 0.934 & 0.936 & 0.204 \\
 & KNN & 0.945 & 0.952 & 0.945 & 0.944 & 0.938 & 0.928 & 0.005 \\
 & RF & 0.945 & 0.952 & 0.945 & 0.944 & 0.940 & 0.934 & 0.487 \\
 & LR & \textbf{\underline{0.964}} & \underline{0.967} & \textbf{\underline{0.964}} & \underline{0.963} & \textbf{\underline{0.959}} & \underline{0.951} & 0.009 \\
 & DT & 0.900 & 0.912 & 0.900 & 0.900 & 0.891 & 0.893 & 0.039 \\
 \cmidrule{2-9} 
           \multirow{7}{1.2cm}{$\phi_{CCP\_NN}$ (ours)} 
 & SVM & 0.727 & 0.732 & 0.727 & 0.727 & 0.676 & 0.677 & 0.004 \\
 & NB & 0.673 & 0.690 & 0.673 & 0.678 & 0.628 & 0.636 & \textbf{\underline{0.002}} \\
 & MLP & 0.855 & 0.867 & 0.855 & 0.854 & 0.818 & 0.822 & 0.564 \\
 & KNN & \textbf{\underline{0.964}} & \textbf{\underline{0.971}} & \textbf{\underline{0.964}} & \textbf{\underline{0.965}} & \underline{0.955} & \textbf{\underline{0.969}} & 0.005 \\
 & RF & 0.955 & 0.958 & 0.955 & 0.955 & 0.946 & 0.951 & 0.465 \\
 & LR & 0.691 & 0.483 & 0.691 & 0.567 & 0.407 & 0.500 & 0.009 \\
 & DT & 0.945 & 0.950 & 0.945 & 0.945 & 0.932 & 0.935 & 0.045 \\
        
         \bottomrule
         \end{tabular}
}
\end{table}

\section{Conclusion}\label{sec_C}
In this study, we addressed the challenges of analyzing molecular sequence data, which involves a large number of sequences and complex protein structures. We proposed an efficient and fast method called Nearest Neighbor Correlated Clustering and Projection (CCP-NN). The CCP-NN method is based on the original Correlated Clustering and Projection (CCP) technique, but it incorporates an NN search for computing the density map and correlations. 
Through a series of experimental evaluations, we compared the performance of CCP and CCP-NN in classifying molecular sequences. The results demonstrated that CCP-NN outperforms CCP in terms of classification accuracy while also reducing computational runtime. 
Future work includes further enhancements to the CCP-NN framework, such as incorporating additional information or integrating it with other machine-learning methods for comprehensive analysis of sequences.
Exploring other methods to determine the optimal number of neighbors is also an interesting future extension, which could include methods like adaptive neighbor selection based on local density estimation, the Elbow method, and Stability analysis. 
Future research aiming at biological interpretability, such as combining k-NN with interpretable models and using post-hoc interpretability frameworks like SHAP or LIME, may be compelling. 

\bibliographystyle{IEEEtran}
\bibliography{references}


\begin{IEEEbiography}
[{\includegraphics[width=1in,height=1.25in,clip,keepaspectratio]{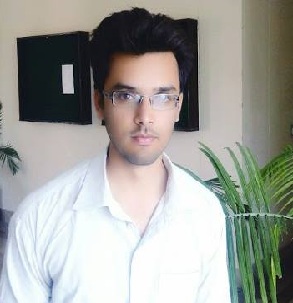}}]
{Sarwan Ali}
is a postdoctoral researcher at Irving Medical Center, Columbia University, New York, USA, working in the fields of Bioinformatics, Data mining, Big data, and Machine Learning. 
\end{IEEEbiography}
\begin{IEEEbiography}
[{\includegraphics[width=1in,height=1.25in,clip,keepaspectratio]{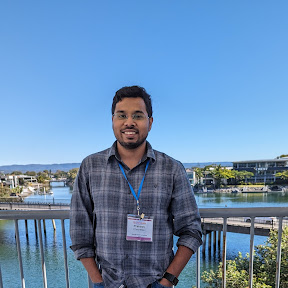}}]
{Prakash Chourasia}
is a Ph.D. student at the Department of Computer Science, Georgia State University, working in the fields of bioinformatics, EHR, Federated Learning, and Machine Learning.
\end{IEEEbiography}
\begin{IEEEbiography}
[{\includegraphics[width=1in,height=1.25in,clip,keepaspectratio]{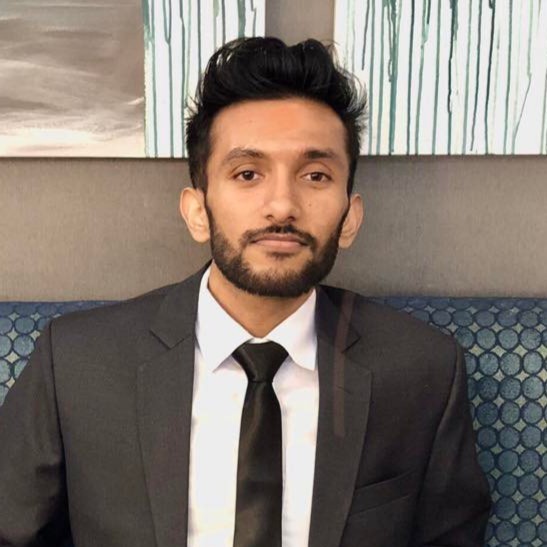}}]
{Bipin Koirala}
is a Ph.D. student at the School of Aerospace Engineering, Georgia Institute of Technology, working in the fields of Statistics, Optimization, and Machine Learning.
\end{IEEEbiography}
\begin{IEEEbiography}
[{\includegraphics[width=1in,height=1.25in,clip,keepaspectratio]{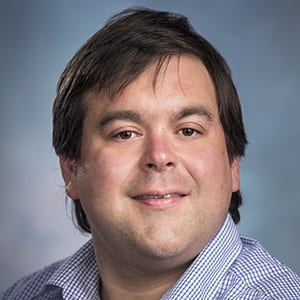}}]
{Murray Patterson}
is an Assistant Professor at Georgia State University working in the fields of Bioinformatics, Computational Biology, Algorithms, and Combinatorics.
\end{IEEEbiography}

\end{document}